\pdfoutput=1
\documentclass{JHEP3}
\usepackage{amsmath}
\usepackage{epsfig}
\usepackage{url}
\usepackage{mathdots}

\def\({\left(}
\def\){\right)}
\def\[{\left[}
\def\]{\right]}
\def\<{\langle}
\def\>{\rangle}
\def\lv{\left|}
\def\rv{\right|}
\def\os{\overset}
\def\und{\underline}
\def\ap{\alpha}
\def\bt{\beta}
\def\gm{\gamma}
\def\de{\delta}
\def\De{\Delta}

\newcommand{\be}{\begin{equation}}
\newcommand{\ee}{\end{equation}}
\newcommand{\bal}{\begin{aligned}}
\newcommand{\eal}{\end{aligned}}

\newcommand{\labell}[1]{\label{#1}}

\title{Positivity, Grassmannian Geometry and Simplex-like Structures of Scattering Amplitudes}

\author{Junjie Rao$^a$\footnote{Email: raojunjie@zju.edu.cn}\\
{$^a$Zhejiang Institute of Modern Physics, Zhejiang University, Hangzhou, 310027, P. R. China}}

\abstract{This article revisits and elaborates the significant role of positive geometry of momentum twistor Grassmannian
for planar $\mathcal{N}\!=\!4$ SYM scattering amplitudes. First we establish the fundamentals of positive Grassmannian
geometry for tree amplitudes, including the ubiquitous Pl\"{u}cker coordinates and the representation of
reduced Grassmannian geometry. Then we formulate this subject,
without making reference to on-shell diagrams and decorated permutations, around these
four major facets: 1. Deriving the tree and 1-loop BCFW recursion relations solely from positivity, after introducing
the simple building blocks called positive components for a positive matrix.
2. Applying Grassmannian geometry and Pl\"{u}cker coordinates to determine the signs in N$^2$MHV homological identities,
which interconnect various Yangian invariants.
It reveals that most of them in fact reflect the secret incarnation of the simple
6-term NMHV identity. 3. Deriving the stacking positivity relation, which is powerful for parameterizing
matrix representatives in terms of positive variables in the $d\log$ form.
It will be used with the reduced Grassmannian geometry representation
to produce the positive matrix of a given geometric configuration,
which is an independent approach besides the combinatoric way involving a sequence of BCFW bridges.
4. Introducing an elegant and highly refined formalism of BCFW recursion relation
for tree amplitudes, which reveals its two-fold simplex-like structures. First, the BCFW contour in terms of (reduced)
Grassmannian geometry representatives is delicately dissected into a triangle-shape sum, as only a small fraction
of the sum needs to be explicitly identified. Second, this fraction can be further
dissected, according to different growing modes with corresponding growing parameters. The growing modes possess
the shapes of solid simplices of various dimensions,
with which infinite number of BCFW cells can be entirely captured by the characteristic objects
called fully-spanning cells. We find that for a given $k$, beyond $n\!=\!4k\!+\!1$ there is no more new
fully-spanning cell, which signifies the essential termination of the recursive growth of BCFW cells. As $n$ increases
beyond the termination point,
the BCFW contour simply replicates itself according to the simplex-like patterns, which enables us to master
all BCFW cells once for all without actually identifying most of them.}


\begin{document}
\maketitle

\section{Introduction}

\subsection{Review}

$\mathcal{N}\!=\!4$ super Yang-Mills theory has been the most understood quantum field theory so far.
In recent years, tremendous progress
on the scattering amplitudes of planar $\mathcal{N}\!=\!4$ SYM was made through its connection to positive Grassmannian,
momentum twistors and on-shell diagrams
\cite{ArkaniHamed:2012nw, ArkaniHamed:2010kv, ArkaniHamed:2009vw, ArkaniHamed:2009dn}.
Among these subjects, the BCFW recursion relation to all loop orders
in momentum twistor space \cite{ArkaniHamed:2010kv}
has been the pillar of this arena, which is an elegant generalization of the well known BCFW
recursion relation \cite{Britto:2004ap, Britto:2005fq} in terms of massless spinors.
The power of this efficient machinery for generating tree amplitudes and loop integrands
is enhanced by the beautiful object known as momentum twistor \cite{Hodges:2009hk},
and the supersymmetric version of momentum twistor manifests
dual superconformal invariance of $\mathcal{N}\!=\!4$ SYM. Besides its formal merits, super momentum twistors
greatly trivialize calculational manipulations that could be tedious in massless spinor space.
Explicitly, at 1-loop and 2-loop orders there exist closed-form formulas of BCFW recursion relation which involve
the ``kermit'' expansion \cite{Bourjaily:2013mma, Bourjaily:2015jna}. In fact, they serve as numerical checking means
of the local expansion of loop integrands which manifests generalized unitarity. The local representation of
loop integrands originates from \cite{ArkaniHamed:2010kv, ArkaniHamed:2010gh}. Its benefits include the manifest locality
as there is a uniform spurious pole only, and the manifest separation of convergent and divergent contributions.
Its price is that generalized unitarity results in the presence of complicated algebraic (or even transcendental)
functions, while loop integrands by BCFW recursion relation are always simply rational.
There are efficient \textsc{Mathematica} packages to implement the results above: ``\verb"positroids"'' for various aspects
of positive Grassmannian cells \cite{Bourjaily:2012gy}, ``\verb"loop amplitudes"'' and ``\verb"two loop amplitudes"''
for 1-loop and 2-loop integrands, amplitudes and their regularization \cite{Bourjaily:2013mma, Bourjaily:2015jna}.

However, in this article we would like to emphasize geometric aspects around the key mathematical object
known as positive Grassmannian, \textit{without} making reference to on-shell diagrams and (decorated) permutations.
In massless spinor space, the state-of-the-art BCFW recursion relation is expressed first via on-shell diagrams, then
compactly encoded by permutations. To convert these combinatoric data back to rational functions of spinors,
the canonical way is to use a sequence of BCFW bridges to decompose the permutation into ``adjacent''
transpositions and the resulting permutation is a (decorated) identity. Then we reverse the chain of BCFW shifts and
construct its matrix representative, starting from this identity.
We will see this way could be replaced by another independent approach that also reaches the same goal,
while it works for momentum twistor Grassmannian directly. What's more, the BCFW recursion relation is now implemented
in the positive matrix form following the default recursion scheme \cite{Bourjaily:2010wh},
which is from geometry to geometry.
In fact, once we have established the bijection between Grassmannian geometric
configurations and positive matrix representatives, we can forget momentum twistors
since Grassmannian geometry has much more advantages for formal purposes. For example,
homological identities are better understood and we can determine their signs more intuitively (even though
a systematic recipe as that in \cite{Olson:2014pfa} still needs further investigation).
This geometric representation can free us from the cumbersome intersection symbols of momentum twistors that result from
solving the orthogonal constraint $C\!\cdot\!Z\!=\!0$, or equivalently, from BCFW recursion relation, neither of which stay
invariant under the GL$(k)$ gauge transformation of Grassmannian.
To not specify the explicit expressions in terms of momentum twistors is a more unbiased choice
so that we can use a unique, invariant and compact representative for each cell. Restoring
the invisible GL$(k)$ invariance proves to be convenient even though the Grassmannian auxiliary variables
have been replaced by the solutions to equations $C\!\cdot\!Z\!=\!0$. But why don't we use permutations?
Indeed, one may use permutations, but those are in massless spinor space and we need to translate them to
permutations in momentum twistor space. Or, one can use the momentum twistor diagrams instead of on-shell diagrams
which will generate the corresponding permutations \cite{Bai:2014cna, Bai:2015qoa}, but that is another story.
In fact, a permutation is slightly less invariant than its geometric counterpart since it actually employs
a special gauge choice to construct the matrix representative, while the latter admits any choice that is not singular
(for example, if $k\!=\!3$ and $(123)\!=\!0$, one cannot fix columns $1,2,3$ to be a unit matrix).
Also, it is much easier to obtain a permutation from the geometry and read off its dimension,
than the reverse operation which is indirectly realized via its matrix representative.

Since we refrain from using diagrams and combinatorics, it is better to develop an independent geometric formalism
in momentum twistor space directly. In that context, the momentum twistor recursion without manifest positivity
is united with the diagrammatic recursion without manifest dual superconformal invariance.
After that, we may only mention geometric configurations, as no quantities depending on spinors are needed and even
momentum twistors show up in very limited occasions. The entire machinery is
\textit{from geometry to geometry}, manifesting positivity, dual superconformal invariance and the auxiliary
GL$(k)$ invariance simultaneously. It will later reveal a new vision of amplitudes consisting of the
elegant simplex-like structures, which are hardly possible to be conceived with solely momentum twistors.
This is an extension continuing the same logic of
\cite{Drummond:2008cr, ArkaniHamed:2009dg, Bourjaily:2010kw}.

Let's stop digressing from the review. The most recent geometric interpretation of amplitudes is the
amplituhedron \cite{Arkani-Hamed:2013jha}, which goes further than the Grassmannian geometry.
There, positivity is supposed to completely replace
recursion for calculating amplitudes, but so far its ambition has been only realized for
4-particle MHV integrands to all loop orders \cite{Arkani-Hamed:2013kca, Franco:2014csa, Galloni:2016iuj}.
When we say it is complete, it means that there positivity is able to reproduce what recursion has brought us in
a more efficient way. Even if conceptually it is enlightening, its generalization to generic $n$ and $k$
cannot be yet considered as satisfying, since we can achieve all of what it promises in the more
``primitive'' momentum twistor recursion without even talking about Grassmannian.
Besides summing all the projective ``volumes'' of non-overlapping positive regions,
positivity can be used in the other way
which nullifies all possible spurious poles \cite{Arkani-Hamed:2014dca}.
The amplituhedron can be also understood from the genuine perspective of volumes \cite{Ferro:2015grk}
(and its connection to some special class of differential equations),
which extends the original proposal of polytopes \cite{ArkaniHamed:2010gg}.
But in this article, we will follow a slightly more conservative direction that only takes Grassmannian into account.

There is a great difference of simplicity and calculational capability between a formulation that uses momentum twistors
and that does not. Without this beautiful object, further developments, such as the amplituhedron, would be hardly possible.
For $\mathcal{N}\!=\!4$ SYM, a large amount of formal advances
including the use of momentum twistor crucially relies on its planarity (or the color ordering),
nevertheless, there is still remarkable progress of the generalizations to non-planar $\mathcal{N}\!=\!4$ SYM
\cite{Arkani-Hamed:2014via, Arkani-Hamed:2014bca, Franco:2015rma, Chen:2014ara, Chen:2015qna, Chen:2015bnt,
Bern:2015ple, Bourjaily:2016mnp}
and even to $\mathcal{N}\!=\!8$ supergravity \cite{Bern:2014kca, Heslop:2016plj, Herrmann:2016qea},
as well as $\mathcal{N}\!=\!6$ ABJM theory \cite{Elvang:2014fja}
and $\mathcal{N}\!<\!4$ SYM theories \cite{Benincasa:2015zna, Benincasa:2016awv}.

For reviews and foundations of $\mathcal{N}\!=\!4$ SYM scattering amplitudes,
readers may refer to \cite{Elvang:2013cua, Feng:2011np, Henn:2014}.
For more mathematical aspects of on-shell diagrams and positive Grassmannian, readers may refer to \cite{Franco:2013nwa}
and \cite{Williams:2003, Postnikov:2006kva, Knutson:2011, Postnikov:2013, Lam:2015uma,
Karp:2015duv, Karp:2016uax, Griffiths:1978}.
Finally, we will also mention the Narayana numbers when counting tree BCFW cells,
and the relevant combinatorics can be found in \cite{Petersen:2015}.

\subsection{Overview as an introductory quick tour}

As one of the simplest examples, the well known NMHV $n\!=\!6$ amplitude (or Yangian invariant precisely)
is represented by
\be
Y^1_6=[12345]+[12356]+[13456],
\ee
where each term is called a 5-bracket, which manifests dual superconformal invariance via making use of super
momentum twistors. However, we have surprisingly found that, it is much more advantageous to rewrite amplitudes
in terms of ``empty slots'', namely those null (or absent) labels. Explicitly, we have
\be
Y^1_6=[6]+[4]+[2]. \labell{eq-1}
\ee
For instance, above we write $[6]$ in stead of $[12345]$ as the latter's complement, so naturally these brackets
are different from the old 5-brackets. The philosophy of characterizing amplitudes or cells by null objects can be
extended to all N$^k$MHV sectors. A further explicit example is the N$^2$MHV $n\!=\!7$ amplitude
\be
Y^2_7=[7]+[5]+[2]+(23)(45)+(23)(67)+(45)(71), \labell{eq-2}
\ee
note that $[7]$ in fact denotes $Y^2_6$, a top cell after removing the 7th column, and so is $[5]$ or $[2]$,
except for a different label of the null column. For the latter three cells, we impose two vanishing constraints on the
$2\!\times\!2$ minors. For example, $(23)(45)$ imposes $(23)\!=\!(45)\!=\!0$ to ensure the dimension of this cell is $4k$,
which geometrically means columns $2,3$ and columns $4,5$ are linearly dependent.
Now, the ``null objects'' are the vanishing minors while the unmentioned content is free as it is,
so the rest ordered minors remain positive.
We name such a representation of cells as the \textit{Grassmannian geometry}.

In terms of Grassmannian geometry representatives, it is convenient to express homological identities.
Compared to (decorated) permutations in either massless spinor or momentum twistor space, it is transparent to read off the
dimensions of all relevant cells, and to judge whether they belong to the boundaries of the
$(4k\!+\!1)$-dimensional cell that generates the identity. An example is the N$^2$MHV $n\!=\!9$ identity
\be
\bal
0=\partial(1234)_2(567)_2=&-[1](234)_2(567)_2+(134)_2[2](567)_2-(124)_2[3](567)_2+(123)_2[4](567)_2\\
&-(1234)_2(5678)_2+(1234)_2(567)_2(89).
\eal
\ee
One can easily check that, all six terms above are indeed the boundaries of generator $(1234)_2(567)_2$
with correct dimensions. More curiously, as we will explore by using Pl\"{u}cker coordinates and choosing
different gauges of Grassmannian matrices for different boundaries,
this identity is in fact the secret incarnation of the simple 6-term NMHV identity
\be
0=-\,[1]+[2]-[3]+[4]-[5]+[6],
\ee
which manifests the cyclic invariance of NMHV $n\!=\!6$ amplitude \eqref{eq-1}, and it is the only distinct NMHV identity.
Note that both identities share the same pattern of six alternating signs.

For $k\!\geq\!3$, it is beneficial to introduce the \textit{reduced Grassmannian geometry}. For example, one of the
N$^3$MHV $n\!=\!9$ BCFW cells is
\be
(4\,\os{6}{\os{|}{5}}\,7)\,(8\,\os{1}{\os{|}{9}}\,2) \labell{eq-13}
\ee
which is equivalent to the more unwieldy representation
\be
\(\begin{array}{c}
(91)(2)(3)(4)(56)(7)(8) \\
(23)(34)(4567)(78)(8912)
\end{array}\).
\ee
In words, the reduced geometry separates vanishing maximal minors and degenerate linear dependencies.
For example, $(457)$ is a vanishing maximal minor while $(56)$ denotes the degenerate linear dependence of columns $5,6$.
With the reduced geometry, we do not have to specify every layer of linear dependencies as above, since all the information
is encapsulated in this compact notation.

The reduced Grassmannian geometry is useful for both checking the dimension of a given cell of $k\!\geq\!3$
and parameterizing its matrix representative in terms of positive variables in the $d\log$ form.
The latter is realized by the \textit{stacking positivity relation}, which literally tells:\\

\textbf{For a fixed $n$, a $k$-row positive Grassmannian matrix can be constructed by stacking a row on top of
a $(k\!-\!1)$-row positive Grassmannian matrix, and then imposing all $(n\!-\!k\!+\!1)$ consecutive
$k\!\times\!k$ minors to be positive.}\\

This relation provides a recursive approach to forge ``larger positivity'' from ``smaller positivity'',
by stacking more rows on a known positive matrix and imposing positivity at each step.
It turns out to be a practical recipe for parameterizing the positive matrix representative,
with a proper gauge chosen. As we will see, the reduced geometry is mandatory for its realization.

Combining the reduced Grassmannian geometry and the stacking positivity relation,
we successfully establish a bijection between Grassmannian geometric configurations and positive matrix representatives.
The latter can be converted to functions of super momentum twistors by solving the equations $C\!\cdot\!Z\!=\!0$ row-wise.
Though we can use the traditional recursion to identify them without mentioning Grassmannian geometry,
it is yet desirable to connect these two seemingly different aspects.
We will later elaborate how positivity itself helps derive
the geometric recursion relation in the matrix form
for generating all BCFW cells of tree amplitudes and 1-loop integrands.

For understanding the \textit{global} structure of tree BCFW contour beyond the well known factorization limits,
we can first honestly generate all BCFW cells (following the default recursion scheme).
To have a peek into this beautiful structure,
one can already find evidence in the simple examples \eqref{eq-1} and \eqref{eq-2}.
For \eqref{eq-1}, the triple $(6,4,2)$
represents the quadratic \textit{growing mode} or 2-mode for short, and $6,4,2$ are its
\textit{growing parameters}. To see how the NMHV amplitudes ``grow'',
for instance, let's rearrange all ten cells of the NMHV $n\!=\!8$ amplitude as
\be
Y^1_8=
\(\begin{array}{cccc}
{} & {} & {} & [234] \\
{} & {} & [238] & [236] \\
{} & [278] & [258] & [256] \\
{[678]} & [478] & [458] & [456]
\end{array}\), \labell{eq-5}
\ee
where $Y^1_8$ is the sum of all entries in the ``triangle'' above. Obviously,
this reflects a simple pattern which will be
later identified as a solid 2-simplex, and the triple $(6,4,2)$ uniquely determines the NMHV
growing pattern for any $n$. This triangle-like dissection of amplitudes can be generalized to all N$^k$MHV sectors.
A further explicit example is the N$^2$MHV $n\!=\!9$ amplitude
\be
Y^2_9=
\(\begin{array}{cccc}
{} & {} & {} & \!\!\!\!\!\![234]~~~~~~~~ \\
{} & {} & \!\!\!\!\![239]~~~~~~ & \!\!\!\!\!\![23]I_{7,3}~~~ \\
{} & \!\!\!\![289]~~~~~ & \!\!\!\!\![29]I_{7,2}~ & \!\!\!\!\!\![2]I_{8,2}~~ \\
\,{[789]}~~~\, & \!\!\!\![89]I_{7,1} & \!\!\!\!\![9]I_{8,1} & \!\!\!\!\!I_{9,1}
\end{array}\!\!\!\!\!\!\!\!\!\), \labell{eq-3}
\ee
where quantities in the bottom row, namely $I_{i,1}$, are the only essential objects to be identified.
Once $I_{i,1}$ is known, it is trivial to obtain $I_{i,1+j}$ by performing a \textit{partial} cyclic shift $i\!\to\!i\!+\!j$
except that label 1 is fixed, for all BCFW cells within $I_{i,1}$. From \eqref{eq-2} we have already known
$I_{7,1}$, namely
\be
I_{7,1}=
\(\begin{array}{cc}
\left\{\begin{array}{c}
\!(45)(71) \\
\!{[5]}~~~~
\end{array} \right. &
(23)\left\{\begin{array}{c}
\!(67) \\
\!(45)~~~~~~
\end{array} \right.
\end{array}\!\!\!\!\!\!\!\),
\ee
here and in \eqref{eq-3}, we have implied the multiplication rule of vanishing constraints or linear dependencies,
which is simply a superposition of all constraints. According to the shift above, we immediately get
\be
I_{7,2}=
\(\begin{array}{cc}
\left\{\begin{array}{c}
\!(56)(81) \\
\!{[6]}~~~~
\end{array} \right. &
(34)\left\{\begin{array}{c}
\!(78) \\
\!(56)~~~~~~
\end{array} \right.
\end{array}\!\!\!\!\!\!\!\)
\ee
for example. It is easy to imagine that, this simple rule saves a large amount of repetitive calculation of the honest
BCFW recursion relation.

On the other hand, from \eqref{eq-2}, though it is obvious that $Y^2_7$ is a ``sub-triangle'' of $Y^2_9$
decorated with additional empty slots $[89]$,
one can also manipulate this sum of BCFW cells to reveal another different structure. It is
the class of anti-NMHV amplitudes for which $n\!=\!k\!+\!5$. Since anti-MHV amplitudes are of $n\!=\!k\!+\!4$
and for each $k$ there is only one BCFW cell (a top cell), the first non-trivial amplitudes are naturally
of $n\!=\!k\!+\!5$. To see its structure, let's rearrange
\be
Y^2_7=
\(\begin{array}{cc}
{} & [2] \\
{[7]} & I_{7,1}
\end{array}\) \labell{eq-4}
\ee
where $I_{7,1}$ packs up four terms, into the ``anti-NMHV triangle'' form
\be
Y^2_7=
\(\begin{array}{ccc}
[2]~ &
(23)\left\{\begin{array}{c}
\!(67) \\
\!(45)~~~~~~
\end{array} \right. &
\!\!\!\!\!\!\left\{\begin{array}{c}
\,~~~[7] \\
\!(45)(71) \\
\!{[5]}~~~~
\end{array} \right.
\end{array}\!\!\). \labell{eq-32}
\ee
For comparison, we write the N$^3$MHV $n\!=\!8$ amplitude in the similar form
\be
Y^3_8=
\(\begin{array}{cccc}
[2]~ &
(23)\left\{\begin{array}{c}
\!(678) \\
\!(456)~~~~~~
\end{array} \right. &
\!\!\!\!\!\!(234)\left\{\begin{array}{c}
~~~(78) \\
\!(456)(781)~~ \\
\!\!(56)~~~~~~~
\end{array} \right. &
\!\!\!\!\left\{\begin{array}{c}
~~~[8] \\
\!(456)(81)~~ \\
\,(56)(812) \\
\!\!{[6]}~~~~
\end{array} \right.
\end{array}\!\!\!\), \labell{eq-11}
\ee
and this trend continues for all anti-NMHV amplitudes. It is not surprising that both NMHV and anti-NMHV amplitudes
have triangle-shape patterns since they are parity conjugate,
even though this conjugation is not manifest in momentum twistor space.

As expected, $Y^3_8$ in the form above can be also rearranged similarly as \eqref{eq-4}, namely
\be
Y^3_8=
\(\begin{array}{cc}
{} & [2] \\
{[8]} & J_{8,1}
\end{array}\), \labell{eq-44}
\ee
where
\be
J_{8,1}=
\(\begin{array}{ccc}
(23)\left\{\begin{array}{c}
\!(678) \\
\!(456)~~~~~~
\end{array} \right. &
\!\!\!\!\!\!(234)\left\{\begin{array}{c}
~~~(78) \\
\!(456)(781)~~ \\
\!\!(56)~~~~~~~
\end{array} \right. &
\!\!\!\!\left\{\begin{array}{c}
\!(456)(81)~~ \\
\,(56)(812) \\
\!\!{[6]}~~~~
\end{array} \right.
\end{array}\!\!\!\),
\ee
and the N$^3$MHV triangle grows similarly as \eqref{eq-3}. Now an amazing fact is that objects after the
triangle-like dissection, such as $I_{i,1}$ in \eqref{eq-3}, can be further dissected! There is a huge redundancy
among the cells within $I_{i,1}$ because each time we increase $i$ by one, only the objects called
\textit{fully-spanning cells}, or full cells for short, need to be identified. In fact, it can be shown that
\be
I_{i,1}=(\textrm{descendent cells from }I_{i-1,1})+(\textrm{new full cells for }n\!=\!i),
\ee
where the cells descend from the preceding $I_{i-1,1}$ follow simple patterns of the ``solid simplices'',
which are generalizations of the solid 2-simplex in \eqref{eq-5}.
The numbers of full cells are finite since beyond $n\!=\!4k\!+\!1$ there exists no more new topology of such cells.
After all full cells are identified with their growing modes and parameters known,
the entire family of amplitudes for any $n$ is understood once for all!

In brief, for a given $k$ we start with the anti-NMHV amplitude, then we identify all full cells
and their simplex-like growing patterns as $n$ increases, until $n\!=\!4k\!+\!1$. After that, amplitudes of any larger $n$
can be quickly produced, according to the elegant two-fold simplex-like structures: the triangle-like dissection and
the simplex-like growing patterns of full cells. We no more need the laborious yet
repetitive recursion relation by factorization limits.

\subsection{Outline}

The preceding overview provides an introductory quick tour of the article. Now we will outline the major
four facets of this subject, as well as the overall organization.
\\ \\
Section \ref{sec2}: We establish the fundamentals of positive Grassmannian geometry
for tree amplitudes without making reference to on-shell diagrams and decorated permutations,
including the extensive use of Pl\"{u}cker coordinates and the reduced Grassmannian geometry representation.\\
Section \ref{sec3}: We introduce the positive components as simple building blocks for a positive matrix,
and then derive the tree and 1-loop BCFW recursion relations solely from positivity.\\
Section \ref{sec4}: We apply Grassmannian geometry and Pl\"{u}cker coordinates to determine most of the signs in
N$^2$MHV homological identities. They are shown to reflect the incarnation of the simple NMHV identity.\\
Section \ref{sec5}: We derive the stacking positivity relation, and then demonstrate how to use it to
parameterize the matrix representatives in terms of positive variables,
with the aid of reduced Grassmannian geometry. This serves as an independent approach of generating
positive matrices for BCFW cells. We also confirm its correctness
by numerically checking an N$^3$MHV $n\!=\!8$ identity derived in the previous section.\\
Section \ref{sec6}: We introduce a refined formalism for the tree BCFW recursion relation,
which possesses the exotic two-fold simplex-like structures: the triangle-like dissection of BCFW contour
and the simplex-like growing patterns of full cells.
There exists a termination point of full cells at $n\!=\!4k\!+\!1$,
which signifies all essential objects have been identified,
and since then the amplitudes of a given $k$ are known for any $n$.
Two classes of amplitudes are elaborated for demonstration of this formalism:
the N$^2$MHV family which terminates at $n\!=\!9$, and the N$^3$MHV family which terminates at $n\!=\!13$.
\\ \\
In brief, the major four facets of positive Grassmannian geometry are:\\
1. How to deduce tree and 1-loop BCFW recursion relations from positivity.\\
2. How to (at least partly) determine the signs in homological identities.\\
3. How to parameterize positive matrices of (reduced) Grassmannian geometry representatives.\\
4. The two-fold simplex-like structures of tree BCFW contour.
\\ \\
\textbf{Having a glance at the fundamentals of positive Grassmannian geometry in section \ref{sec2},
readers may disregard the order above and skip to any interested
section of \ref{sec3}, \ref{sec4}, \ref{sec5} or \ref{sec6},
as the organization of this article is tailored such that each specific section is as self-contained as possible.}

\newpage
\section{Fundamentals of Positive Grassmannian Geometry}
\label{sec2}

This section establishes the minimal ingredients of positive Grassmannian geometry for tree amplitudes,
providing the fundamental concepts and techniques necessary for the posterior four specific sections.

\subsection{Momentum twistor Grassmannian, Yangian invariants and their geometric avatars}

The super momentum twistor $\mathcal{Z}_a^I=(Z_a^I,\eta_a^A)$ generalizes
the bosonic momentum twistor $Z_a^I$ to fit it into a
theory which enjoys the dual superconformal invariance,
and $\eta_a^A$ is a fermionic object in the fundamental representation
of $SU(4)$ for $\mathcal{N}\!=\!4$ SYM in our context \cite{ArkaniHamed:2009vw}.
For the generic N$^k$MHV $n$-particle sector, we introduce an auxiliary Grassmannian $C_{\ap a}\!\in\!G(k,n)$
which by definition owns the GL$(k)$ gauge invariance.
Now, we want to describe an N$^k$MHV $n$-particle dual superconformally invariant object
in terms of super momentum twistors through its auxiliary avatar $C$.
First, to make $C$ interact with the kinematical data,
all $4k$-dimensional cells of $C$ must obey the orthogonal constraint
\be
C_{\ap a}Z_a^I=0, \labell{eq-6}
\ee
which has in total $4k$ equations.
Then, each particular cell corresponds to a \textit{Yangian invariant} \cite{ArkaniHamed:2012nw}
via the Grassmannian contour integration
\be
\int_\textrm{some contour}
\frac{d^{k\times n}C}{\textrm{vol(GL($k$))}}\frac{1}{(12\ldots k)(23\ldots k\!+\!1)\ldots(n\,1\ldots k\!-\!1)}
\times\de^{4k|4k}(C_{\ap a}\mathcal{Z}_a) \labell{eq-7}
\ee
where `some contour' imposes $k(n\!-\!k\!-\!4)$ constraints, so that this integral has zero degree of freedom.
Now, each cell is assigned with a geometric configuration by the contour, see \eqref{eq-2} for example.
The rest $4k$-dimensional actual integration can be done by solving \eqref{eq-6},
which might however become cumbersome because of
the resulting intersection symbols $\cap$ of momentum twistors. To see this, let's first review the
building blocks known as \textit{5-brackets}.

A 5-bracket is the simplest Yangian invariant, but it only manifests dual superconformal invariance in fact.
For $k\!=\!1$, $n\!=\!5$, \eqref{eq-7} can be trivially solved and integrated, the resulting quantity is
\be
\frac{\de^{0|4}(\<1234\>\eta_5+\<2345\>\eta_1+\<3451\>\eta_2+\<4512\>\eta_3+\<5123\>\eta_4)}
{\<1234\>\<2345\>\<3451\>\<4512\>\<5123\>}\equiv[12345].
\ee
It is antisymmetric, which makes it cyclic, and under color reflection it is even:
\be
[12345]=-\,[21345]\Longrightarrow[12345]=[23451]=[54321].
\ee
It is projective, namely invariant under the rescaling of any single super momentum twistor:
\be
[\,(t\,1)\,2345]=[12345],
\ee
and more importantly, it is dual superconformally invariant:
\be
\[1'\,2'\,3'\,4'\,5'\,\]=[12345]~~\textrm{for}~~\(\mathcal{Z}_a'\)^I=M^I_{~J}\mathcal{Z}_a^J
\ee
where $M$ is a non-degenerate GL$(4)$ matrix.
Its invariance results from the integral \eqref{eq-7} which manifests both dual superconformal
and the auxiliary GL$(k)$ invariance.

From 5-brackets
we can construct more nontrivial Yangian invariants, such as the one for $(23)(67)$ of
N$^2$MHV $n\!=\!7$ amplitude in \eqref{eq-2}, as readers may refer to
{\S 12} of \cite{ArkaniHamed:2012nw} for relevant details.
One form of its corresponding Yangian invariant is (this expression is not unique)
\be
[\,1\,(123)\cap(45)\,5\,6\,7\,]\,[12345], \labell{eq-8}
\ee
of which the matrix representative is
\be
C=\(\begin{array}{ccccccc}
\ap_{234} & \ap_{34}\ap_5 & \ap_{34}\ap_6 & \ap_4\ap_7 & 0 & -1 & -\ap_1 \\
1 & \ap_5 & \ap_6 & \ap_7 & \ap_8 & 0 & 0
\end{array}\), \labell{eq-9}
\ee
where $\ap_{i\ldots j}\!=\!\ap_i\!+\!\ldots\!+\!\ap_j$ and $\ap_i$'s are the positive variables with integration measures
in the $d\log$ form. Here, except $(23)\!=\!(67)\!=\!0$, the rest ordered minors are manifestly positive.
Note that in our context, all vanishing minors are \textit{consecutive}!
This is related to positivity, which we will further discuss in sections \ref{sec4} and \ref{sec5}.

It is easy to fix all $\ap_i$'s above by first solving the second row, then using the ``intersected'' momentum twistors
such as $(123)\!\cap\!(45)$ to solve the first row. The $\ap_i$ integration gives the exact result \eqref{eq-8},
and we see that $(23)(67)$ has automatically encoded the specific kinematical information of \eqref{eq-8}
prior to solving for $\ap_i$'s.
As $k$ grows, such intersection symbols will soon become overwhelming. For example, one form of the Yangian invariant
for cell $[8]$ of N$^3$MHV $n\!=\!8$ amplitude in \eqref{eq-11} is
\be
\bal
&[34567]\,[\,2\,3\,(34)\cap(567)\,(345)\cap(67)\,7\,]\\
&\times[\,1\,2\,\,(23)\cap((34)\cap(567)\,(345)\cap(67)\,7)\,\,(2\,3\,(34)\cap(567))\cap((345)\cap(67)\,7)\,\,7\,],
\labell{eq-25}
\eal
\ee
which is lengthy and dazing. In contrast, the compact notation $[8]$ means it is simply a top cell with
the 8th column removed. What's more, $[8]$ is a manifestly GL$(k)$ invariant representation
while the expression above is not.
This is partly due to the row operations: for a matrix representative,
one can always shift any row by a constant times another row, without causing any difference in the
contour integration \eqref{eq-7}.
Equivalent matrix representatives up to row operations
will then lead to equivalent but seemingly different Yangian invariants,
while $[8]$ is inert to any GL$(k)$ transformation.

Besides the row operations, one can further choose different
matrix representatives for the same cell.
Back to \eqref{eq-8}, we may choose another one for $(23)(67)$:
\be
C=\(\begin{array}{ccccccc}
0 & \,1 & \,\ap_1 & \ap_{234} & \ap_{34}\ap_5 & \ap_4\ap_6 & \ap_4\ap_7 \\
-\ap_8 & \,0 & \,0 & 1 & \ap_5 & \ap_6 & \ap_7
\end{array}\),
\ee
of which the Yangian invariant is
\be
[\,2\,3\,4\,(45)\cap(671)\,1\,]\,[14567]. \labell{eq-10}
\ee
Comparing them with \eqref{eq-8} and \eqref{eq-9}, we find that
neither these two Yangian invariants nor their matrix representatives obviously match,
but both of them encode the same geometric configuration $(23)(67)$.
It should be understood that different matrix representatives are equivalent up to a GL$(k)$ transformation,
and the row operations belong to a special type of such manipulations.

We have not explained how to read off the order of super momentum twistors in 5-brackets
from the corresponding $\ap_i$'s, which could cause a sign ambiguity.
Let's postpone the discussion of this subtlety to the parameterization of positive
matrix representatives in section \ref{sec5}.

In summary, to characterize Yangian invariants in a more invariant way, which manifests both dual superconformal
invariance and the hidden but inevitable GL$(k)$ invariance, it is better to use the
Grassmannian geometric configurations. The Grassmannian is not a luxurious optional tool,
but a mandatory bridge towards the GL$(k)$ invariance concealed by solely using momentum twistors.
It is analogous to the BRST symmetry in gauge theory, which introduces auxiliary fields as a must to completely
describe this hidden symmetry.

In the introduction we have discussed the advantages of Grassmannian geometry representatives over
decorated permutations. Practically, one may use the efficient \textsc{Mathematica} package ``\verb"positroids"''
\cite{Bourjaily:2012gy} to generate
a list of cells in terms of permutations in massless spinor space,
then translate them to those in momentum twistor space
and finally obtain their corresponding geometry representatives.
The formula for translating permutations is \cite{ArkaniHamed:2012nw}
\be
\sigma_\textrm{MT}(a)=\sigma(a\!-\!1)-1, \labell{eq-42}
\ee
which is useful as a manual check.
In fact, for a permutation given by the on-shell diagrammatic recursion \cite{ArkaniHamed:2012nw}, the command
``\verb"//dualGrassmannian//permToGeometry"'' directly outputs its Grassmannian geometry representative.

\subsection{Pl\"{u}cker coordinates and Pl\"{u}ckerian integral formulation}

As we are motivated to promote the positive Grassmannian geometry to a primary machinery for better understanding
amplitudes, it is then necessary to extensively use the \textit{Pl\"{u}cker coordinates},
as done in \cite{Franco:2014csa} for 4-particle MHV integrands to all loop orders.
Relevant mathematical aspects can be found in \cite{Postnikov:2013, Griffiths:1978}.
Pl\"{u}cker coordinates are the $\binom{\,n\,}{\,k\,}$ ordered minors of the Grassmannian $C\!\in\!G(k,n)$,
which have in total $k(n\!-\!k)$ degrees of freedom.
They are SL($k$) invariant and projective with the GL(1) redundancy,
as they rescale uniformly under the GL($k$) transformation. More importantly, Pl\"{u}cker coordinates are positive.

They obey some homogenous quadratic identities called the \textit{Pl\"{u}cker relations},
which are also known as \textit{Schouten identities} for physicists.
Without a proper counting method, Pl\"{u}cker relations appear to be highly redundant.
Nevertheless, if we fix the ordered minor $(i_1\ldots i_k)$ to be a non-vanishing value,
we can expand another column, say $(a_1|$, in a basis of $k$ linearly independent vectors which are columns
$i_1,\ldots,i_k$ in this case. The Cramer's rule tells
\be
0=(i_1\ldots i_k)(a_1|+(-)^k(i_2\ldots i_k\,a_1)(i_1|+(-)^{2k}(i_3\ldots i_k\,a_1\,i_1)(i_2|+\ldots
+(-)^{k^2}(a_1\,i_1\ldots i_{k-1})(i_k|, \labell{eq-12}
\ee
to complete the determinant contraction one can freely choose $(k\!-\!1)$ columns to fill its RHS above.
The first choice is $|i_2\ldots i_k)$ and we get a 2-term identity,
which is a trivial statement of the antisymmetry of minors.
Let's denote the $(n\!-\!k)$ unfixed columns by $a_j$'s. Now the first nontrivial type which is a 3-term identity,
can be obtained by contracting \eqref{eq-12} with $|a_2\,i_3\ldots i_k)$, namely
\be
\bal
0&=\((i_1\ldots i_k)(a_1|+(-)^k(i_2\ldots i_k\,a_1)(i_1|
+(-)^{2k}(i_3\ldots i_k\,a_1\,i_1)(i_2|+\ldots\)\times|a_2\,i_3\ldots i_k)\\
&=(i_1\ldots i_k)(a_1\,a_2\,i_3\ldots i_k)+(-)^k(i_2\ldots i_k\,a_1)(i_1\,a_2\,i_3\ldots i_k)
+(-)^{2k}(i_3\ldots i_k\,a_1\,i_1)(i_2\,a_2\,i_3\ldots i_k).
\eal
\ee
Following this trend, we can increase the number of $a_j$'s in the $(k\!-\!1)$ columns,
and hence increase the non-vanishing terms upon contraction. The second type (a 4-term identity) is
\be
\bal
0=\,&\,(i_1\ldots i_k)(a_1\,a_2\,a_3\,i_4\ldots i_k)+(-)^k(i_2\ldots i_k\,a_1)(i_1\,a_2\,a_3\,i_4\ldots i_k)\\
&+(-)^{2k}(i_3\ldots i_k\,a_1\,i_1)(i_2\,a_2\,a_3\,i_4\ldots i_k)
+(-)^{3k}(i_4\ldots i_k\,a_1\,i_1\,i_2)(i_3\,a_2\,a_3\,i_4\ldots i_k),
\eal
\ee
and we can enumerate all of them until the last type
\be
\bal
0=\,&\,(i_1\ldots i_k)(a_1\,a_2\ldots a_k)+(-)^k(i_2\ldots i_k\,a_1)(i_1\,a_2\ldots a_k)
+(-)^{2k}(i_3\ldots i_k\,a_1\,i_1)(i_2\,a_2\ldots a_k)\\
&+\ldots+(-)^{k^2}(a_1\,i_1\ldots i_{k-1})(i_k\,a_2\ldots a_k),
\eal
\ee
which has $(k\!+\!1)$ non-vanishing terms.

Every distinct type of identity consists of $2k$ labels, among which $k$ labels are fixed to be $i_1,\ldots,i_k$,
and the rest $k$ labels may contain $(k\!-\!j)$ ones from $i_1,\ldots,i_k$ and $j$ ones from $a_1,\ldots,a_{n-k}$.
Since $j$ varies from 2 to $k$, we reach the matching of degrees of freedom as below
\be
\binom{n}{k}-\sum_{j=2}^k\binom{k}{k\!-\!j}\binom{n\!-\!k}{j}-1=k(n-k),
\ee
where the LHS counts the number of independent Pl\"{u}cker coordinates modulo Pl\"{u}cker relations and the GL$(k)$
rescaling, and the RHS simply counts the degrees of freedom of $C\!\in\!G(k,n)$.

This can be seen more straightforwardly:
if we gauge fix columns $1,\ldots,k$ of $C$ to be a unit matrix,
it is easy to find the bijection between
independent Pl\"{u}cker coordinates and the entries of $C$ as
\be
C_{\ap a}=(-)^{k-\ap}(1\ldots\widehat{\ap}\ldots k\,a)\equiv(-)^{k-\ap}\De_{1\,\ldots\,\widehat{\ap}\,\ldots\,k\,a}
~~\textrm{for}~~a\!=\!k\!+\!1,\ldots,n
\ee
where $\widehat{\ap}$ means the label $\ap$ is removed. From now on, we will use $\De$'s as independent variables
to express the rest ordered minors as derivative quantities.
A simple example is the top cell $C\!\in\!G(3,8)$ with
\be
C=\(\begin{array}{cccccccc}
1\, & 0\, & 0\, & \De_{234} & \De_{235} & \De_{236} & \De_{237} & \De_{238} \\
0\, & 1\, & 0\, & -\De_{134} & -\De_{135} & -\De_{136} & -\De_{137} & -\De_{138} \\
0\, & 0\, & 1\, & \De_{124} & \De_{125} & \De_{126} & \De_{127} & \De_{128}
\end{array}\),
\ee
and all derivative Pl\"{u}cker coordinates are simply expressed in terms of $\De$'s by definition, such as
\be
(567)=\frac{1}{(\De_{123})^2}\lv\begin{array}{ccc}
\De_{235} & \De_{236} & \De_{237} \\
-\De_{135} & -\De_{136} & -\De_{137} \\
\De_{125} & \De_{126} & \De_{127}
\end{array}\rv.
\ee
Although $\De_{123}\!=\!1$, writing it explicitly can make $(567)$ correctly rescale under the GL($k$) transformation.
Therefore, choosing a gauge is equivalent to fixing the set of independent Pl\"{u}cker coordinates.

To purely formulate in terms of Pl\"{u}cker coordinates, we propose the \textit{Pl\"{u}ckerian integral}
through its equivalence to the familiar Grassmannian counterpart (with cyclicly consecutive minors)
\be
\bal
&\int\frac{d^{k\times n}C}{\textrm{vol(GL($k$))}}\frac{1}{(12\ldots k)(23\ldots k\!+\!1)\ldots(n\,1\ldots k\!-\!1)}\\
=&\int\frac{d^{\,\binom{n}{k}}\De}{\textrm{vol(GL(1))}}
\frac{1}{\De_{12\,\ldots\,k}\,\De_{23\,\ldots\,k+1}\ldots\De_{n\,1\,\ldots\,k-1}}\times
J\,\de^{\,p}\(\sum\pm\De\De\), \labell{eq-17}
\eal
\ee
where
\be
J=(\De_{i_1\,\ldots\,i_k})^{p+n-k(n-k)-1},~~p=\binom{n}{k}-k(n-k)-1.
\ee
$\sum\pm\De\De$ in $p$ delta functions denote all
independent Pl\"{u}cker relations, with respect to the chosen gauge fixing columns $i_1,\ldots,i_k$,
where $p$ is the number of independent Pl\"{u}cker relations.
The Jacobian $J$ is fixed such that the integral remains invariant under the uniform GL(1) rescaling,
since it helps compensate the rescaling weight.
It is easy to check
that this new integral reduces to the familiar one if columns $1,\ldots,k$ are fixed to be a unit matrix,
without loss of generality. In fact, we have done nothing new but renaming
$C_{\ap a}\!=\!(-)^{k-\ap}\De_{1\,\ldots\,\widehat{\ap}\,\ldots\,k\,a}$ for $a\!=\!k\!+\!1,\ldots,n$.

In the Pl\"{u}ckerian integral formulation, its gauge fixing is simplified to
choosing a Pl\"{u}cker coordinate to be any non-vanishing value (usually, it is unity),
rather than fixing a $k\!\times\!k$ sub-matrix of $C$. It is more convenient to switch the gauge in order
to suitably characterize each cell now. We will use this technique extensively in section \ref{sec4}, which proves to be
efficient for determining the signs in homological identities.

One may ask how we express the super delta function in \eqref{eq-7}, in terms of Pl\"{u}cker coordinates as
\be
\de^{4k|4k}(C_{\ap a}(\De)\mathcal{Z}_a),
\ee
the general answer is unknown, but for the canonical gauge above, we can simply follow the substitution
$C_{\ap a}\!=\!(-)^{k-\ap}\De_{1\,\ldots\,\widehat{\ap}\,\ldots\,k\,a}$ as a trivial renaming.

\subsection{Reduced Grassmannian geometry}

While the Grassmannian geometric configurations can be explicitly
parameterized by Pl\"{u}cker coordinates, for plenty of cells
it is practical to characterize them in a more compact form, with the price of suppressing their parameterizations.
These notations are the (reduced) Grassmannian geometry representatives, as we have met in the introduction.
Here, a more systematic exposition will be presented.

First, we use ``empty slots'', such as $[i]$, to denote removed columns. It is worth emphasizing that
all these symbols only make sense when $k$ and $n$ are specified. For example, when $k\!=\!1$, $n\!=\!6$, $[6]$ means
the 6th entry is absent in the Yangian invariant, so the resulting quantity is a 5-bracket $[12345]$
(to avoid confusion, we will stress it when a 5-bracket shows up). They are multiplicative,
for instance, $[6][7]\!=\![67]$. As we have mentioned, such a product is simply a superposition of all constraints.

Then, for $k\!\geq\!2$ we use $(ij)$ to denote that columns $i,j$ are proportional,
which is also multiplicative. For the example $(23)(67)$ with \eqref{eq-8}, it is now nontrivial to read off its
Yangian invariant but of course we refrain from doing so.
Still we must at least ensure that it has the correct dimension, which in this case is $k(n\!-\!k)\!-\!2\!=\!4k\!=\!8$.
When $k\!\geq\!3$, the dimension counting is not transparent in general, and that's one of the motivations
to introduce the reduced Grassmannian geometry.

The first example is the cell $(23)(456)$ of N$^3$MHV $n\!=\!8$ amplitude in \eqref{eq-11}, where columns $2,3$ are
proportional and columns $4,5,6$ are ``minimally'' linearly dependent. We see that $(23)$ leads to $(123)\!=\!(234)\!=\!0$,
but the latter does not necessarily lead to $(23)$. For both clarity and conciseness, we will use
distinct notations for these two different constraints. This is called the reduced Grassmannian geometry,
where its term ``reduced'' means that we will only extract the essential geometric information
and nothing more than that. For example, when $k\!=\!4$,
$(34)$ will lead to $(234)\!=\!(345)\!=\!(1234)\!=\!(2345)\!=\!(3456)\!=\!0$,
still we should write $(34)$ only. Note that $(34)$ also denotes the vanishing of an arbitrary $2\!\times\!2$ minor
of columns $3,4$, and $(234)$ denotes the vanishing of an arbitrary $3\!\times\!3$ minor of columns $2,3,4$.

In general, for $j\!\leq\!k$, $(i_1\ldots i_j)$ denotes the vanishing of an arbitrary $j\!\times\!j$ minor of
columns $i_1,\ldots,i_j$ and its implicit vanishing constraints involving more columns are suppressed.
All minors here are \textit{cyclicly consecutive}, which is sufficient for planar amplitudes.
It is also convenient to use the abbreviated notation $(i_1\ldots i_j\,i_{j+1}\ldots i_{j+l-1})_j$ to denote the
cyclicly adjacent product of consecutive minors
\be
(i_1\ldots i_j)(i_2\ldots i_{j+1})\ldots(i_l\ldots i_{j+l-1}),
\ee
note the subscript $j$ makes a difference, for instance, $(345)_2\!=\!(34)(45)$ is clearly different from $(345)$
when $k\!=\!3$. Without an explicit subscript, the size of a vanishing minor is the number of columns it involves.

\begin{figure}
\begin{center}
\includegraphics[width=0.3\textwidth]{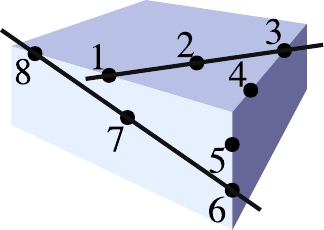}
\caption{Geometric configuration of $k\!=\!4$, $n\!=\!8$ labeled by permutation \{3,7,6,10,9,8,13,12\}
in \cite{Bourjaily:2012gy}.} \label{fig-1}
\end{center}
\end{figure}

For the example $(23)(456)$, since $(23)$ and $(456)$ have no overlapped column, we can write it trivially
as a product. However, for a ``nested'' object such as \eqref{eq-13} in the introduction, one needs a more artistic way
to represent it concisely. Now let's use a second example for demonstration:
in \cite{Bourjaily:2012gy}, a 9-dimensional geometric configuration of $k\!=\!4$, $n\!=\!8$
labeled by permutation $\{3,7,6,10,9,8,13,12\}$ is characterized by the representation
(its pictorial sketch is shown in figure \ref{fig-1})
\be
\(\begin{array}{c}
(1)(2)(3)(4)(5)(6)(7)(8) \\
(123)(34)(45)(56)(67)(678)(81) \\
(3456)(56781)(81234)
\end{array}\),
\ee
which specifies every layer of linear dependencies, but most of them are trivial or can be derived from
the upstair ones. To encode the geometric information more compactly, we will instead use
\be
(3\,4\,5\,6)\,(5\,\os{7}{\os{\,\,/}{6}\,\os{\backslash\,\,}{8}}\,1)\,(8\,\os{2}{\os{\,\,/}{1}\,\os{\backslash\,\,}{3}}\,4)
\ee
which contains nested constraints $(123)$ and $(678)$. Similar to \eqref{eq-13},
each connection between two column labels stands for one degree of freedom, for instance, column 2 is spanned by
columns $1,3$ as above. This column is diagrammatically divalent, and its two degrees of freedom can be regarded as
additional on top of columns $1,3,4,5,6,8$.

Now it is trivial to count its dimension from the perspective that
separates vanishing maximal minors and degenerate linear dependencies.
For the reduced Grassmannian made of columns $1,3,4,5,6,8$, there are three constraints, namely
$(3456)(5681)(8134)$. In addition, the two columns attached on it altogether contribute four valencies.
In total, we have
\be
4(6-4)-3+4=9
\ee
degrees of freedom, which is harder to be inferred from the un-reduced geometry, let alone the permutation.
This reduced geometry also has a clear pictorial meaning: in figure \ref{fig-1}, we first construct the
3-dimensional (projectively) skeleton made of edges $34,56,81$ and faces $3456,5681,8134$,
then add points $2$ and $7$ to the configuration. Each of the latter has two degrees of freedom, one of which is
the translation along a line and the other is suppressed by the projective sketching.

In general, the dimension of a reduced Grassmannian geometry representative is given by
\be
d=k(n'\!-\!k)-(\textrm{\# vanishing maximal minors})+(\textrm{\# degenerate valencies}),
\ee
where $n'$ is the width of the reduced Grassmannian. This formula also works for former example $(23)(456)$
since $(23)$ can be regarded as a degenerate valency,
then its dimension is $3(7\!-\!3)\!-\!1\!+\!1\!=\!12$. Note that we always
list all vanishing constraints in the cyclic order, and all of them are cyclicly consecutive in the reduced sense.
It is also equivalent to choose any columns of the degenerate
constraint to be put upstairs, although for convenience, we have chosen columns $2,7$ above as in that way we
have a minimally reduced Grassmannian of $n'\!=\!6$.

\subsection{Tree BCFW recursion relation in the positive matrix form}

Now we will show how the (reduced) Grassmannian geometry representatives of tree amplitudes are generated,
by the matrix version of BCFW recursion relation \cite{ArkaniHamed:2010kv}.
The idea first appeared in \cite{Bourjaily:2010wh},
and here we will use a more geometric form that manifests positivity, as presented in the following.

\begin{figure}
\begin{center}
\includegraphics[width=0.6\textwidth]{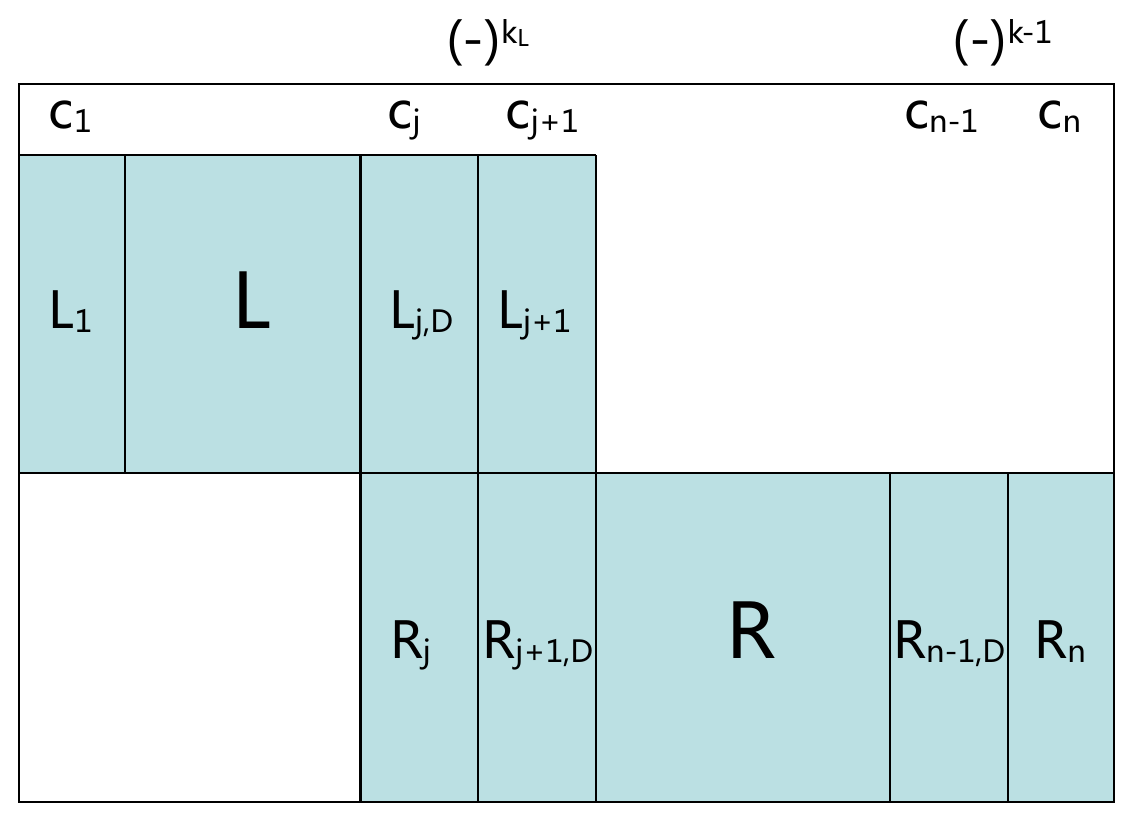}
\caption{Matrix version of tree BCFW recursion relation.} \label{fig-2}
\end{center}
\end{figure}

In figure \ref{fig-2}, the first row represents a 5-bracket $[1\,j\,j\!+\!1\,n\!-\!1\,n]$
and below it, we merge two positive matrices
$Y_\textrm{L}$ and $Y_\textrm{R}$ in a way such that the entire matrix is positive
(we abuse $Y$'s for the corresponding matrices of Yangian invariants).
Positivity also demands us to add sign factors $(-)^{k_\textrm{L}}$ to $c_j$ and $c_{j+1}$,
as well as $(-)^{k-1}$ to $c_{n-1}$ and $c_n$, where $k\!=\!k_\textrm{L}\!+\!k_\textrm{R}\!+\!1$ and
$n\!=\!n_\textrm{L}\!+\!n_\textrm{R}\!-\!2$.
When $n_\textrm{L}\!=\!3,4$, $Y_\textrm{L}$ trivially disappears as $k_\textrm{L}\!=\!0$,
and so does $Y_\textrm{R}$ for $n_\textrm{R}\!=\!4$.
All the rest blank areas are filled with zeros.
Note that although the $c_i$'s are taken to be positive, it is not the essence of positivity,
as will be explained soon.

The deformed sub-columns with subscript `D' are given by
\be
L_{j,\,\textrm{D}}=L_j+\frac{c_j}{c_{j+1}}L_{j+1},~~R_{j+1,\,\textrm{D}}=R_{j+1}+\frac{c_{j+1}}{c_j}R_j,~~
R_{n-1,\,\textrm{D}}=R_{n-1}+\frac{c_{n-1}}{c_n}R_n, \labell{eq-14}
\ee
where $L_j$, $R_{j+1}$ and $R_{n-1}$ are the un-deformed columns of $Y_\textrm{L}$ and $Y_\textrm{R}$.
After all $c_i$'s find their solutions on the support of the bosonic delta function in \eqref{eq-7},
we find, for instance:
\be
L_{j,\,\textrm{D}}\,\mathcal{Z}_j+L_{j+1}\mathcal{Z}_{j+1}=L_j\mathcal{Z}_j+L_{j+1}\widehat{\mathcal{Z}}_{j+1},
\ee
where $\widehat{\mathcal{Z}}_{j+1}\!=\!(\,j\,j\!+\!1)\cap(n\!-\!1\,n\,1)$ is the deformed super momentum twistor.
It is similar for $R_{j+1,\,\textrm{D}}$ and $R_{n-1,\,\textrm{D}}$, so we reach the familiar tree BCFW product of
Yangian invariants
\be
[1\,j\,j\!+\!1\,n\!-\!1\,n]\,Y_\textrm{L}(1,\ldots,j,I)\,Y_\textrm{R}(I,j\!+\!1,\ldots,n\!-\!1,\widehat{n})
\ee
with $\mathcal{Z}_I\!=\!\widehat{\mathcal{Z}}_{j+1}\!=\!\widehat{\mathcal{Z}}_j\!=\!(\,j\,j\!+\!1)\cap(n\!-\!1\,n\,1)$
and $\widehat{\mathcal{Z}}_n\!=\!(n\!-\!1\,n)\cap(1\,j\,j\!+\!1)$.

We will derive the matrix in figure \ref{fig-2} from positivity in section \ref{sec3}, which is exactly equivalent to
the original BCFW recursion relation in momentum twistor space in \cite{ArkaniHamed:2010kv}.
It is worth emphasizing that in this article, positivity is only related to the auxiliary Grassmannian
and it has nothing to do with momentum twistors, which is different from the amplituhedron formulation
in \cite{Arkani-Hamed:2013jha}. It is appealing to set the ordered minors of all $n$ momentum twistors to be positive,
but it contradicts with the principle of this article. To see why, one can consider a simplest example:
NMHV $n\!=\!5$ amplitude in the matrix form. Figure \ref{fig-2} gives its positive matrix
prior to solving the kinematical constraint as a single row
\be
(c_1~c_2~c_3~c_4~c_5)
\ee
without signs. If we set up positive kinematical data $Z_1,Z_2,Z_3,Z_4,Z_5$ so that any $\<ijkl\>$
for $i\!<\!j\!<\!k\!<\!l$ is positive, the bosonic delta function in \eqref{eq-7} imposes
\be
c_1=\<2345\>,~c_2=\<3451\>,~c_3=\<4512\>,~c_4=\<5123\>,~c_5=\<1234\>,
\ee
now $c_2$ and $c_4$ have to be negative. Therefore, we will adopt the perspective that positivity should be
only used for determining the auxiliary Grassmannian. Once we know the
explicit parameterization, positivity is then \textit{forgotten} in the next step of solving for the positive
parameters. This actually coincides with the idea of manipulating geometric configurations instead of
momentum twistors as previously mentioned, if we stick to positivity as the primary guiding principle.

From the mathematical perspective, positivity does not mean literally being positive, but the essence
of subtraction-free expressions. For example, consider a subtraction-free polynomial $(x\!+\!y\!+\!z)$,
there is no way to render it vanish without setting $x\!=\!y\!=\!z\!=\!0$,
while for $(x\!+\!y\!-\!z)$ this is not true. The same property also holds for $-(x\!+\!y\!+\!z)$, which is
literally negative but subtraction-free as well.
Furthermore, neither the Grassmannian nor momentum twistors should be genuinely taken to be positive, because the former
has contour integrations to be done, while the latter originate from complex massless spinors.
It is intuitive and convenient to pretend that the positive matrix uses genuinely positive variables,
of which the zero limits represent some kinds of singularities. However, we don't have positive parameterization
or variables for momentum twistors, which are purely external kinematical data.

The deformed sub-columns in figure \ref{fig-2} will affect the corresponding Pl\"{u}cker sub-coordinates,
as those involved are deformed accordingly:
\be
\De^\textrm{L,\,D}_{\ldots\,j\,\ldots}=\De^\textrm{L}_{\ldots\,j\,\ldots}
+\frac{c_j}{c_{j+1}}\De^\textrm{L}_{\ldots\,j+1\,\ldots}\,
\ee
for the left Pl\"{u}cker sub-coordinates, and
\be
\De^\textrm{R,\,D}_{\ldots\,j+1\,\ldots}=\De^\textrm{R}_{\ldots\,j+1\,\ldots}
+\frac{c_{j+1}}{c_j}\De^\textrm{R}_{\ldots\,j\,\ldots}\,,~~
\De^\textrm{R,\,D}_{\ldots\,n-1\,\ldots}=\De^\textrm{R}_{\ldots\,n-1\,\ldots}
+\frac{c_{n-1}}{c_n}\De^\textrm{R}_{\ldots\,n\,\ldots}\,
\ee
for the right counterparts. Pl\"{u}cker coordinates of the entire matrix are then given by the schematic form
\be
\De=\sum c\times\De^\textrm{L}\De^\textrm{R},
\ee
where $\De^\textrm{L}$ and $\De^\textrm{R}$ may or may not be deformed,
depending on whether they contain the deformed sub-columns.
For any ordered minor, the RHS polynomial above is positive (or subtraction-free precisely),
which exactly justifies the sign factors in figure \ref{fig-2}.

Finally, let's demonstrate figure \ref{fig-2} with a detailed example. For the N$^3$MHV $n\!=\!10$ amplitude, there is
a set of BCFW cells that belong to the sub-class generated by its characteristic 5-bracket $[1789\,10]$,
for which $k_\textrm{L}\!=\!2$. They are given by the matrix
\be
C=\(\begin{array}{cccccccccc}
c_1 & 0 & 0 & 0 & 0 & 0 & c_7 & c_8 & \,c_9 & \,c_{10} \\[+0.5em]
L_1 & L_2 & L_3 & L_4 & L_5 & L_6 & \(L_7\!+\!\dfrac{c_7}{c_8}L_8\) & L_8 & \,0 & \,0
\end{array}\),
\ee
where $L_i$'s are 2-vectors of the sub-amplitude $(Y_\textrm{L})^2_8$. One of its sub-cells is
$(234)_2(678)_2$ which will result in the N$^3$MHV $n\!=\!10$ cell
\be
(\!\!\os{2\,\,\,\,3}{\os{\backslash\,/}{4}}\!5\,6)\,(6\,7\,8\os{10}{\os{|}{9}})_3\,,
\ee
this can be seen from its matrix form
\be
C=\(\begin{array}{cccccccccc}
c_1 & 0 & \,0 & \,0 & 0 & 0 & c_7 & c_8 & c_9 & \,c_{10} \\[+0.5em]
L_1 & L_2 & \,\ap_3L_2 & \,\ap_4L_2 & L_5 & L_6 & \(\ap_7\!+\!\dfrac{c_7}{c_8}\ap_8\)\!L_6 & \ap_8L_6 & 0 & \,0
\end{array}\),
\ee
where the positive $\ap_i$'s help manifest linear dependencies $(234)_2(678)_2$ of the sub-cell.

It is crucial to realize that, reduced Grassmannian geometry requires us to extract the essential geometric information
only, which means, we must ensure that it can be no more reduced. For the example above, naively we find
$(123)\!=\!(234)\!=\!(345)\!=\!(456)\!=\!0$ which could be ambiguous,
hence we will remove an arbitrary row and repeat the analysis
of linear dependencies until its configuration is precisely identified. This procedure is common
since the blank areas filled with zeros in figure \ref{fig-2} can bring significant amount of such ambiguities
resulting from several cyclicly adjacent vanishing consecutive minors.
We will discuss further relevant aspects as well as examples in sections \ref{sec5} and \ref{sec6}.

\newpage
\section{Tree and 1-loop BCFW Recursion Relations from Positivity}
\label{sec3}

This section extends the idea of deriving locality and unitarity as a consequence of positivity
\cite{Arkani-Hamed:2013jha}, making explicit contact with tree and 1-loop BCFW recursion relations.
Only a minimal amount of knowledge of momentum twistors and specific matrix structures is needed
for this deductive reformulation.

\subsection{Tree BCFW recursion relation from positivity}

\begin{figure}
\begin{center}
\includegraphics[width=0.7\textwidth]{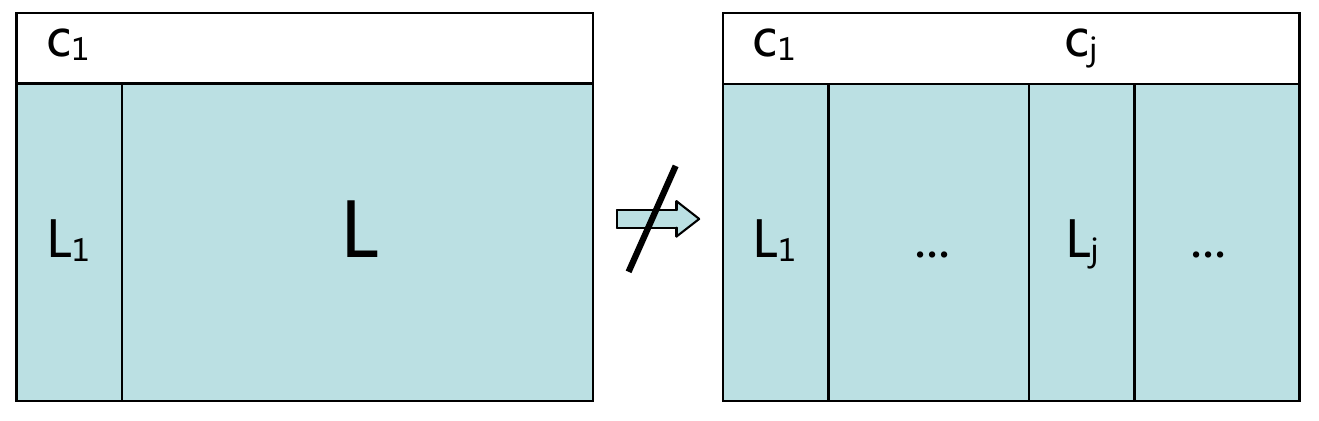}
\caption{Positivity forces $c_i$'s of the top row to align adjacently (in the cyclic sense).} \label{fig-3}
\end{center}
\end{figure}

The matrix version of tree BCFW recursion relation in figure \ref{fig-2} in fact reflects some simple mathematical
structures which will be named as the \textit{positive components}.
To deduce figure \ref{fig-2} with as few assumptions as possible, let's imagine that there is a positive matrix
$Y_\textrm{L}$, and we want to build a more nontrivial positive matrix from it.
We may stack one more row on its top, of which only the first entry is nonzero, as shown in
the left of figure \ref{fig-3}.

There are only two choices of where to place the nonzero entry: the leftmost or the rightmost corner,
otherwise positivity will be violated, as one can easily verified. In fact, they are equivalent up to a cyclic shift
which will induce a sign factor $(-)^{k_\textrm{L}}$,
and this is known as the \textit{twisted cyclicity} \cite{ArkaniHamed:2012nw}.
Without loss of generality, we have chosen the leftmost one. Now we add a second entry in the same row,
which also has to obey positivity, then it is pushed to either the leftmost or the rightmost corner.
Again, we choose the leftmost one adjacent to the first entry. More entries can be added in the same way,
and positivity forces all of them to align adjacently in the cyclic sense,
as demonstrated in the right of figure \ref{fig-3}.

\begin{figure}
\begin{center}
\includegraphics[width=0.8\textwidth]{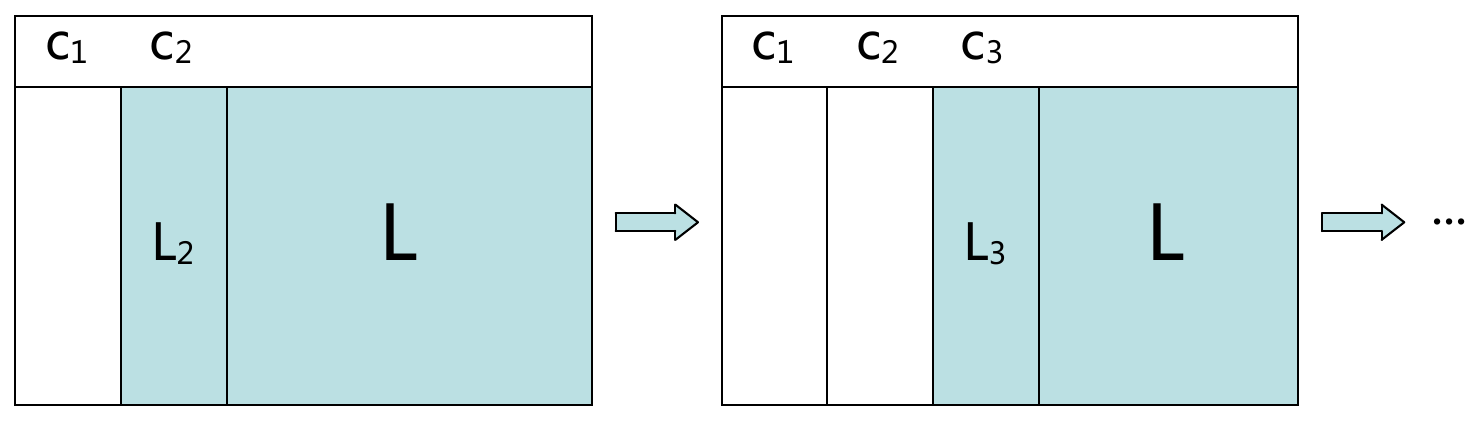}
\caption{First pattern of positive components: column-wise elimination.} \label{fig-4}
\end{center}
\end{figure}

\begin{figure}
\begin{center}
\includegraphics[width=0.8\textwidth]{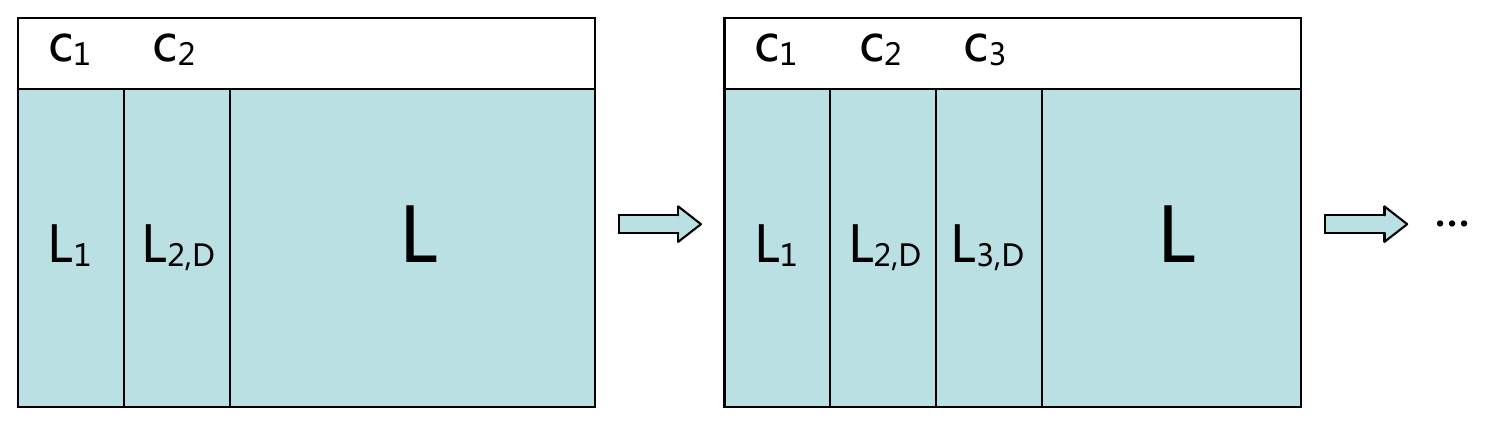}
\caption{Second pattern of positive components: column-wise deformation.} \label{fig-5}
\end{center}
\end{figure}

But that's not enough yet: with two or more $c_i$'s in the top row, those columns below them take part in positivity
in a nontrivial way, namely, now there are cross terms in the involved minors of the form
\be
c_i(L_{i+1}\ldots)-c_{i+1}(L_i\ldots),
\ee
where $\ldots$ denotes the rest $(k_\textrm{L}\!-\!1)$ columns. It is not manifestly positive, and to render it so
we have two choices of manipulation. The resulting matrix structures are shown in figures
\ref{fig-4} and \ref{fig-5}.

The first one is the column-wise elimination: we remove all columns but the rightmost one below the $c_i$'s
so that positivity is still preserved, but ``shortened''. The second one is the column-wise deformation,
which is more interesting: we deform all columns but the leftmost one below the $c_i$'s, according to
\be
L_{i+1,\,\textrm{D}}=L_i+\frac{c_{i+1}}{c_i}L_{i+1,\,\textrm{D}},
\ee
hence in figure \ref{fig-5} we have
\be
L_{2,\,\textrm{D}}=L_2+\frac{c_2}{c_1}L_1,~~L_{3,\,\textrm{D}}=L_3+\frac{c_3}{c_2}L_2+\frac{c_3}{c_1}L_1,
\ee
for example. These are the two patterns of positive components. As we will soon reveal, tree and 1-loop BCFW
recursion relations can emerge from such simple building blocks.
For tree amplitudes, we need the physical knowledge of momentum twistors and factorization limits,
where the former tells that each row must have five degrees of freedom (including its GL(1) redundancy)
so we need exactly five $c_i$'s in the top row,
and the latter determines which types of and how many positive components to choose
so we know how to place the $c_i$'s to induce the deformed columns.

\begin{figure}
\begin{center}
\includegraphics[width=0.75\textwidth]{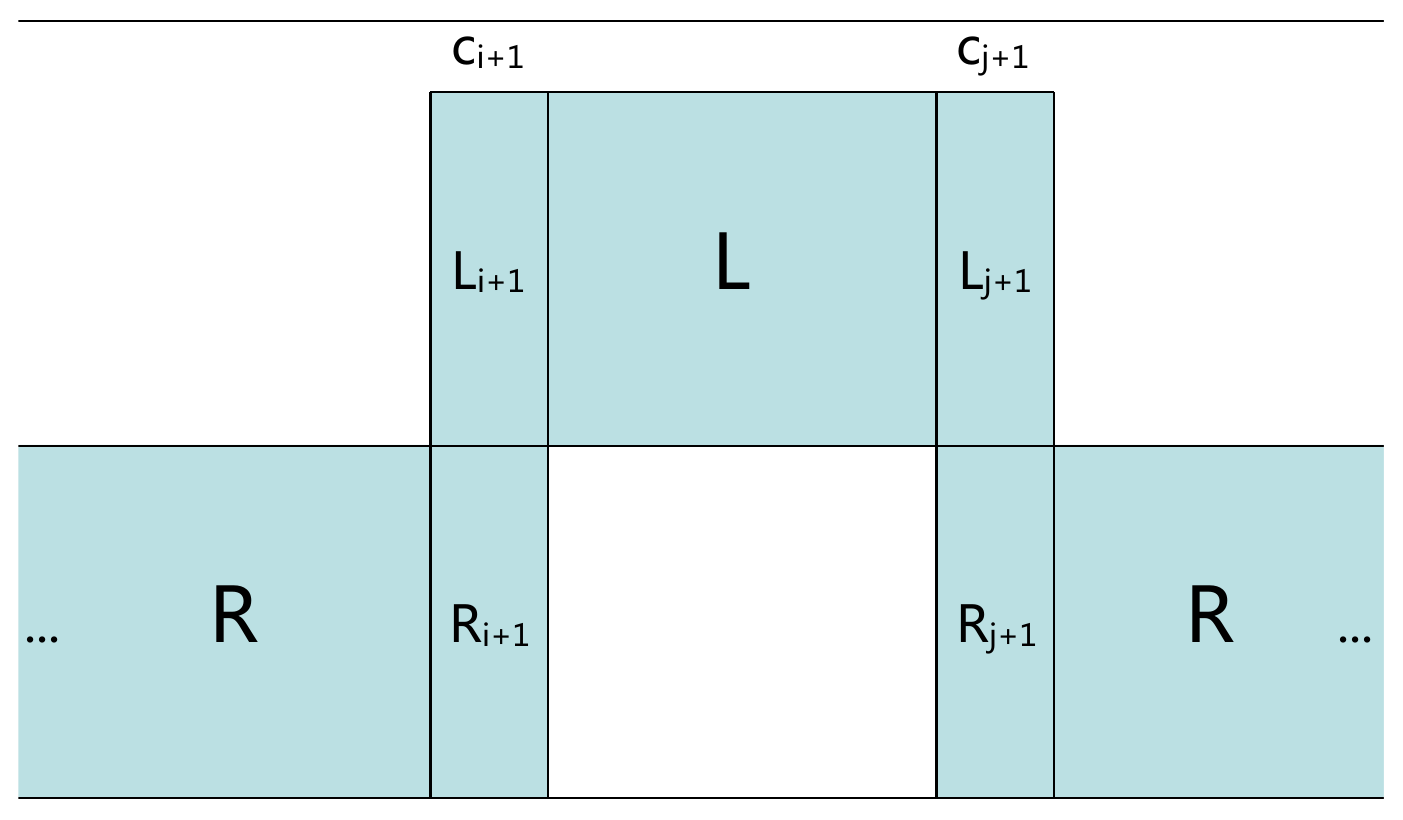}
\caption{Overlapped positive matrices $Y_\textrm{L}$ and $Y_\textrm{R}$ with $c_{i+1}$ and $c_{j+1}$ on top of them.}
\label{fig-6}
\end{center}
\end{figure}

Imagine that there are two positive matrices $Y_\textrm{L}$ and $Y_\textrm{R}$, and we want to merge them to build
a larger positive matrix without modifying themselves. Besides the trivial direct sum
$Y_\textrm{L}\!\oplus\!Y_\textrm{R}$, it is better to make them have two overlapped columns for positivity
to find its arena, as demonstrated in figure \ref{fig-6}.
Then we can repeat the reasoning of how to stack one more row on its top, similar to figure \ref{fig-3}.

First, as there are two positive sub-matrices, we can equally place two nonzero entries on top of the leftmost columns
of $Y_\textrm{L}$ and $Y_\textrm{R}$ respectively.
Be aware of that, we have not chosen which label to denote the first column, in other words, the entire matrix is
wrapped on a cylinder so that cyclicity is manifest.
Therefore, the overlapped sub-columns upon which $c_{i+1}$ and $c_{j+1}$
are placed are labeled by $i\!+\!1$ and $j\!+\!1$ with $i$ and $j$ unspecified. In this way, there is no need to
specify the sign factors of $c$'s as they depend on which label denotes the first column of this cyclic matrix.

\begin{figure}
\begin{center}
\includegraphics[width=0.75\textwidth]{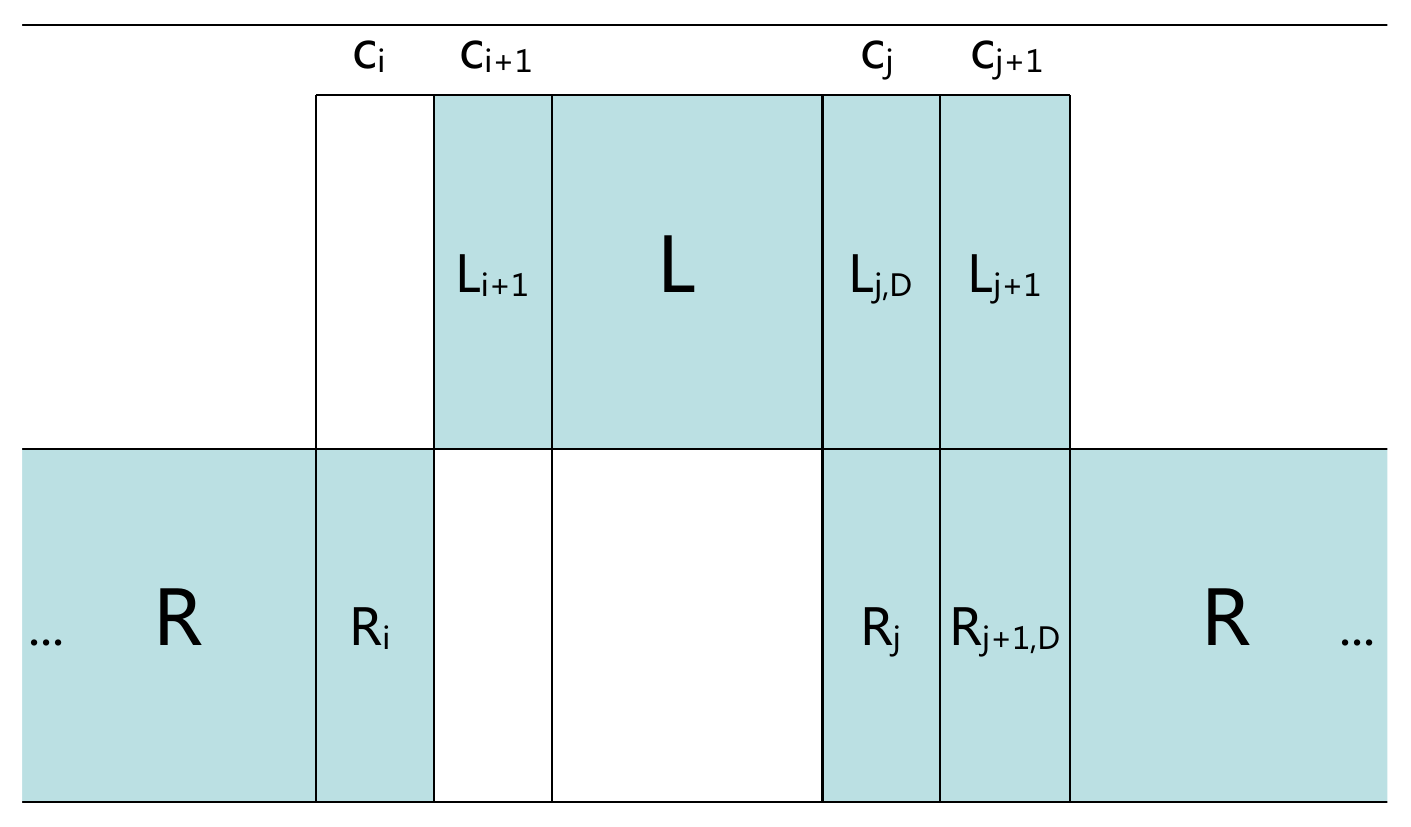}
\caption{Factorization limits apply both patterns of positive components.} \label{fig-7}
\end{center}
\end{figure}

We could be content with this matrix structure and next consider how to output its relevant physics. However,
momentum twistors demand us to go further as there should be three more $c$'s in the top row. Again, let's equally place
two $c$'s adjacent to $c_{i+1}$ and $c_{j+1}$, denoted by $c_i$ and $c_j$, as shown in figure \ref{fig-7}.
Note that different from figure \ref{fig-3}, now there are two pairs of adjacent $c$'s which are
essentially separated by the splitting matrix structure.
Then, factorization limits require that there is only one pair of internal legs attached to the two sub-amplitudes
respectively, which leads to the BCFW-like structure for $L_j$ and $R_{j+1}$ in \eqref{eq-14}.
Without loss of generality, we choose columns $j,j\!+\!1$ to implement this pattern, while in columns $i,i\!+\!1$
one overlapped sub-column must be removed. This requirement nicely applies both of the two patterns of
positive components in figures \ref{fig-4} and \ref{fig-5}: elimination in columns $i,i\!+\!1$ and deformation in
columns $j,j\!+\!1$, as we remove $R_{i+1}$ and add $R_j$ to maintain the symmetry between $Y_\textrm{L}$
and $Y_\textrm{R}$. Starting with the symmetric matrix structure in
figure \ref{fig-6}, factorization limits further carve out the more physical profile in figure \ref{fig-7}.

\begin{figure}
\begin{center}
\includegraphics[width=0.75\textwidth]{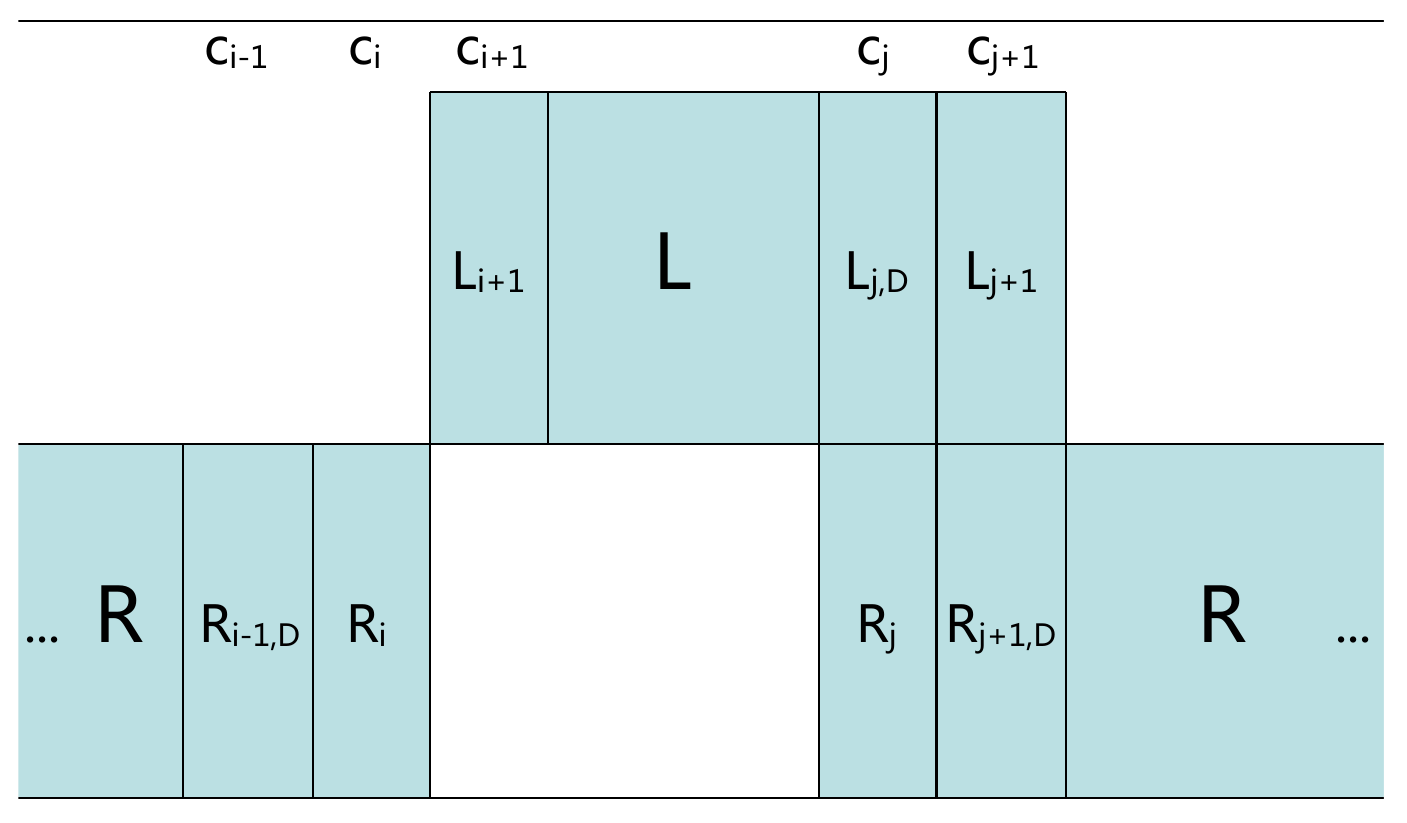}
\caption{$c_{i-1}$ is placed adjacent to $c_i$ as a regulator.} \label{fig-8}
\end{center}
\end{figure}

The last step is to put the ``regulator'', namely a fifth $c$, in the top row without removing or adding any sub-column,
otherwise factorization limits are violated. We choose to place it adjacent to $c_i$ in figure \ref{fig-8},
so it should be $c_{i-1}$ which will also induce a BCFW deformation for $R_{i-1}$.
Without the regulator $c_{i-1}$, there would be a physical singularity $c_{i-1}\!=\!\<i\,i\!+\!1\,j\,j\!+\!1\>\!\to\!0$
on the support of the bosonic delta function in terms of momentum twistors in \eqref{eq-7}.

\begin{figure}
\begin{center}
\includegraphics[width=0.6\textwidth]{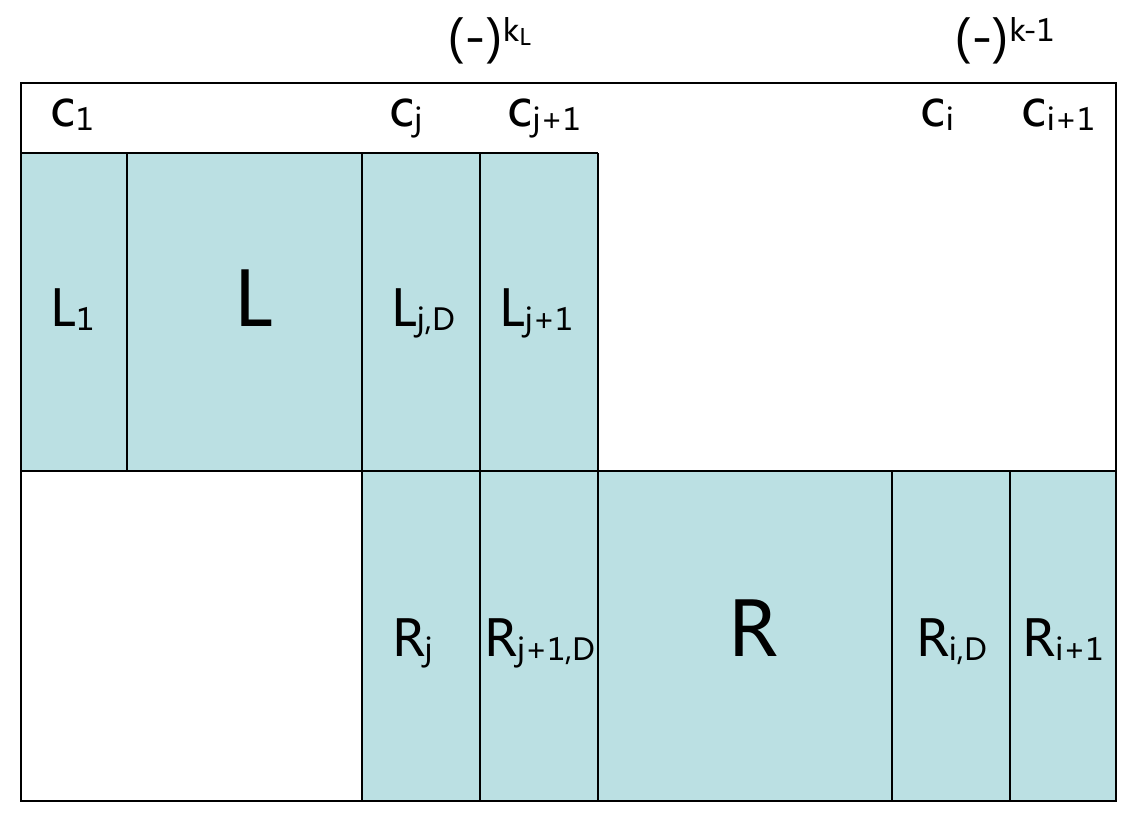}
\caption{Generic matrix configuration of tree BCFW recursion relation labeled by $i,j$.} \label{fig-9}
\end{center}
\end{figure}

Conventionally, we fix $i\!=\!n$ so that figure \ref{fig-8} is identical to figure \ref{fig-2}. Of course,
once label $i\!+\!1\!=\!1$ is designated to be the first column, it is mandatory to add the sign factors in
figure \ref{fig-2}, which reflect the twisted cyclicity.
Summing all possible factorization limits, we can express a general tree amplitude as
\be
Y^k_n=\sum_{i=k+3}^{n-1}\sum_{j=2}^{i-2}\,\textrm{FAC}_{\,i,\,j}\,, \labell{eq-38}
\ee
where $\textrm{FAC}_{\,i,\,j}$ represents the generic matrix configuration of tree BCFW recursion relation,
as shown in figure \ref{fig-9}. Each matrix configuration consists of a subset of BCFW cells, or more explicitly,
Grassmannian geometry representatives, of various $k_\textrm{L}$ and $k_\textrm{R}$ satisfying
$k_\textrm{L}\!+\!k_\textrm{R}\!=\!k\!-\!1\!\geq\!0$ and
$0\!\leq\!k_\textrm{L,\,R}\!\leq\!n_\textrm{L,\,R}\!-\!4$, with $k_\textrm{L}\!=\!0$ for
$n_\textrm{L}\!=\!3$ as the only special case.

Finally, we discuss the arbitrariness of $c_{i-1}$'s location. While $c_i,c_{i+1}$ and $c_j,c_{j+1}$ are placed
symmetrically with respect to $Y_\textrm{L}$ and $Y_\textrm{R}$, with their induced matrix structure
fixed by factorization limits, $c_{i-1}$ seems to be placed with more uncertainty.
If we place it adjacent to $c_{i+1}$ so it now should be $c_{i+2}$,
we can easily see this is equivalent to the original choice,
up to a color reflection which swaps $Y_\textrm{L}$ and $Y_\textrm{R}$.
What about the choices of $c_{j-1}$ and $c_{j+2}$?
Since to maintain the matrix structure of factorization limits, we can no longer remove or add any sub-column,
so $c_{j-1}$ may induce a BCFW deformation for $L_{j-1}$ and nothing else, and similarly for $c_{j+2}$.
It is unclear what physics this matrix structure conceals yet, as we will leave it to the future work.

\subsection{1-loop BCFW recursion relation from positivity}

\begin{figure}
\begin{center}
\includegraphics[width=0.65\textwidth]{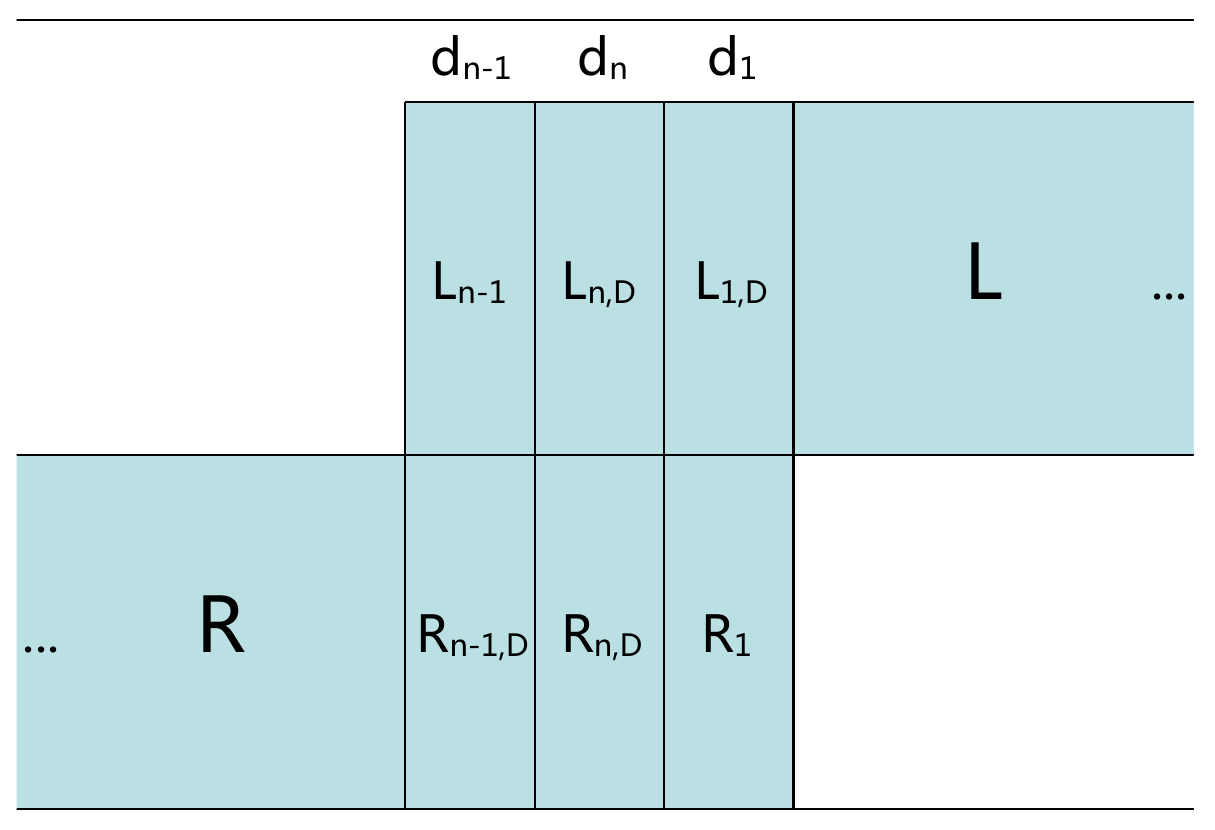}
\caption{Not manifestly positive matrix structure containing three overlapped sub-columns.} \label{fig-10}
\end{center}
\end{figure}

The structure represented by columns $j,j\!+\!1$ in figure \ref{fig-9} is a typical sector of BCFW deformations,
while the latter will be generalized to loop level. As we have already demonstrated,
the relevant two overlapped sub-columns encode the characteristic of factorization limits.
It is then natural to consider the structure that has three overlapped sub-columns as its simplest generalization,
which nicely happens to encode the characteristic of forward limits.
For this purpose, in figure \ref{fig-10}, we construct a matrix structure with
\be
L_{n,\,\textrm{D}}=L_n+\frac{d_n}{d_{n-1}}L_{n-1},~~
L_{1,\,\textrm{D}}=L_1+\frac{d_1}{d_n}L_n+\frac{d_1}{d_{n-1}}L_{n-1},
\ee
and similarly
\be
R_{n,\,\textrm{D}}=R_n+\frac{d_n}{d_1}R_1,~~
R_{n-1,\,\textrm{D}}=R_{n-1}+\frac{d_{n-1}}{d_n}R_n+\frac{d_{n-1}}{d_1}R_1,
\ee
note that although column 1 has been specified, we neglect the sign factors for now.
However, we immediately find that it is not manifestly positive, unlike the case with two overlapped sub-columns.

\begin{figure}
\begin{center}
\includegraphics[width=0.65\textwidth]{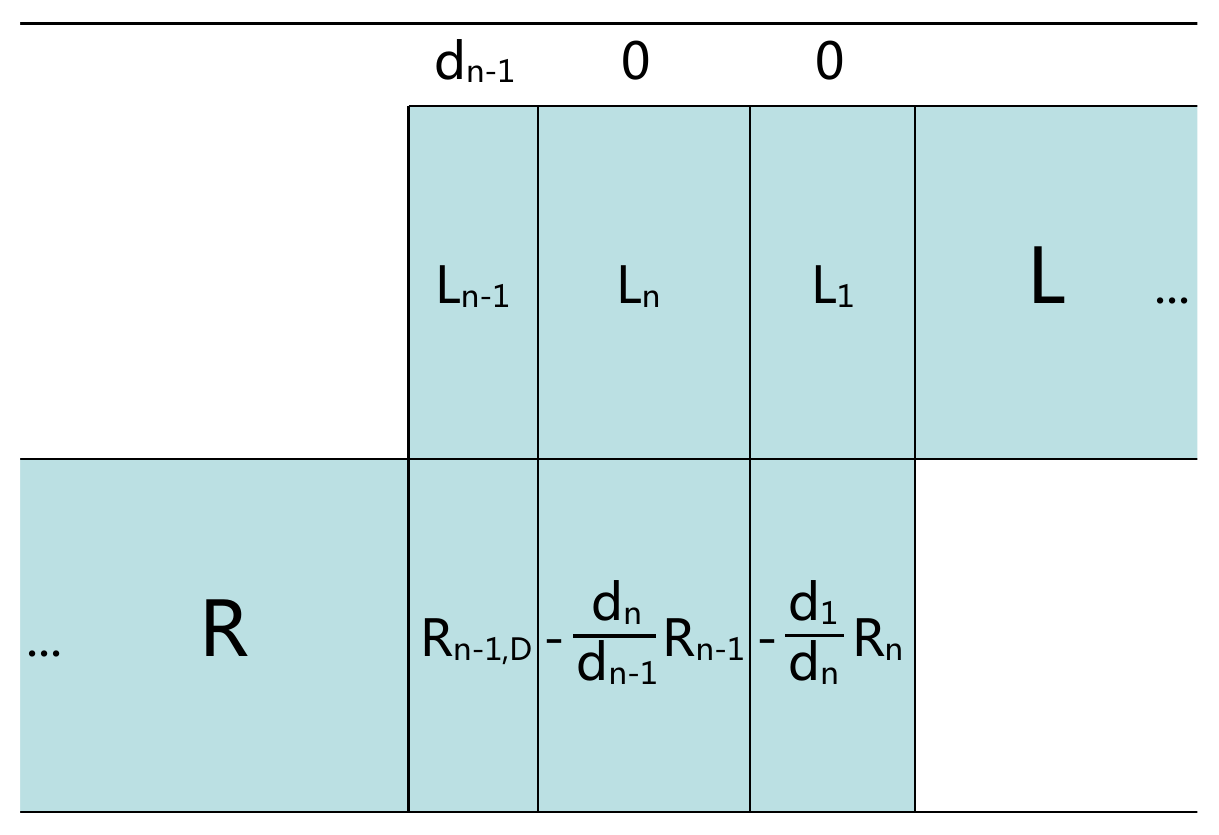}
\caption{For any minor containing columns $n\!-\!1,n,1$,
$d_n$ and $d_1$ can be removed by shifting columns $n,1$.} \label{fig-11}
\end{center}
\end{figure}

To see why, one may consider any minor involving all or part of these three columns. The nontrivial cases are
those involving two and three columns, while the former is similar to that in figure \ref{fig-9}, the latter
needs further manipulation for ensuring positivity. Explicitly, for a minor containing columns $n\!-\!1,n,1$,
we can shift columns $n,1$ by proper constants times column $n\!-\!1$ and obtain the configuration in figure \ref{fig-11},
where $d_n$ and $d_1$ in the top row are removed by this shift.

Now there are cross terms in the minor, proportional to
\be
-\frac{d_1}{d_n}(L_n\ldots)(R_n\ldots)+\frac{d_n}{d_{n-1}}(L_1\ldots)(R_{n-1}\ldots),
\ee
where $\ldots$ denote the rest $(k_\textrm{L}\!-\!1)$ and $(k_\textrm{R}\!-\!1)$ columns repectively.
To render it subtraction-free, we can turn off one of $L_n$, $L_1$, $R_{n-1}$ and $R_n$, while the choice here is $L_n$.
Choosing one sub-column of $Y_\textrm{L}$ or $Y_\textrm{R}$ in fact makes no difference, since we treat them symmetrically.
Then, between $L_n$ and $L_1$ it is physically more natural to preserve $L_1$ for $Y_\textrm{L}$, as this structure will be
engineered to reproduce forward limits.

\begin{figure}
\begin{center}
\includegraphics[width=0.65\textwidth]{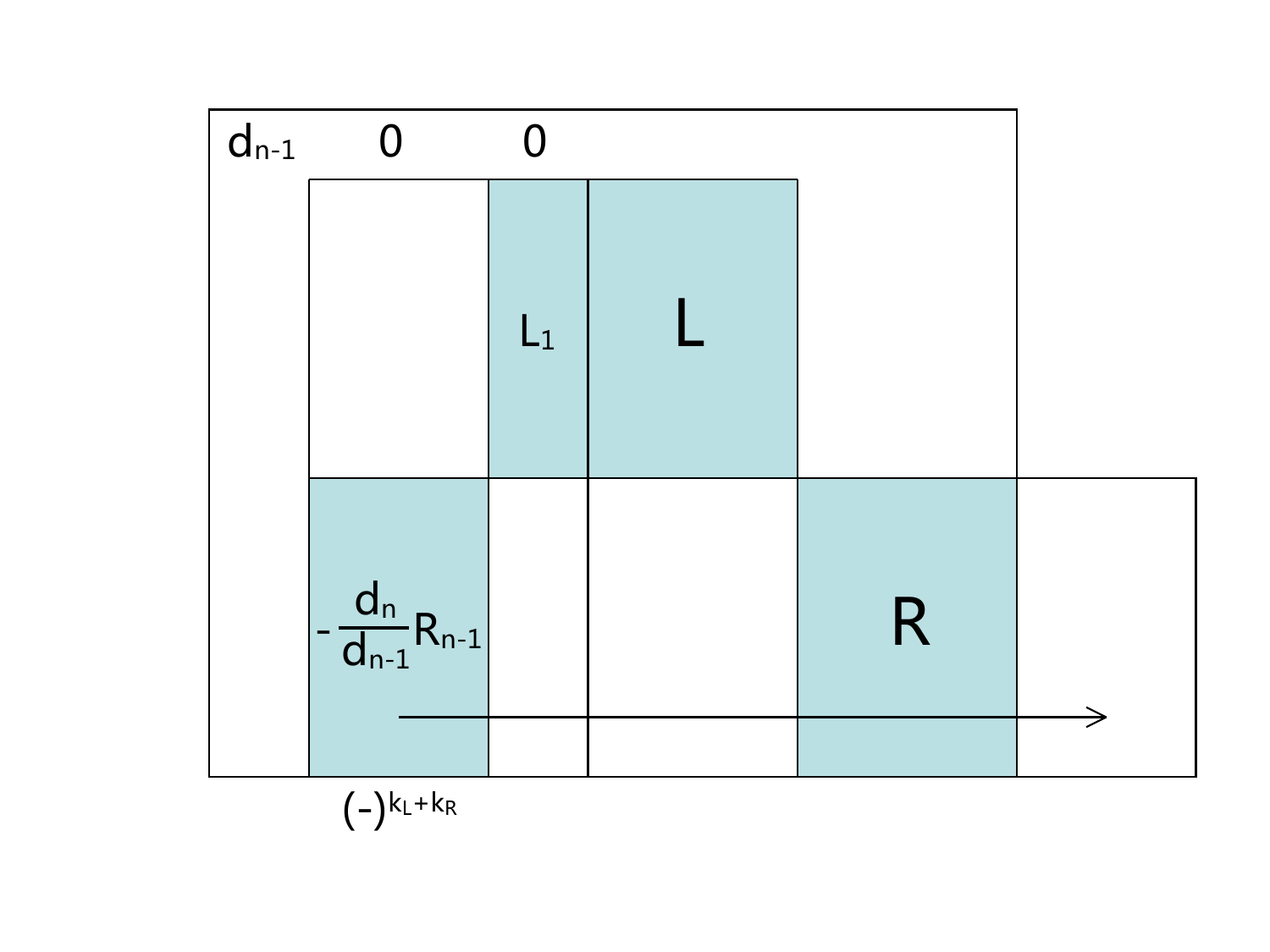}
\caption{A particular minor containing columns $n\!-\!1,n,1$ after $L_n$ is turned off.} \label{fig-12}
\end{center}
\end{figure}

After positivity is ensured we need to fix the sign factors, as shown in figure \ref{fig-12}.
There, we consider a particular minor containing columns $n\!-\!1,n,1$,
and since $L_n$ is turned off, $R_n$ does not contribute to this determinant.
Then we choose $(k_\textrm{L}\!-\!1)$ sub-columns besides $L_1$ and $(k_\textrm{R}\!-\!1)$ sub-columns
besides $R_{n-1}$ for the corresponding sub-minors, but $R_{n-1}$ must be
pulled through $(k_\textrm{L}\!+\!k_\textrm{R}\!-\!1)$ columns on its right
to ensure the sub-minor of $Y_\textrm{R}$ is correctly ordered. To offset its induced sign factor
$(-)^{k_\textrm{L}+k_\textrm{R}-1+1}$, where the extra minus sign comes from $-d_n/d_{n-1}$,
we should add a same factor to $R_{n-1}$.
To maintain the legitimacy of shifting columns $n,1$, the same factor should also be added to
$R_n$ and $R_1$. Therefore, we require the sign factor $(-)^{k_\textrm{L}+k_\textrm{R}}$ for $R_{n-1}$,
$R_n$ and $R_1$ in figure \ref{fig-10}.

\begin{figure}
\begin{center}
\includegraphics[width=0.65\textwidth]{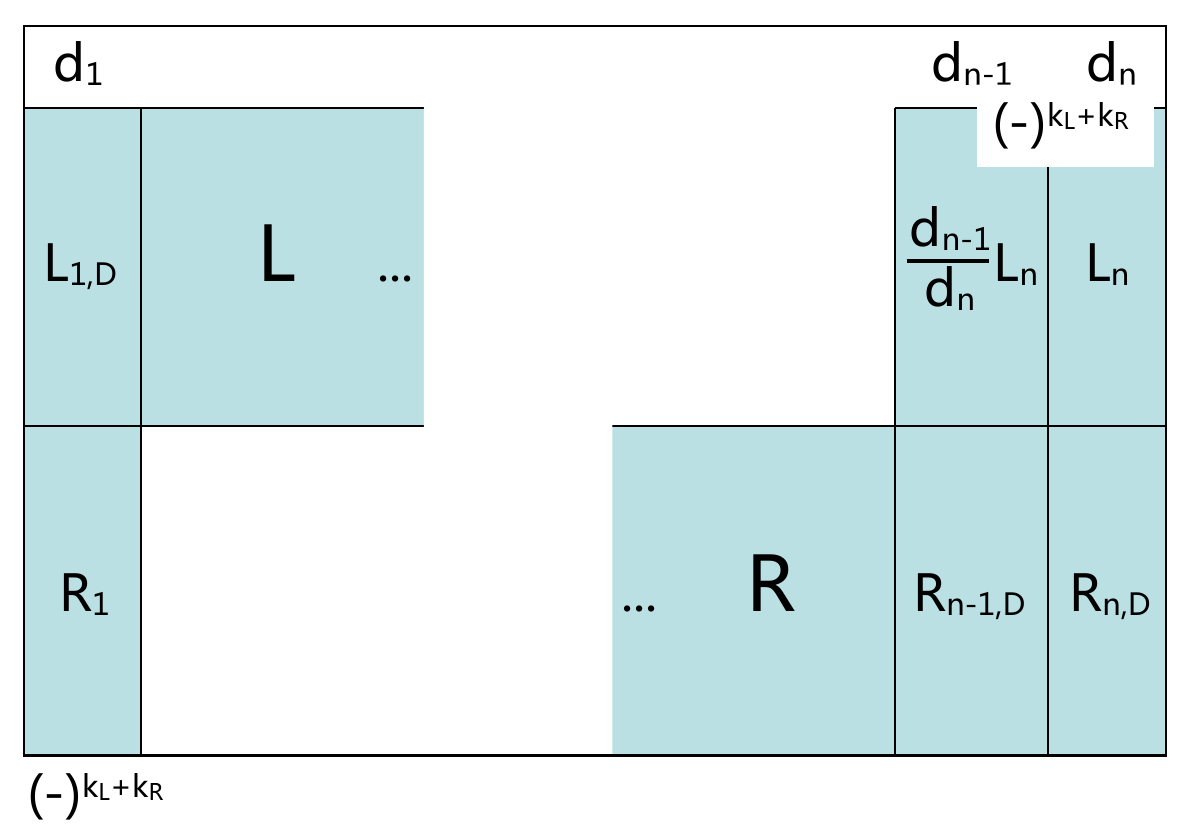}
\caption{Manifestly positive matrix structure containing three overlapped sub-columns.} \label{fig-13}
\end{center}
\end{figure}

However, in figures \ref{fig-10} and \ref{fig-11}, label 1 is not designated to be the first column, since we
intentionally use this nonstandard ordering to make columns $n\!-\!1,n,1$ align adjacently.
If we switch to the standard one, the sign factors of $R_{n-1,\,\textrm{D}}$ and $R_{n,\,\textrm{D}}$ are offset,
but $d_{n-1}$, $d_n$, $L_{n-1}$ and $L_{n,\,\textrm{D}}$ acquire the same factor due to twisted cyclicity!
In brief, positivity demands us to add factor $(-)^{k_\textrm{L}+k_\textrm{R}}$
to $d_{n-1}$, $d_n$, $L_n$ and $R_1$, as shown in figure \ref{fig-13}.
There, a trivial replacement $L_{n-1}\!\to\!d_{n-1}/d_n\!\cdot\!L_n$ is used to make the sub-columns of $Y_\textrm{L}$
consecutive. We must emphasize that, the ``larger positivity'' demands
$Y_\textrm{L}$ and $Y_\textrm{R}$ to be positive with respect to orderings $(n,1,\ldots)$
and $(\ldots,n\!-\!1,n,1)$ of the corresponding sub-columns.

\begin{figure}
\begin{center}
\includegraphics[width=0.65\textwidth]{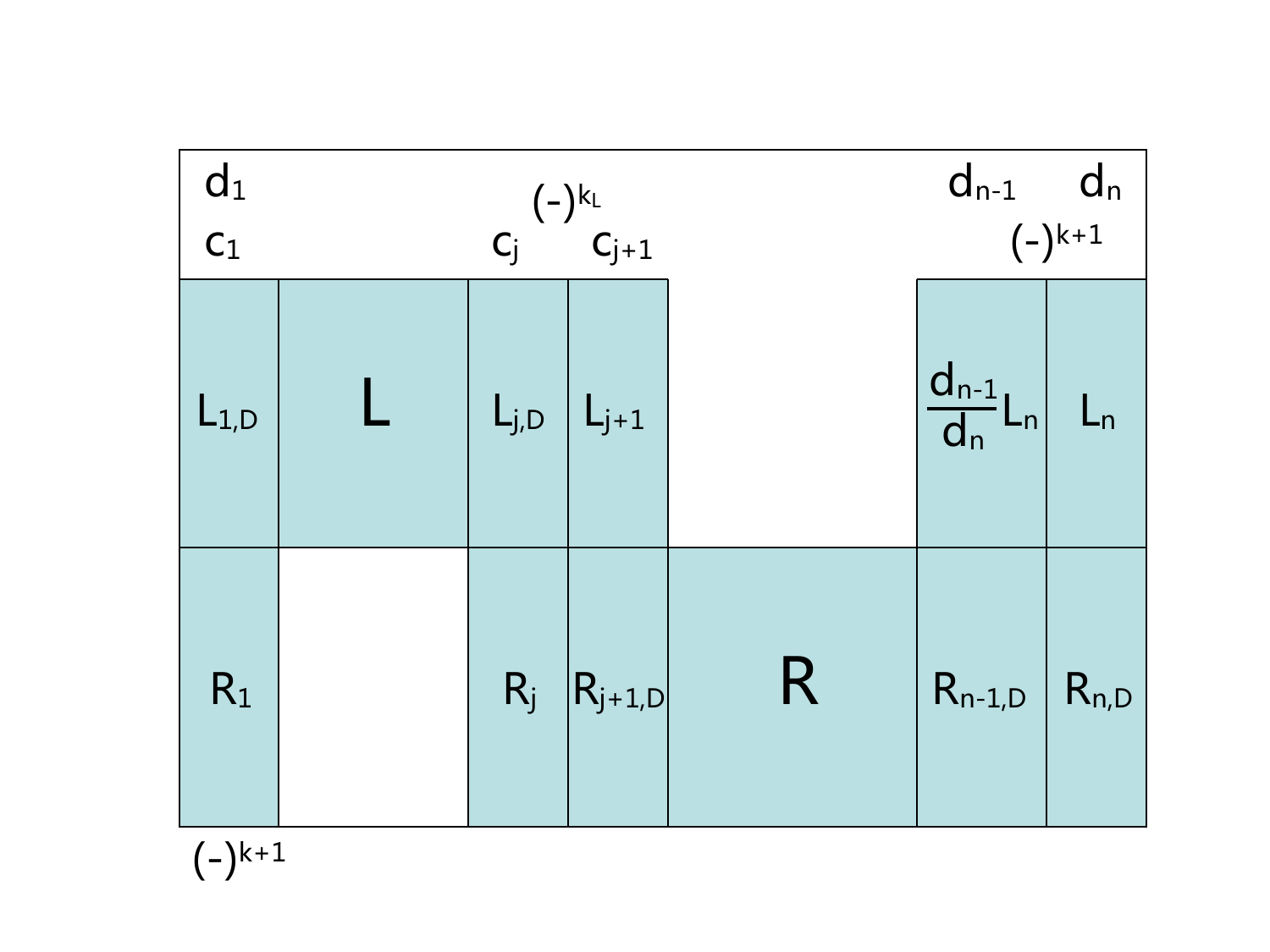}
\caption{Matrix version of 1-loop BCFW forward-limit terms.} \label{fig-14}
\end{center}
\end{figure}

\begin{figure}
\begin{center}
\includegraphics[width=0.5\textwidth]{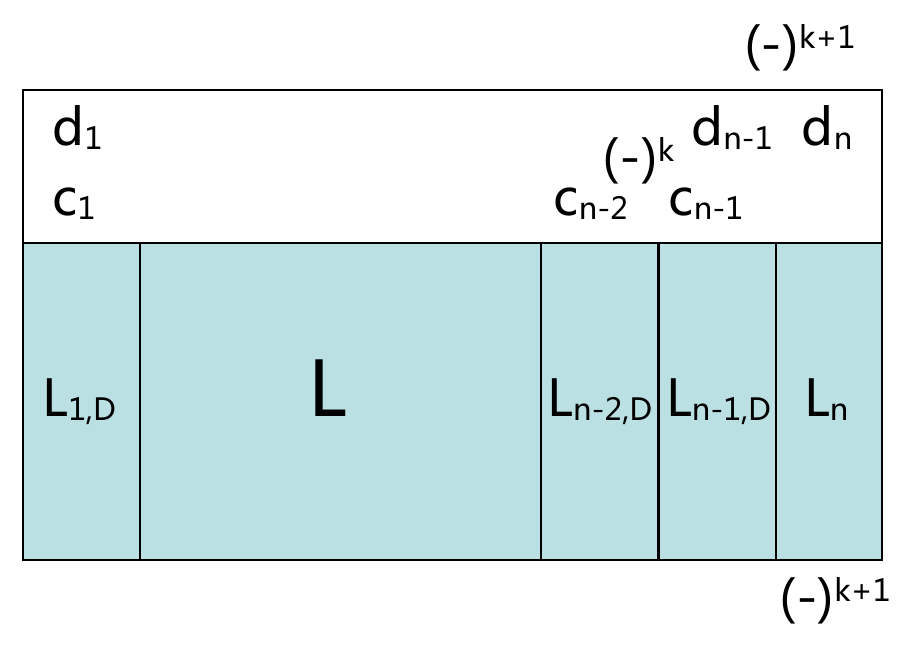}
\caption{Nontrivial boundary case $j\!=\!n\!-\!2$ for which column $j\!+\!1$ coincides with $n\!-\!1$.} \label{fig-15}
\end{center}
\end{figure}

Now, let's clarify the unspecified internal area of figure \ref{fig-13}. According to the 1-loop kermit expansion
in \cite{Bourjaily:2013mma}, we should place a matrix structure encoding factorization limits in this area,
since the closed-form formula for 1-loop forward-limit terms is expressed in terms of (deformed) tree amplitudes.
The resulting matrix structure is presented in figure \ref{fig-14}, with
\be
L_{1,\,\textrm{D}}=L_1+\frac{d_1}{d_n}L_n,~~R_{n,\,\textrm{D}}=R_n+\frac{d_n}{d_1}R_1,~~
R_{n-1,\,\textrm{D}}=R_{n-1}+\frac{d_{n-1}}{d_n}R_n+\frac{d_{n-1}}{d_1}R_1,
\ee
and
\be
L_{j,\,\textrm{D}}=L_j+\frac{c_j}{c_{j+1}}L_{j+1},~~R_{j+1,\,\textrm{D}}=R_{j+1}+\frac{c_{j+1}}{c_j}R_j.
\ee
We see the additional row $(c_1,\ldots,c_j,c_{j+1},\ldots)$ is the same as that in figure \ref{fig-9},
including the sign factors. Slightly different from figure \ref{fig-13}, the uniform sign factor of
$d_{n-1}$, $d_n$, $L_n$ and $R_1$ is now $(-)^{k+1}$ instead of $(-)^{k}$
where $k\!=\!k_\textrm{L}\!+\!k_\textrm{R}$, due to twisted cyclicity with one additional row.
More explicitly, $Y_\textrm{L}$ and $Y_\textrm{R}$ are positive with respect to orderings $(n,1,\ldots,j,j\!+\!1)$
and $(\,j,j\!+\!1,\ldots,n\!-\!1,n,1)$ separately.
Since $Y_\textrm{L}$ and $Y_\textrm{R}$ represent tree amplitudes, their special ``boundary'' cases
are similar to those of figure \ref{fig-9}:
for $n_\textrm{L}\!=\!4$, $Y_\textrm{L}$ trivially disappears as $k_\textrm{L}\!=\!0$,
and so does $Y_\textrm{R}$ for $n_\textrm{R}\!=\!4$, with $n\!=\!n_\textrm{L}\!+\!n_\textrm{R}\!-\!4$.
The latter case is nontrivial of which $j\!=\!n\!-\!2$, so that column $j\!+\!1$ coincides with $n\!-\!1$.
As shown in figure \ref{fig-15}, we have
\be
L_{n-1,\,\textrm{D}}=L_{n-1}+(-)^{k+1}\frac{d_{n-1}}{d_n}L_n,
\ee
where columns $j\!+\!1,n\!-\!1$ trivially merge as one. Naively, one may find the sign factor $(-)^{k+1}$
incorrect, but keep in mind that the positive ordering of $Y_\textrm{L}$ is $(n,1,\ldots,j,j\!+\!1)$!
This sign factor exactly preserves positivity after $L_n$ is pulled through $(k\!-\!1)$ columns on its left
for a non-vanishing minor.

We have not explained the sudden appearance of $c_1$. It can be also regarded as a regulator, similar to
that in figure \ref{fig-8}. More essentially, it is mandatory for the 1-loop kermit expansion:
the two rows made of $d_1,d_{n-1},d_n$ and $c_1,c_j,c_{j+1}$ form a $2\!\times\!n$ $D$-matrix
\cite{Arkani-Hamed:2013jha}, for which
\be
D_{\ap a}Z_a=(AB)\cap X_\ap,~~X_\ap=((1\,n\!-\!1\,n),(1\,j\,j\!+\!1)) \labell{eq-15}
\ee
such that, up to sign factors
\be
d_1=\<AB\,n\!-\!1\,n\>,~d_{n-1}=\<AB\,n\,1\>,~d_n=\<AB\,1\,n\!-\!1\>,
\ee
and similarly
\be
c_1=\<AB\,j\,j\!+\!1\>,~c_j=\<AB\,j\!+\!1\,1\>,~c_{j+1}=\<AB\,1\,j\>,
\ee
here $A,B$ are momentum twistors of loop variables which have four degrees of freedom modulo the GL(2) redundancy.
It is clear that without $c_1$ there would be a physical singularity $c_1\!=\!\<AB\,j\,j\!+\!1\>\!\to\!0$,
on the support of a novel delta function imposing \eqref{eq-15} (which will be investigated in the future work).

We may also discuss the arbitrariness of $c_1$'s location. The kermit expansion only allows two options
that are adjacent to $A,B$: $c_n$ and $c_1$, since conventionally, $A,B$ are inserted between columns $n,1$.
If we place it adjacent to $A$ so it should be $c_n$, positivity will be violated.
To see why, let's consider a minor containing columns $1,n$, positivity induces a sign factor $(-)^k$ for $c_n$.
Then for a minor containing columns $n\!-\!1,n$, positivity induces a sign factor $(-)^{k+1}$ for $c_n$,
which contradicts with the former.

Knowing the physical significance of $d_i$'s and $c_i$'s,
it is easy to reproduce the familiar 1-loop BCFW forward-limit product of Yangian invariants
\be
Y_\textrm{L}(\widehat{A},1,\ldots,j,I)\,Y_\textrm{R}(I,j\!+\!1,\ldots,\widehat{n},\widehat{A}\,)
\ee
with $\widehat{\mathcal{Z}}_I\!=\!(\,j\,j\!+\!1)\cap(1AB)$, $\widehat{\mathcal{Z}}_n\!=\!(n\!-\!1\,n)\cap(1AB)$
and $\widehat{A}\!=\!(AB)\cap(n\!-\!1\,n\,1)$.
But its corresponding integrand should also include the kermit function, namely
\be
K[(1\,j\,j\!+\!1);(1\,n\!-\!1\,n)]=\frac{-\,\<AB|(1\,j\,j\!+\!1)\cap(1\,n\!-\!1\,n)\>^2}
{\<AB\,1\,j\>\<AB\,j\,j\!+\!1\>\<AB\,j\!+\!1\,1\>\<AB\,1\,n\!-\!1\>\<AB\,n\!-\!1\,n\>\<AB\,n\,1\>},
\ee
which is a rational function of $d_i$'s and $c_i$'s. Due to the GL(1) redundancies from forward limits, in which
$d_i$'s and $c_i$'s originate from two 5-brackets, we can fix one of $d_i$'s to be unity
and similarly for $c_i$'s, hence they have only four degrees of freedom.
We will leave the formulation of the kermit part from positivity to the future work,
for now let's be content with the product of Yangian invariants above.

\begin{figure}
\begin{center}
\includegraphics[width=0.65\textwidth]{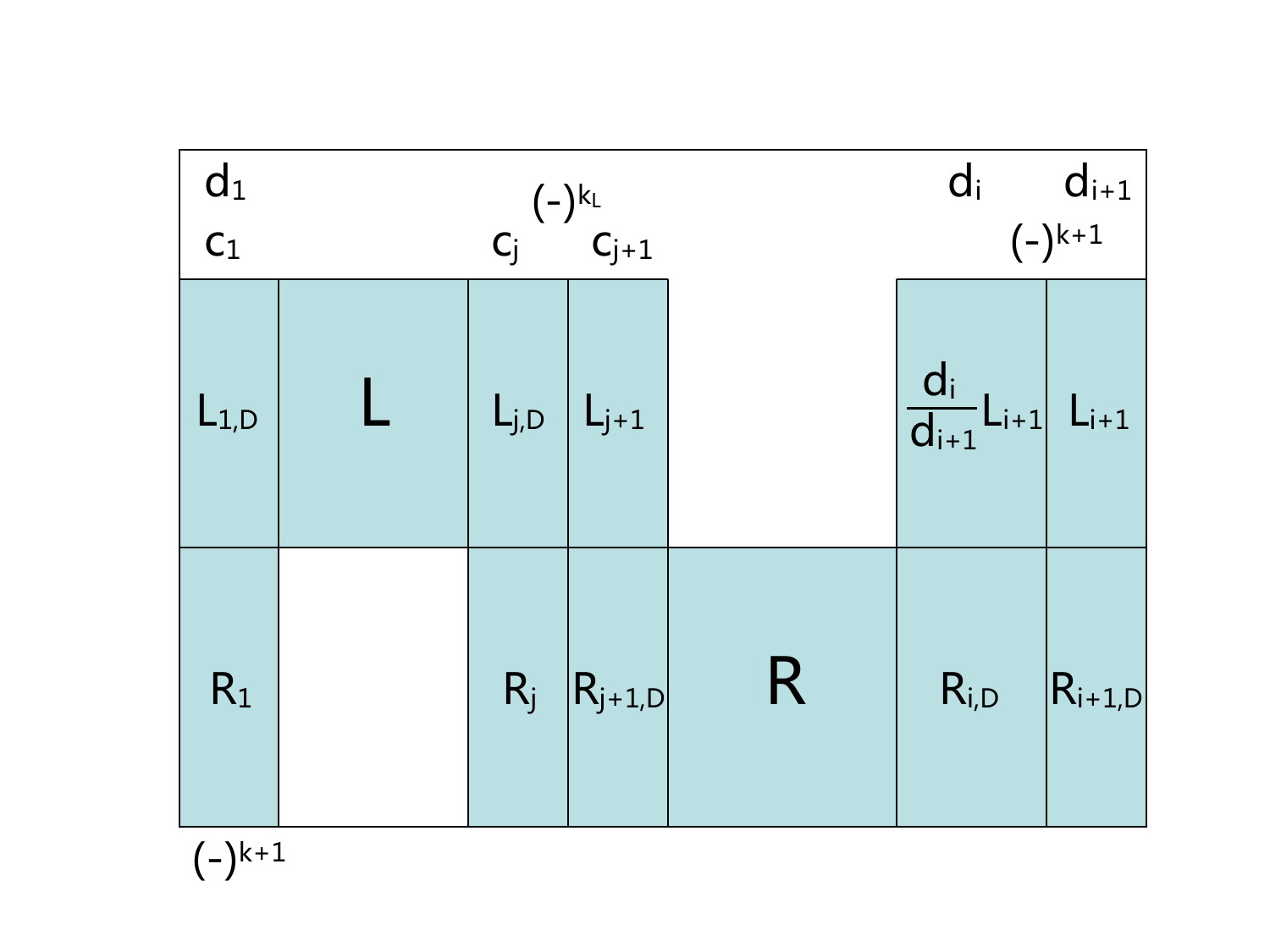}
\caption{Generic matrix configuration of 1-loop BCFW forward-limit terms labeled by $i,j$.} \label{fig-16}
\end{center}
\end{figure}

Summing all possible forward limits and 1-loop factorization limits, we can express a general 1-loop integrand as
\be
Y^k_{n,\,l=1}=\sum_{i=k+3}^{n-1}\sum_{j=2}^{i-1}\,\textrm{FWD}_{\,i,\,j}
+\sum_{i=k+3}^{n-1}\sum_{j=2}^{i-2}\,\textrm{FAC}^{\,\textrm{1-loop}}_{\,i,\,j},
\ee
where $\textrm{FWD}_{\,i,\,j}$ represents the generic matrix configuration of 1-loop BCFW forward-limit terms
as shown in figure \ref{fig-16}, and $\textrm{FAC}^{\,\textrm{1-loop}}_{\,i,\,j}$ represents that of
1-loop BCFW factorization-limit terms.
Each forward-limit term consists of a subset of BCFW cells
of various $k_\textrm{L}$ and $k_\textrm{R}$, similarly satisfying
$k_\textrm{L}\!+\!k_\textrm{R}\!=\!k\!\geq\!0$ and $0\!\leq\!k_\textrm{L,\,R}\!\leq\!n_\textrm{L,\,R}\!-\!4$.
Each 1-loop factorization-limit term is identical to its tree counterpart in figure \ref{fig-9},
except that one of two sub-amplitudes is a 1-loop integrand expressed by FWD terms, of which $n\!\geq\!4$
as there is no 1-loop $n\!=\!3$ integrand. Be aware of the difference between these two types of contributions:
forward-limit terms start with $k\!=\!0$, while 1-loop factorization-limit terms start with $k\!=\!1$.

The BCFW cells (or Grassmannian geometry representatives) of forward-limit terms in fact
must be decomposed into two sectors: the kermit part (or $D$-matrix), and the product of Yangian invariants.
Only the latter takes part in the Grassmannian geometry strictly speaking, and its matrix structure below
the $D$-matrix is also positive due to $(-)^{k+1}\!=\!(-)^{k-1}$! In fact, for both FAC and FWD terms,
the $D$-matrix can be freely shifted up and down as a single row with positivity preserved.

In appendix \ref{app1}, we present the 1-loop NMHV $n\!=\!6$ integrand in terms of matrix representatives
as one example. It has 16 terms, of which the pure $n\!=\!6$ contribution is separated from the $n\!=\!5$ part, and
the FAC and FWD terms are distinguished as well.

In passing, the number of BCFW terms in 1-loop integrands is given by
\be
N^k_{n,\,l=1}=\binom{n\!-\!2}{k}\binom{n\!-\!2}{k\!+\!2}.
\ee
For comparison, the number of BCFW terms in tree amplitudes is given by
\be
N^k_n=\frac{1}{n\!-\!3}\binom{n\!-\!3}{k}\binom{n\!-\!3}{k\!+\!1}, \labell{eq-36}
\ee
which is the \textit{Narayana number} $N(n\!-\!3,k\!+\!1)$. This number will be useful in section \ref{sec6}, when we
further dissect tree amplitudes to reveal deeper structures.

\newpage
\section{Applications to Homological Identities}
\label{sec4}

This section applies positive Grassmannian geometry and Pl\"{u}cker coordinates
to homological identities, including identifying all relevant
boundaries from a given generator and determining their relative signs.
The simple 6-term NMHV identity proves to be essential for most of the signs, as they are associated to
BCFW-like cells only, while the rest ones are associated to more complicated leading singularities.

\subsection{NMHV identity and cyclicity of NMHV amplitudes}

In terms of empty slots denoting removed entries, the NMHV $n\!=\!6$ identity is expressed as
\be
0=-\,[1]+[2]-[3]+[4]-[5]+[6], \labell{eq-16}
\ee
which manifests the cyclicity of NMHV $n\!=\!6$ amplitude. This is the only distinct NMHV identity, and in fact it accounts
for the cyclicity of NMHV amplitudes for any $n$. To prove it, naturally we need to write NMHV amplitudes also
in terms of empty slots, in the triangle form as \eqref{eq-5}. Let's directly take \eqref{eq-5} as an example,
which is copied below
\be
Y^1_8=
\(\begin{array}{cccc}
{} & {} & {} & [234] \\
{} & {} & [238] & [236] \\
{} & [278] & [258] & [256] \\
{[678]} & [478] & [458] & [456]
\end{array}\).
\ee
Now we subtract its cyclicly shifted (by $+1$) counterpart from itself as
\be
Y^1_8-Y^1_{8,+1}=
\(\begin{array}{cccc}
{} & {} & {} & [\textbf{2}34] \\
{} & {} & [238] & [236] \\
{} & [278] & [258] & [256] \\
{[678]} & [478] & [458] & [456]
\end{array}\)-
\(\begin{array}{cccc}
{} & {} & {} & [34\textbf{5}] \\
{} & {} & [34\textbf{1}] & [34\textbf{7}] \\
{} & [381] & [361] & [367] \\
{[781]} & [581] & [561] & [567]
\end{array}\),
\ee
the four terms containing bold labels above are
\be
[234]-[341]-[345]-[347]=[34]\,([2]-[1]-[5]-[7])=[34]\,(-\,[6]-[8]),
\ee
where we have used the NMHV identity for entries 1,2,5,6,7,8. Now we have
\be
Y^1_8-Y^1_{8,+1}=
\(\begin{array}{cccc}
{} & {} & {} & 0 \\
{} & {} & [238] & [\textbf{2}36] \\
{} & [278] & [258] & [256] \\
{[678]} & [478] & [458] & [456]
\end{array}\)-
\(\begin{array}{cccc}
{} & {} & {} & 0 \\
{} & {} & [348] & [3\textbf{4}6] \\
{} & [381] & [36\textbf{1}] & [36\textbf{7}] \\
{[781]} & [581] & [561] & [567]
\end{array}\),
\ee
again
\be
[236]-[361]-[346]-[367]=[36]\,([2]-[1]-[4]-[7])=[36]\,(-\,[5]-[8]),
\ee
and then
\be
\bal
Y^1_8-Y^1_{8,+1}&=
\(\begin{array}{cccc}
{} & {} & {} & 0 \\
{} & {} & [238] & 0 \\
{} & [278] & [258] & [\textbf{2}56] \\
{[678]} & [478] & [458] & [\textbf{4}56]
\end{array}\)-
\(\begin{array}{cccc}
{} & {} & {} & 0 \\
{} & {} & [348] & 0 \\
{} & [381] & [368] & [\textbf{3}65] \\
{[781]} & [581] & [56\textbf{1}] & [56\textbf{7}]
\end{array}\)\\
&=
\(\begin{array}{cccc}
{} & {} & {} & 0 \\
{} & {} & [238] & 0 \\
{} & [278] & [258] & 0 \\
{[678]} & [478] & [458] & 0
\end{array}\)-
\(\begin{array}{cccc}
{} & {} & {} & 0 \\
{} & {} & [348] & 0 \\
{} & [381] & [368] & 0 \\
{[781]} & [581] & [56\textbf{8}] & 0
\end{array}\)\\
&=
\(\begin{array}{ccc}
{} & {} & [238] \\
{} & [278] & [258] \\
{[678]} & [478] & [458]
\end{array}\)-
\(\begin{array}{ccc}
{} & {} & [348] \\
{} & [318] & [368] \\
{[718]} & [518] & [568]
\end{array}\)\\
&=[8]\,(Y^1_7-Y^1_{7,+1}),
\eal
\ee
which nicely reduces proving the cyclicity of $Y^1_8$ to that of $Y^1_7$. In general, one can easily show that
\be
Y^1_n-Y^1_{n,+1}=[n]\,(Y^1_{n-1}-Y^1_{n-1,+1}),
\ee
hence the cyclicity of NMHV amplitudes for any $n$ can be proved by induction, as long as the $n\!=\!6$ case holds.
However, one may doubt its validity even with respect to the simplest $n\!=\!6$ identity.
It is a subtle matter to determine relative signs without explicitly
analyzing the corresponding Yangian invariants, as we prefer to use their geometric avatars.
But curiously, we find this straightforward approach more than just tedious calculation. In terms of 5-brackets,
\eqref{eq-16} returns to its traditional form
\be
0=[12345]-[12346]+[12356]-[12456]+[13456]-[23456].
\ee
To prove it, we use the BCFW deformation $\mathcal{Z}_5\!\to\!\mathcal{Z}_5\!+\!z\mathcal{Z}_6$, under which
$[12345]$ has four poles at finite locations: $\<2345\>$, $\<3451\>$, $\<4512\>$ and $\<5123\>$.
Plus the poles at zero and infinity, we get six terms as
\be
\sum_\textrm{finite}\oint\frac{dz}{z}\,[1234\,(5\!+\!z\,6)\,]=-\,[23456]+[13456]-[12456]+[12356],
\ee
and
\be
\oint_0\frac{dz}{z}\,[1234\,(5\!+\!z\,6)\,]=[12345],~-\oint_\infty\frac{dz}{z}\,[1234\,(5\!+\!z\,6)\,]=-\,[12346],
\ee
altogether, we have
\be
0=\oint\frac{dz}{z}\,[1234\,(5\!+\!z\,6)\,]
=\(-\oint_\infty+\oint_0+\sum_\textrm{finite}\oint\)\frac{dz}{z}\,[1234\,(5\!+\!z\,6)\,],
\ee
which matches the NMHV $n\!=\!6$ identity above. Its signs vary simply due to the definition of 5-bracket
so that we can circumvent the ``orientations'' of Grassmannian cells, which could be ambiguous.
In the next subsection, we will see an immediate generalization of
this approach, as it more nontrivially leads to the validity of an N$^2$MHV $n\!=\!7$ identity.

\subsection{N$^2$MHV identities}

As $k$ goes up, homological identities become much more intricate. While there is only one distinct
NMHV identity, there are 24 distinct N$^2$MHV identities spanning from $n\!=\!7$ to $n\!=\!11$ and
some of the involved Yangian invariants do not correspond to simple BCFW-like cells,
which means the identities encode more information than solely the cyclicity of N$^2$MHV amplitudes.

The first example is the N$^2$MHV $n\!=\!7$ identity
\be
0=\partial(12)=-\,[1]+[2]-(12)(34)+(12)(45)-(12)(56)+(12)(67), \labell{eq-18}
\ee
where the \textit{generator} $(12)$ is a 9-dimensional cell that generates the vanishing identity as a
consequence of the residue theorem. Besides empty slots, we also have nontrivial Grassmannian geometry
representatives as boundaries of $(12)$. This relation can be output
by the \textsc{Mathematica} package ``\verb"positroids"'',
and one needs to translate the permutations to Grassmannian geometry representatives as mentioned in section \ref{sec2}.
There are two more vanishing terms $(712)_2$ and $(123)_2$, which fail to have kinematical supports.
From the geometric perspective, it is trivial to explain this. For example, one form of the matrix representative
for $(123)_2$ is
\be
C=\(\begin{array}{ccccccc}
1\, & \De_{24} & \De_{34} & \,0\, & -\De_{45} & -\De_{46} & -\De_{47} \\
0\, & 0 & 0 & \,1\, & \De_{15} & \De_{16} & \De_{17}
\end{array}\),
\ee
where the second row only has three degrees of freedom signifying a singularity, but the first row has
five degrees of freedom signifying an extra integration. Note that we have used Pl\"{u}cker coordinates
rather than positive variables in the $d\log$ form, so not all positive conditions are manifest.

Since $(12)$ is the only possible boundary generator up to a cyclic shift at $n\!=\!7$, it must guarantee
the cyclicity of N$^2$MHV $n\!=\!7$ amplitude. Explicitly, we find
\be
Y^2_7-Y^2_{7,+1}=-\,\partial(23)-\partial(56)-\partial(71),
\ee
where the expression of $Y^2_7$ in \eqref{eq-2}, and its cyclicly shifted (by $+1$) counterpart are used.

One may ask whether $(13)$ can be a generator, the answer is no because of positivity.
Let's write one of its matrix representatives as
\be
C=\(\begin{array}{ccccccc}
1\, & \De_{24} & \pm\De_{34} & \,0\, & -\De_{45} & -\De_{46} & -\De_{47} \\
0\, & \De_{12} & 0 & \,1\, & \De_{15} & \De_{16} & \De_{17}
\end{array}\),
\ee
now no matter what sign $\De_{34}$ takes, minors $(23)$ and $(34)$ cannot be positive simultaneously.
In this sense, positivity demands all vanishing constraints to be consecutive. If we insist on imposing
$(13)\!=\!0$, the only legitimate choice is $(123)_2$ minimally.

To manifest positivity (or consecutiveness), we modify the Pl\"{u}ckerian integral \eqref{eq-17} slightly
so that up to an overall sign factor, its denominator part becomes
\be
\frac{1}{\De_{12\,\ldots\,k}\,\De_{23\,\ldots\,k+1}\ldots\De_{1\,\ldots\,k-1\,n}}\,, \labell{eq-19}
\ee
where all $\De$'s are positive ordered minors. When one of them vanishes, a singularity appears and Pl\"{u}cker
relations may \textit{factorize}. Therefore, although there are only cyclicly consecutive Pl\"{u}cker coordinates in the
denominator, we may use factorized Pl\"{u}cker relations to reveal more singularities!

Let's finish proving \eqref{eq-18}, of which the denominator part is
\be
\frac{1}{\De_{12}\De_{23}\De_{34}\De_{45}\De_{56}\De_{67}\De_{17}}\,,
\ee
upon $\De_{12}\!=\!0$, we find a factorized Pl\"{u}cker relation
\be
\De_{23}\De_{17}=\De_{27}\De_{13}-\De_{37}\De_{12}=\De_{27}\De_{13},
\ee
if we fix $\De_{17}\!=\!1$, $\De_{23}\!=\!\De_{27}\De_{13}$ contains two singularities: $[2]$ and $(123)_2$.
For $[2]$, $\De_{12}\!=\!\De_{27}\!=\!0$ implies $[2]$ or $(712)_2$, but the latter is excluded by $\De_{17}\!=\!1$.
For $(123)_2$, $\De_{12}\!=\!\De_{13}\!=\!0$ implies $[1]$ or $(123)_2$, but the former is excluded by $\De_{17}\!=\!1$.
On the other hand, if we fix $\De_{23}\!=\!1$, the similar reasoning tells that
$\De_{17}\!=\!\De_{27}\De_{13}$ also contains two singularities: $(712)_2$ and $[1]$.

Since choosing different gauges can reveal different singularities, in general there are more boundaries from a given
generator, than those apparently show up in \eqref{eq-19}. Such implicit singularities are known as
``composite residues''. For the example \eqref{eq-18}, upon $\De_{12}\!=\!0$ the remaining poles are in fact
\be
\De_{23}\De_{34}\De_{45}\De_{56}\De_{67}\De_{17}\to
\(\frac{\De_{27}\De_{13}}{\De_{17}}\)\times\De_{34}\De_{45}\De_{56}\De_{67}\times
\(\frac{\De_{27}\De_{13}}{\De_{23}}\),
\ee
which clearly exhibit all eight boundaries including two vanishing cells:
\be
[2],~(123)_2,~(12)(34),~(12)(45),~(12)(56),~(12)(67),~(712)_2,~[1],
\ee
of course, they cannot be revealed only in one gauge, as we at least need $\De_{17}\!=\!1$ and $\De_{23}\!=\!1$.
It is useful to switch the gauge for suitably characterizing each cell, and then identify all possible boundary cells,
by using Pl\"{u}cker coordinates to trivialize this manipulation. Next, we need to determine their relative signs,
and what we have done in the NMHV case provides a hint for this purpose.
One can easily prove that the Yangian invariant of $(12)$ can be expressed as
\be
[\,1\,(2\!+\!z\,3)\,(34)\cap(567)\,(345)\cap(67)\,7\,]\,[34567],
\ee
where $z$ stands for the extra degree of freedom, from its matrix representative
\be
C=\(\begin{array}{ccccccc}
1\, & \ap_1 & (\ap_{234}\!+\!z)\ap_5 & \ap_{234}\ap_6 & \ap_{34}\ap_7 & \ap_4\ap_8 & \,0 \\
0\, & 0 & \ap_5 & \ap_6 & \ap_7 & \ap_8 & \,1
\end{array}\).
\ee
Again, the residue theorem gives
\be
0=\(\oint_\infty-\oint_0-\sum_\textrm{finite}\oint\)
\frac{dz}{z}\,[\,1\,(2\!+\!z\,3)\,(34)\cap(567)\,(345)\cap(67)\,7\,]\,[34567],
\ee
where explicitly
\be
\oint_\infty\frac{dz}{z}\,[\,1\,(2\!+\!z\,3)\,(34)\cap(567)\,(345)\cap(67)\,7\,]\,[34567]=[2],
\ee
\be
-\oint_0\frac{dz}{z}\,[\,1\,(2\!+\!z\,3)\,(34)\cap(567)\,(345)\cap(67)\,7\,]\,[34567]=-\,(12)(34),
\ee
\be
-\sum_\textrm{finite}\oint\frac{dz}{z}\,[\,1\,(2\!+\!z\,3)\,(34)\cap(567)\,(345)\cap(67)\,7\,]\,[34567]
=-\,[1]+(12)(45)-(12)(56)+(12)(67),
\ee
note that we have mapped the Yangian invariants back to their Grassmannian geometry representatives.
To check these results straightforwardly could be not convenient, hence we will use a better technique
by implementing the multivariate residue theorem, instead of the univariate one which uses $z$ only.
Still, we see the essence of this trick is to incarnate the NMHV $n\!=\!6$ identity into the given N$^2$MHV identity,
on the kinematical support of another simpler 5-bracket, which is $[34567]$ in this case.

Using the multivariate residue theorem for $(12)$, we may fix $\De_{14}\!=\!1$ so that its matrix is
\be
C=\(\begin{array}{ccccccc}
\os{-}{1}\, & \os{+}{\De_{24}} & \os{-}{\De_{34}} & \,0\, & -\os{+}{\De_{45}} & -\os{-}{\De_{46}} & -\os{+}{\De_{47}} \\
0\, & 0 & \De_{13} & \,1\, & \De_{15} & \De_{16} & \De_{17}
\end{array}\),
\ee
where the first row is associated with six alternating signs, then we can conceive an NMHV identity with respect to
these six entries. The entry 1 also has a sign because it can be traded for a Pl\"{u}cker coordinate,
if we fix $\De_{24}\!=\!1$ instead so that the matrix becomes
\be
C=\(\begin{array}{ccccccc}
\os{-}{\De_{14}} & \,\os{+}{1}\, & \os{-}{\De_{34}} & \,0\, & -\os{+}{\De_{45}} & -\os{-}{\De_{46}} & -\os{+}{\De_{47}} \\
0 & \,0\, & \De_{23} & \,1\, & \De_{25} & \De_{26} & \De_{27}
\end{array}\).
\ee
From these six entries, which will be individually removed, we read off part of the relative signs as
\be
-\,[1]+[2]-(12)(34)+(12)(45),
\ee
while $\De_{46}$ and $\De_{47}$ do not appear as singularities in cyclicly consecutive minors $(56)$, $(67)$ or $(17)$,
unlike $\De_{14}$, $\De_{24}$, $\De_{34}$ and $\De_{45}$ in minors $(17)$, $(23)$, $(34)$ and $(45)$ respectively.
Now, we switch to fix $\De_{16}\!=\!1$ so that the matrix is
\be
C=\(\begin{array}{ccccccc}
\os{-}{1}\, & \os{+}{\De_{26}} & \os{-}{\De_{36}} & \os{+}{\De_{46}} & \os{-}{\De_{56}} & \,0\, & -\os{+}{\De_{67}} \\
0\, & 0 & \De_{13} & \De_{14} & \De_{15} & \,1\, & \De_{17}
\end{array}\),
\ee
and this part of relative signs is given by
\be
-\,[1]+[2]-(12)(56)+(12)(67),
\ee
hence we have proved \eqref{eq-18}. Note that it is a convention to assign $[1]$ with $-1$, but once this is fixed,
we must assign $[2]$ with $+1$, for instance. The full identity contains two pieces, as the pivot part $-[1]\!+\![2]$
is crucial for connecting sign conventions of the rest boundary cells.

The second example is an N$^2$MHV $n\!=\!8$ identity
\be
\bal
0=\partial(123)_2(45)=&-[1](23)(45)+(13)[2](45)-(12)[3](45)\\
&+(123)_2(456)_2-(123)_2(45)(67)+(123)_2(45)(78), \labell{eq-20}
\eal
\ee
and the denominator part of generator $(123)_2(45)$ is
\be
\frac{1}{\und{\De_{12}\De_{23}}\De_{34}\und{\De_{45}}\De_{56}\De_{67}\De_{78}\De_{18}},
\ee
where the underlined minors are vanishing constraints, however, we need to further specify them by using a
factorized Pl\"{u}cker relation. Upon $\De_{12}\!=\!\De_{45}\!=\!0$, the remaining poles are
\be
\und{\De_{23}}\De_{34}\De_{56}\De_{67}\De_{78}\De_{18}\to\(\frac{\De_{28}}{\De_{18}}\times\und{\De_{13}}\)
\(\frac{\De_{38}}{\De_{18}}\times\frac{\De_{15}\De_{46}}{\De_{56}}\)\(\frac{\De_{35}\De_{46}}{\De_{34}}\)\times
\De_{67}\De_{78}\times\(\frac{\De_{38}\De_{14}}{\De_{34}}\),
\ee
where $\De_{23}$ is replaced by the factorized Pl\"{u}cker relation, in which $\De_{13}\!=\!0$. Since $\De_{28}$ is
non-vanishing in this choice, $\De_{12}\!=\!\De_{23}\!=\!\De_{13}\!=\!0$ unambiguously implies $(123)_2$ rather than $[2]$.
Then, all non-vanishing numerators above exhibit ten boundaries including four vanishing cells:
\be
\bal
&(13)[2](45),~(12)[3](45),~(12345)_2,~(123)_2[4],~(123)_2[5],~(123)_2(456)_2,\\
&(123)_2(45)(67),~(123)_2(45)(78),~(8123)_2(45),~[1](23)(45).
\eal
\ee
To determine the relative signs of non-vanishing boundary cells, for the generator $(123)_2(45)$,
we may fix $\De_{16}\!=\!1$  so that its matrix is
\be
C=\(\begin{array}{cccccccc}
\os{-}{1}\, & \os{+}{\De_{26}} & \os{-}{\De_{36}} & ~\os{+}{\De_{56}}\dfrac{\De_{14}}{\De_{15}}~ & \os{+}{\De_{56}}
& \,0\, & -\os{-}{\De_{67}} & -\os{+}{\De_{68}} \\[+0.5em]
0\, & 0 & 0 & \De_{14} & \De_{15} & \,1\, & \De_{17} & \De_{18}
\end{array}\),
\ee
now there is a new feature: $\De_{56}$ shows up twice in the first row for parameterizing $(45)\!=\!0$. Of course,
we can also choose the matrix
\be
C=\(\begin{array}{cccccccc}
\os{-}{1}\, & \os{+}{\De_{26}} & \os{-}{\De_{36}} & \os{+}{\De_{46}} & ~\os{+}{\De_{46}}\dfrac{\De_{15}}{\De_{14}}~
& \,0\, & -\os{-}{\De_{67}} & -\os{+}{\De_{68}} \\[+0.5em]
0\, & 0 & 0 & \De_{14} & \De_{15} & \,1\, & \De_{17} & \De_{18}
\end{array}\),
\ee
in which $\De_{46}$ shows up twice instead. But both of them give the same relative signs as
\be
-\,[1](23)(45)+(13)[2](45)-(12)[3](45)+(123)_2(456)_2-(123)_2(45)(67).
\ee
Now, we switch to fix $\De_{17}\!=\!1$ so that the matrix is
\be
C=\(\begin{array}{cccccccc}
\os{-}{1}\, & \os{+}{\De_{27}} & \os{-}{\De_{37}} & ~\os{+}{\De_{57}}\dfrac{\De_{14}}{\De_{15}}~ & \os{+}{\De_{57}}
& \os{-}{\De_{67}} & \,0\, & -\os{+}{\De_{78}} \\[+0.5em]
0\, & 0 & 0 & \De_{14} & \De_{15} & \De_{16} & \,1\, & \De_{18}
\end{array}\),
\ee
and this part of relative signs is
\be
-\,[1](23)(45)+(13)[2](45)-(12)[3](45)-(123)_2(45)(67)+(123)_2(45)(78),
\ee
hence we have proved \eqref{eq-20}. Note that in the row which lists accessible boundaries, there are always
six alternating signs, so we need to appropriately parameterize the vanishing constraints.
Such a parameterization depends on which row is under consideration, as we will immediately see.

The third example is a special N$^2$MHV $n\!=\!8$ identity
\be
\bal
0=\partial(12)(34)(56)=\,&-[1](34)(56)+[2](34)(56)-(12)[3](56)+(12)[4](56)-(12)(34)[5]+(12)(34)[6]\\
&-(12)(34)(567)_2+(812)_2(34)(56)-(12)(34)(56)(78),
\eal
\ee
which involves a \textit{quadratic} cell $(12)(34)(56)(78)$, and this cell is also the most general leading singularity
under the quad-cut \cite{ArkaniHamed:2012nw, Bourjaily:2013mma}. The denominator part of generator $(12)(34)(56)$ is
\be
\frac{1}{\und{\De_{12}}\De_{23}\und{\De_{34}}\De_{45}\und{\De_{56}}\De_{67}\De_{78}\De_{18}}.
\ee
Again, upon $\De_{12}\!=\!\De_{34}\!=\!\De_{56}\!=\!0$, the remaining poles are
\be
\De_{23}\De_{45}\De_{67}\De_{78}\De_{18}\to
\(\frac{\De_{28}}{\De_{18}}\times\frac{\De_{14}\De_{35}}{\De_{45}}\)\(\frac{\De_{24}\De_{36}}{\De_{23}}\times
\frac{\De_{57}}{\De_{67}}\)\(\frac{\De_{46}\De_{57}}{\De_{45}}\)\times\De_{78}\times\(\frac{\De_{28}\De_{13}}{\De_{23}}\),
\ee
of which all numerators exhibit 11 boundaries including two vanishing cells:
\be
\bal
&[2](34)(56),~(1234)_2(56),~(12)[3](56),~(12)[4](56),~(12)(3456)_2,~(12)(34)[5],\\
&(12)(34)[6],~(12)(34)(567)_2,~(12)(34)(56)(78),~(812)_2(34)(56),~[1](34)(56).
\eal
\ee
Next, we fix $\De_{15}\!=\!1$ so that the matrix is
\be
C=\(\begin{array}{cccccccc}
\os{-}{1}\, & \os{+}{\De_{25}} & \os{-}{\De_{35}} & \os{+}{\De_{45}} & \,0\, & 0
& -\os{-}{\De_{57}} & -\os{+}{\De_{58}} \\[+0.5em]
0\, & 0 & \De_{13} & ~\De_{13}\dfrac{\De_{45}}{\De_{35}}~ & \,1\, & \De_{16} & \De_{17} & \De_{18}
\end{array}\),
\ee
from which we read off part of the relative signs as
\be
-\,[1](34)(56)+[2](34)(56)-(12)[3](56)+(12)[4](56)-(12)(34)(567)_2.
\ee
Now switching to the second row, we get
\be
C=\(\begin{array}{cccccccc}
1\, & \De_{25} & \De_{35} & ~\De_{35}\dfrac{\De_{14}}{\De_{13}}~ & \,0\, & 0 & -\De_{57} & -\De_{58} \\[+0.5em]
0\, & 0 & \os{-}{\De_{13}} & \os{+}{\De_{14}} & \,\os{-}{1}\, & \os{+}{\De_{16}} & \os{-}{\De_{17}} & \os{+}{\De_{18}}
\end{array}\),
\ee
and this part of relative signs is
\be
-\,(12)[3](56)+(12)[4](56)-(12)(34)[5]+(12)(34)[6]+(812)_2(34)(56).
\ee
Note that for each row under consideration, we parameterize the vanishing constraints
so that its entries are always associated with six alternating signs, while the other row always has
four degrees of freedom to form the 5-bracket which provides kinematical support.

But $(12)(34)(56)(78)$ has not been associated with a specific sign yet! This is because there is no way for an
NMHV $n\!=\!6$ identity to connect such a quadratic cell, as its sign should be determined by some more nontrivial
residue theorem and we will leave this problem to the future work. To better understand the quadratic
property, let's explicitly parameterize its matrix representative as
\be
C=\(\begin{array}{cccccccc}
1\, & \ap_1 & \,0\, & 0 & -\ap_2\ap_5 & -\ap_2\ap_6 & -\ap_{23}\ap_7 & -\ap_{23}\ap_8 \\
0\, & 0 & \,1\, & \ap_4 & \ap_5 & \ap_6 & \ap_7 & \ap_8
\end{array}\),
\ee
so $C\!\cdot\!Z\!=\!0$ imposes $2\!\times\!4$ equations which can only be solved quadratically.
Relevant details can be found in \cite{ArkaniHamed:2012nw, Bourjaily:2013mma}. Moreover, the resulting Yangian invariant
is made of not only the product of two 5-brackets, but also a factor that involves square roots of
momentum twistor contractions.

A similar but simpler cell is $(123)_2(456)_2(789)_2$, which starts to show up in N$^2$MHV $n\!=\!9$ identities.
One form of its matrix representative is
\be
C=\(\begin{array}{ccccccccc}
1\, & \ap_1 & \ap_2 & \,0\, & 0 & 0 & -\ap_3\ap_6 & -\ap_3\ap_7 & -\ap_3\ap_8 \\
0\, & 0 & 0 & \,1\, & \ap_4 & \ap_5 & \ap_6 & \ap_7 & \ap_8
\end{array}\),
\ee
then $C\!\cdot\!Z\!=\!0$ imposes $2\!\times\!4$ equations which can only be solved as a linear system of equations,
and the resulting Yangian invariant contains a factor that involves ratios of
momentum twistor contractions \cite{ArkaniHamed:2012nw}.
We will classify it as a \textit{composite-linear} cell.
In contrast, BCFW-like cells are made of 5-brackets without extra non-unity factors, since we can solve for
the $4k$ positive variables \textit{row-wise} via $C\!\cdot\!Z\!=\!0$, which will be elaborated in section \ref{sec5}.

In appendix \ref{app2}, we list all distinct N$^2$MHV homological identities, and specify those non-BCFW-like cells
with underlines, of which the signs cannot be determined by incarnating the NMHV $n\!=\!6$ identity.
It is an interesting future problem to explore the residue theorems that can connect these more intricate objects
with BCFW-like cells.

\subsection{N$^3$MHV $n\!=\!8$ identities}

To understand more intriguing aspects of homological identities, let's have a peek into the N$^3$MHV sector.
Even at $n\!=\!8$, namely the minimal case, we have found four distinct identities.
The first one is generated by the cell $(23)$,
of which we have\footnote{It is peculiar that, the \textsc{Mathematica} package ``positroids'' does not
output the identical results as below. Some relative signs are assigned to vanishing boundary cells, while some
non-vanishing ones are assigned with zeros. Nevertheless, in the next section, we will numerically confirm
the correctness of an N$^3$MHV $n\!=\!8$ identity derived by our approach.}
\be
\textrm{VBC:}~~(\os{2}{\os{|}{3}}\,4\,5),~(8\,1\,\os{3}{\os{|}{2}}),~(1\,\os{3}{\os{|}{2}}\,4),
\ee
\be
0=-\,[3]+(23)(456)-(23)(567)+(23)(678)-(23)(781)+[2], \labell{eq-26}
\ee
where `VBC' denotes the vanishing boundary cells. The second one is generated by $(123)(345)$ as the two
vanishing constraints ``move away'' from each other, of which we have
\be
\textrm{VBC:}~~(\os{2}{\os{|}{3}}\,4\,5),~(123)(3456)_3,~(8123)_3(345),~(12345)_3,~(1\,2\,\os{4}{\os{|}{3}}),
\ee
\be
0=-\,[3]+(123)(45)-(123)(345)(567)+(123)(345)(678)-(123)(345)(781)+(12)(345). \labell{eq-21}
\ee
The third one is generated by $(123)(456)$, of which we have
\be
\textrm{VBC:}~~(123)(3456)_3,~(123)(4567)_3,~(8123)_3(456),~(1234)_3(456),
\ee
\be
0=-\,(23)(456)+(123)(45)-(123)(56)+(123)(456)(678)-(123)(456)(781)+(12)(456). \labell{eq-27}
\ee
And the last one is generated by $(123)(567)$, of which we have
\be
\textrm{VBC:}~~(123)(4567)_3,~(123)(5678)_3,~(8123)_3(567),~(1234)_3(567),
\ee
\be
0=-\,(23)(567)+(123)(345)(567)-(123)(56)+(123)(67)-(123)(567)(781)+(12)(567).
\ee
One may naively consider a further generator $(123)(678)$, but it is simply $(123)(456)$ after cyclicly shifted by $+3$.
As the two vanishing constraints reach the ``maximal cyclic distance'', we have covered all distinct
N$^3$MHV $n\!=\!8$ identities (while the generator $(1234)_3$ itself has no kinematical support already).

Now we will pick one example for demonstration: identity \eqref{eq-21} generated by $(123)(345)$, as the rest cases are
in fact similar. Again, its denominator part is
\be
\frac{1}{\und{\De_{123}}\De_{234}\und{\De_{345}}\De_{456}\De_{567}\De_{678}\De_{178}\De_{128}},
\ee
and upon $\De_{123}\!=\!\De_{345}\!=\!0$, the remaining poles are
\be
\De_{234}\De_{456}\De_{567}\De_{678}\De_{178}\De_{128}\to
\(\frac{\De_{235}\De_{134}}{\De_{135}}\)\(\frac{\De_{356}\De_{145}}{\De_{135}}\)\times
\De_{567}\De_{678}\De_{178}\times\(\frac{\De_{138}\De_{125}}{\De_{135}}\). \labell{eq-22}
\ee
It is more transparent to confirm the factorized Pl\"{u}cker relations above, if we fix the gauge as $\De_{135}\!=\!1$,
of which the matrix representative is
\be
C=\(\begin{array}{cccccccc}
1\, & \De_{235} & \,0\, & 0 & \,0\, & \De_{356} & \De_{357} & \De_{358} \\
0\, & \os{+}{\De_{125}} & \,\os{-}{1}\, & \os{+}{\De_{145}} & \,0\,
& -\os{-}{\De_{156}} & -\os{+}{\De_{157}} & -\os{-}{\De_{158}} \\
0\, & 0 & \,0\, & \De_{134} & \,1\, & \De_{136} & \De_{137} & \De_{138}
\end{array}\),
\ee
then from $\De_{125}$ and $\De_{145}$ in the second row, we read off part of the relative signs as
\be
+\,(12)(345)+(123)(45),
\ee
and from $\De_{235}$, $\De_{134}$, $\De_{356}$ and $\De_{138}$, we find four vanishing boundary cells:
\be
(\os{2}{\os{|}{3}}\,4\,5),~(1\,2\,\os{4}{\os{|}{3}}),~(123)(3456)_3,~(8123)_3(345).
\ee
Now we switch to fix $\De_{246}\!=\!1$ with
\be
C=\(\begin{array}{cccccccc}
\De_{146} & 1\, & \De_{346} & \,0 & -\De_{456} & 0\, & \De_{467} & \De_{468} \\
-\De_{126} & 0\, & \De_{236} & \,1 & \De_{256} & 0\, & -\De_{267} & -\De_{268} \\[+0.5em]
~\De_{234}\dfrac{\De_{126}}{\De_{236}}~ & 0\, & -\De_{234} & \,0 & ~\De_{234}\dfrac{\De_{456}}{\De_{346}}~
& 1\, & \De_{247} & \De_{248}
\end{array}\),
\ee
which gives the vanishing boundary cell $(12345)_3$ for $\De_{234}\!=\!0$.
Next, we can use a less obvious factorized Pl\"{u}cker relation, namely
\be
\De_{234}\to\frac{\De_{356}}{\De_{256}}\times\frac{\De_{245}}{\De_{456}}\times
\frac{\De_{125}\De_{246}-\De_{126}\De_{245}}{\De_{125}},
\ee
which is equivalent to fixing $\De_{256}\!=\!1$ so that the matrix is
\be
C=\(\begin{array}{cccccccc}
\os{-}{\De_{156}} & \,\os{+}{1} & \os{-}{\De_{356}} & \os{+}{\De_{456}} & \,0 & \,0\,
& \os{-}{\De_{567}} & \os{+}{\De_{568}} \\[+0.5em]
-\De_{126} & \,0 & ~\De_{356}\dfrac{\De_{245}}{\De_{456}}\dfrac{\De_{126}}{\De_{125}}~ & \De_{246}
& \,1 & \,0\, & -\De_{267} & -\De_{268} \\[+1.0em]
\De_{125} & \,0 & -\De_{356}\dfrac{\De_{245}}{\De_{456}} & -\De_{245} & \,0 & \,1\, & \De_{257} & \De_{258}
\end{array}\),
\ee
then from $\De_{356}$ and $\De_{567}$ in the first row, we read off
\be
-\,[3]-(123)(345)(567).
\ee
Finally, for $\De_{567}$, $\De_{678}$ and $\De_{178}$ in \eqref{eq-22}, we can fix $\De_{356}\!=\!1$ so that
the matrix is
\be
C=\(\begin{array}{cccccccc}
\os{-}{\De_{156}} & \os{+}{\De_{256}} & \os{-}{1}\, & \os{+}{\De_{456}} & \,0 & \,0\,
& \os{-}{\De_{567}} & \os{+}{\De_{568}} \\[+0.5em]
-\De_{136} & ~\De_{136}\dfrac{\De_{235}}{\De_{135}}~ & 0\, & \De_{346} & \,1 & \,0\, & -\De_{367} & -\De_{368} \\[+0.5em]
\De_{135} & \De_{235} & 0\, & 0 & \,0 & \,1\, & \De_{357} & \De_{358}
\end{array}\),
\ee
from which we immediately read off
\be
+\,(123)(45)-(123)(345)(567),
\ee
pushing the gauge-fixed columns $5,6$ forward, it is trivial to get
\be
+\,(123)(45)-(123)(345)(567)+(123)(345)(678)-(123)(345)(781),
\ee
since similar to $\De_{456}$ and $\De_{567}$, adjacent Pl\"{u}cker coordinates $\De_{567}$ and $\De_{678}$
are of opposite signs, and so are $\De_{678}$ and $\De_{178}$. Therefore,
we have proved \eqref{eq-21} after combining all pieces above.

Although there are four distinct identities at $n\!=\!8$, it is interesting that not all of them are directly related to
the cyclicity of N$^3$MHV $n\!=\!8$ amplitude. Explicitly, we find
\be
Y^3_8-Y^3_{8,+1}=\partial(23)+\partial(67)+\partial(81)+\partial(234)(567)+\partial(567)(812)+\partial(781)(234),
\ee
where
\be
\bal
Y^3_8=[8]+[6]+[2]&+(23)(456)+(23)(678)+(56)(234)+(56)(812)\\
&+(78)(234)+(81)(456)+(234)(456)(781),
\eal
\ee
is a recast form of \eqref{eq-11}. We see that its cyclicity only relies on identities of two types:
those generated by $(23)$ and $(123)(456)$, namely \eqref{eq-26} and \eqref{eq-27} respectively.
This can be regarded as an indirect consistency check of these two identities.

\newpage
\section{Positive Parameterization}
\label{sec5}

This section derives the stacking positivity relation, and then uses it to
parameterize the positive matrix representatives with the aid of reduced Grassmannian geometry.
This independent approach is confirmed by applying it to
numerically check an N$^3$MHV $n\!=\!8$ identity proved in the last section, now in terms of Yangian invariants.

\subsection{Stacking positivity relation}

The powerful stacking positivity relation in fact originates from simple observation of linear algebra and
positive Grassmannian. We can start with the $2\!\times\!4$ Grassmannian matrix as an example:
\be
C=\(\begin{array}{cccc}
c_{11} & c_{12} & c_{13} & c_{14} \\
c_{21} & c_{22} & c_{23} & c_{24}
\end{array}\),
\ee
and the Pl\"{u}cker relation of first three columns gives
\be
(13)\,c_{22}=(12)\,c_{23}+(23)\,c_{21}.
\ee
Now, when $c_{21},c_{22},c_{23}$ and $(12),(23)$ are all positive, $(13)$ must be positive. This relation also holds
if we replace $c_{21},c_{22},c_{23}$ by $c_{11},c_{12},c_{13}$ as which row to choose makes no difference.
Next let's consider all four columns, for which we additionally have
\be
(14)\,c_{23}=(13)\,c_{24}+(34)\,c_{21},~(24)\,c_{23}=(23)\,c_{24}+(34)\,c_{22}.
\ee
It is clear that when the second row is positive
and so are all consecutive minors $(12),(23),(34)$, the rest non-consecutive minors
$(13),(14),(24)$ must be positive, and hence the matrix $C$ is positive. This can be generalized to the case of any $n$.
Each time we increase $n$ by one, we regard all ordered minors of the old matrix as ``consecutive''.
Explicitly, for a generic $n$, the new Pl\"{u}cker relations involving column $n$ are
\be
\bal
(1n)\,c_{2,n-1}&=(1\,n\!-\!1)\,c_{2n}+(n\!-\!1\,n)\,c_{21},\\
(2n)\,c_{2,n-1}&=(2\,n\!-\!1)\,c_{2n}+(n\!-\!1\,n)\,c_{22},\\
&\,\,\vdots\\
(n\!-\!2\,n)\,c_{2,n-1}&=(n\!-\!2\,n\!-\!1)\,c_{2n}+(n\!-\!1\,n)\,c_{2,n-2},\\
\eal
\ee
when all $c_{2,i}$'s are positive, $(1n),(2n),\ldots,(n\!-\!2\,n)$ must be positive,
while $(n\!-\!1\,n)$ is already positive by assumption. We can express these relations
in a more compact form as
\be
(i\,n)\,(n\!-\!1)=(i\,n\!-\!1)\,(n)+(n\!-\!1\,n)\,(i),
\ee
where $i\!=\!1,2,\ldots,n\!-\!2$, and $c_{2,i}$ is replaced by $(i)$, which represents the trivial $1\!\times\!1$ minor of
the $1\!\times\!n$ sub-matrix. This form can be generalized to higher $k$, now let's consider the case of $k\!=\!3$.
As an example, for the $3\!\times\!5$ Grassmannian matrix
\be
C=\(\begin{array}{ccccc}
c_{11} & c_{12} & c_{13} & c_{14} & c_{15} \\
\cdot & \cdot & \cdot & \cdot & \cdot \\
\cdot & \cdot & \cdot & \cdot & \cdot
\end{array}\),
\ee
by recasting the Pl\"{u}cker relations of first four columns, we find
\be
(124)(23)=(123)(24)+(234)(12),~~(134)(23)=(123)(34)+(234)(13),
\ee
where the unspecified dots are entries of the $2\!\times\!5$ sub-matrix, as $(ij)$ is its $2\!\times\!2$ sub-minor.
Again, when the $2\!\times\!2$ sub-minors and $3\!\times\!3$ consecutive minors are positive,
the rest non-consecutive minors must be positive. Next, for all five columns, we additionally have
\be
\bal
&(125)(24)=(124)(25)+(245)(12),\\
&(135)(34)=(134)(35)+(345)(13),~~(235)(34)=(234)(35)+(345)(23),\\
&(145)(34)=(134)(45)+(345)(14),~~(245)(34)=(234)(45)+(345)(24), \labell{eq-52}
\eal
\ee
then the positivity of $(125)$ relies on that of $(245)$. Note the positivity relations for
$(125),(135),(145)$ are in fact duplicate, as we can alternatively choose
\be
\bal
&(125)(23)=(123)(25)+(235)(12),\\
&(135)(23)=(123)(35)+(235)(13),\\
&(145)(24)=(124)(45)+(245)(14), \labell{eq-53}
\eal
\ee
now their positivity also relies on that of $(235)$.
In the LHS of \eqref{eq-52} and \eqref{eq-53}, the middle label of a $3\!\times\!3$ minor can be $i$ or $j$
from the sub-minor $(ij)$, if it is not one of them, such an identity cannot exist.
There is no need to further consider the case of $n\!=\!6$, because one can arbitrarily choose four out of
columns $1,2,3,4,5$, which appear to be ``consecutive'' due to the positivity imposed above on these five columns,
then group them with column 6 to impose positivity again. After taking all possible choices
into account, we can trivially confirm the positivity of this new $3\!\times\!6$ matrix.
It is amazing that consecutive minors can help forge ``larger positivity'' from ``smaller positivity''.

Before presenting the general statement of stacking positivity relation, unfortunately we have to move on to
the case of $k\!=\!4$ in order to understand its duplicate positivity relations as well.
As an example, for the $4\!\times\!6$ Grassmannian matrix
\be
C=\(\begin{array}{cccccc}
c_{11} & c_{12} & c_{13} & c_{14} & c_{15} & c_{16} \\
\cdot & \cdot & \cdot & \cdot & \cdot & \cdot \\
\cdot & \cdot & \cdot & \cdot & \cdot & \cdot \\
\cdot & \cdot & \cdot & \cdot & \cdot & \cdot
\end{array}\),
\ee
by recasting the Pl\"{u}cker relations of first five columns, we find
\be
\bal
&(1235)(234)=(1234)(235)+(2345)(123),\\
&(1245)(234)=(1234)(245)+(2345)(124),\\
&(1345)(234)=(1234)(345)+(2345)(134), \labell{eq-23}
\eal
\ee
as expected. In the LHS above, the middle two labels of a $4\!\times\!4$ minor can be $(2,3)$, $(2,4)$ or $(3,4)$
from the sub-minor $(234)$. Next, for all six columns, we additionally have
\be
\bal
&(1236)(235)=(1235)(236)+(2356)(123),\\
&(1246)(245)=(1245)(246)+(2456)(124),\\
&(1256)(245)=(1245)(256)+(2456)(125),\\
&(1346)(345)=(1345)(346)+(3456)(134),~~(2346)(345)=(2345)(346)+(3456)(234),\\
&(1356)(345)=(1345)(356)+(3456)(135),~~(2356)(345)=(2345)(356)+(3456)(235),\\
&(1456)(345)=(1345)(456)+(3456)(145),~~(2456)(345)=(2345)(456)+(3456)(245),
\eal
\ee
where again, the positivity relations involving column 1 are duplicate.
We find no matter how large $k$ is, for the first $(k\!+\!2)$ columns
there are always two duplicate relations involving columns 1 and $k\!+\!2$, as will be explained soon.
Note the first three relations above rely on the positivity of $(2356)$ and $(2456)$, similar to the case of $k\!=\!3$.
Again, there is no need to further consider $n\!=\!7$, because all necessary ingredients are encoded in the
procedure of increasing $n$ from 5 to 6.

In general, for a $k\!\times\!(k\!+\!2)$ Grassmannian matrix,
by recasting the Pl\"{u}cker relations of first $(k\!+\!1)$ columns, we find
\be
(1\,A_i)_k(A_{k+1})_{k-1}=(1\,A_{k+1})_k(A_i)_{k-1}+(A)_k(1\,A_{i,k+1})_{k-1}, \labell{eq-54}
\ee
where $A$ denotes the set of columns $2,\ldots,k\!+\!1$ and $A_i$ denotes $A$ with column $i$ removed,
with $i\!=\!2,\ldots,k$ so that we have $(k\!-\!1)$ relations above.
We have added subscripts to highlight the sizes of corresponding minors and sub-minors.
This identity simply generalizes \eqref{eq-23}, and it guarantees the positivity of $(k\!-\!1)$ non-consecutive minors.
Next, for all $(k\!+\!2)$ columns, we additionally have
\be
(1\,A_{ij}\,k\!+\!2)_k(A_j)_{k-1}=(1\,A_j)_k(A_{ij}\,k\!+\!2)_{k-1}+(A_j\,k\!+\!2)_k(1\,A_{ij})_{k-1}, \labell{eq-55}
\ee
where $A_{ij}$ denotes $A$ with columns $i,j$ removed, with $i,j\!=\!2,\ldots,k\!+\!1$ and $i\!<\!j$.
The duplicate relation also holds as
\be
(1\,A_{ij}\,k\!+\!2)_k(A_i)_{k-1}=(1\,A_i)_k(A_{ij}\,k\!+\!2)_{k-1}+(A_i\,k\!+\!2)_k(1\,A_{ij})_{k-1},
\ee
which generalizes the observed pattern of $k\!=\!3,4$. For the positivity relations involving
column 2 but not 1, similar to \eqref{eq-54}, we have
\be
\(2\,A'_i\)_k\(A'_{k+2}\)_{k-1}=\(2\,A'_{k+2}\)_k\(A'_i\)_{k-1}+\(A'\)_k\(2\,A'_{i,k+2}\)_{k-1},
\ee
where $A'$ denotes the set of columns $3,\ldots,k\!+\!2$, and $i\!=\!3,\ldots,k\!+\!1$.
Note the positivity of $(A_j\,k\!+\!2)_k$ in \eqref{eq-55} for $j\!=\!3,\ldots,k\!+\!1$ relies on that of $\(2\,A'_i\)_k$.
There is no need to further consider the case of $n\!=\!k\!+\!3$ as previously mentioned.
Now we are ready to present the stacking positivity relation below:\\

\textbf{For a fixed $n$, a $k$-row positive Grassmannian matrix can be constructed by stacking a row on top of
a $(k\!-\!1)$-row positive Grassmannian matrix, and then imposing all $(n\!-\!k\!+\!1)$
consecutive $k\!\times\!k$ minors to be positive.}\\

This is a concise statement while its proof could look tedious. Once we have chosen a proper gauge,
it can be used for parameterizing the positive matrix representative of any Grassmannian geometric configuration,
with the aid of reduced Grassmannian geometry.

\subsection{Parameterization of positive matrix representatives}

We don't need reduced Grassmannian geometry yet when trying to parameterize top cells,
since there is no vanishing constraint. An immediate example is the top cell $C\!\in\!G(3,7)$ with the matrix
\be
C=\(\begin{array}{ccccccc}
1\, & 0\, & 0\, & \De_{234} & \De_{235} & \De_{236} & \De_{237} \\
0\, & 1\, & 0\, & -\De_{134} & -\De_{135} & -\De_{136} & -\De_{137} \\
0\, & 0\, & 1\, & \De_{124} & \De_{125} & \De_{126} & \De_{127}
\end{array}\),
\ee
of which the gauge can be chosen arbitrarily, since it would never be singular (otherwise setting $\De_{123}\!=\!1$ above
will be problematic if $(123)\!=\!0$, for instance). To entirely manifest its positivity,
it is easy to start from the bottom two rows, as we can first write
\be
C=\(\begin{array}{ccccccc}
1\, & 0\, & 0\, & \De_{234} & \De_{235} & \De_{236} & \De_{237} \\
0\, & 1\, & 0\, & -\bt_4\ap_4 & -\bt_{45}\ap_5 & -\bt_{456}\ap_6 & -\bt_{4567}\ap_7 \\
0\, & 0\, & 1\, & \ap_4 & \ap_5 & \ap_6 & \ap_7
\end{array}\).
\ee
Although $(234)\!=\!\De_{234}$ is already positive, for later convenience, let's adopt the parameterization
$\De_{234}\!=\!\gm_4\bt_4\ap_4$ so that the entry's power increases by one each time we move upstairs by a row.
The positivity of $(345)$ requires that in
\be
C=\(\begin{array}{ccccccc}
1\, & 0\, & 0\, & \gm_4\bt_4\ap_4 & (\gm_4\bt_{45}\!+\!x)\ap_5 & \De_{236} & \De_{237} \\
0\, & 1\, & 0\, & -\bt_4\ap_4 & -\bt_{45}\ap_5 & -\bt_{456}\ap_6 & -\bt_{4567}\ap_7 \\
0\, & 0\, & 1\, & \ap_4 & \ap_5 & \ap_6 & \ap_7
\end{array}\),
\ee
$x$ is positive, as $(345)\!=\!0$ when $x\!=\!0$. Again,
we set $x\!=\!\gm_5\bt_5$ so that $\De_{235}$ is a homogeneous polynomial.
Following this logic, the positivity of $(456)$ and $(567)$ unambiguously fixes
\be
C=\(\begin{array}{ccccccc}
1\, & 0\, & 0\, & \gm_4\bt_4\ap_4 & (\gm_4\bt_4\!+\!\gm_{45}\bt_5)\ap_5
& (\gm_4\bt_4\!+\!\gm_{45}\bt_5\!+\!\gm_{456}\bt_6)\ap_6
& (\gm_4\bt_4\!+\!\gm_{45}\bt_5\!+\!\gm_{456}\bt_6\!+\!\gm_{4567}\bt_7)\ap_7 \\
0\, & 1\, & 0\, & -\bt_4\ap_4 & -\bt_{45}\ap_5 & -\bt_{456}\ap_6 & -\bt_{4567}\ap_7 \\
0\, & 0\, & 1\, & \ap_4 & \ap_5 & \ap_6 & \ap_7
\end{array}\), \labell{eq-28}
\ee
and its consecutive minors besides $(123)\!=\!1$ are
\be
(234)=\gm_4\bt_4\ap_4,~~(345)=\bt_4\ap_4\gm_5\bt_5\ap_5,~~(456)=\ap_4\bt_5\ap_5\gm_6\bt_6\ap_6,~~
(567)=\ap_5\bt_6\ap_6\gm_7\bt_7\ap_7,
\ee
which are all monomials due to the proper parameterization adopted. This is in fact a must, because the vanishing of
each consecutive minor above must be triggered by only one variable, and one can easily see that such variables
are the $\gm_i$'s. In contrast, $\ap_i$'s and $\bt_i$'s are illegitimate, as they ``kill too many entries'' when
tending to zero.

One might sensitively find an obstacle for more general parameterizations: the vanishing of each $\ap_i$ or $\bt_i$ above
cannot directly lead to a proper singularity, without redefining some of the variables to maintain the property that
consecutive minors are monomials. For example, if $\ap_4\!=\!0$, naively this leads to $[4]$ and we get a top cell
of $n'\!=\!6$, with column 4 removed. However, now $(235)\!=\!(\gm_4\bt_4\!+\!\gm_{45}\bt_5)\ap_5$,
which is no longer a monomial, such that $(235)\!=\!0$ cannot be triggered by only one variable.

It seems that we can only make consecutive minors vanish in this way of parameterization, but here is where the
reduced Grassmannian geometry finds its arena, which is not introduced solely for economical or aesthetic reasons!
Let's recall the example of figure \ref{fig-1}, for which we have
\be
(3\,4\,5\,6)\,(5\,\os{7}{\os{\,\,/}{6}\,\os{\backslash\,\,}{8}}\,1)\,(8\,\os{2}{\os{\,\,/}{1}\,\os{\backslash\,\,}{3}}\,4)
\labell{eq-24}
\ee
as its geometric configuration. It clearly has the reduced geometric content due to the ``upstair parts''.
If we remove these parts, it becomes $(3456)(5681)(8134)$ of $n'\!=\!6$, with
\be
C'=\(\begin{array}{cccccc}
1\, & 0\, & 0\, & 0\, & 0 & -\De_{3458} \\
0\, & 1\, & 0\, & 0\, & \De_{1456} & \De_{1458} \\
0\, & 0\, & 1\, & 0\, & -\De_{1356} & -\De_{1358} \\
0\, & 0\, & 0\, & 1\, & \De_{1346} & 0
\end{array}\),
\ee
for which $(1568)\!=\!-\De_{1456}\De_{1358}\!+\!\De_{1458}\De_{1356}\!=\!0$. We can parameterize this matrix as
\be
C'=\(\begin{array}{cccccc}
1\, & 0\, & 0\, & 0\, & 0 & -\de_8\bt_8 \\
0\, & 1\, & 0\, & 0\, & \gm_6\bt_6\ap_6 & \gm_6\bt_8 \\
0\, & 0\, & 1\, & 0\, & -\bt_6\ap_6 & -\bt_8 \\
0\, & 0\, & 0\, & 1\, & \ap_6 & 0
\end{array}\), \labell{eq-30}
\ee
now there are two new features: its parameterization does manifest $(1568)\!=\!0$, and even though normally
the entry's power should increase by one each time we move upstairs, the top right entry is $-\de_8\bt_8$
instead of $-\de_8\gm_8\bt_8$. This redefinition is due to the vanishing constraint $(1568)\!=\!0$,
triggered by $\gm_8\!=\!0$ in
\be
C'=\(\begin{array}{cccccc}
1\, & 0\, & 0\, & 0\, & 0 & -\de_8\gm_8\bt_8 \\
0\, & 1\, & 0\, & 0\, & \gm_6\bt_6\ap_6 & \gm_{68}\bt_8 \\
0\, & 0\, & 1\, & 0\, & -\bt_6\ap_6 & -\bt_8 \\
0\, & 0\, & 0\, & 1\, & \ap_6 & 0
\end{array}\),
\ee
moreover, there should be
\be
4(6-4)-3=5
\ee
degrees of freedom in $C'$, but with $\gm_8$ there are six. Four degrees of freedom from its upstair parts
can be added back to the reduced matrix, as columns 2 and 7 are spanned by columns $1,3$ and $6,8$ respectively,
so that we have
\be
C=\(\begin{array}{cccccccc}
1\, & \de_2\gm_2 & \,0\, & 0\, & 0\, & 0 & -\bt_7\ap_7\de_8 & -\de_8\bt_8 \\
0\, & \gm_2 & \,1\, & 0\, & 0\, & \gm_6\bt_6\ap_6 & \gm_6\bt_{67}\ap_7 & \gm_6\bt_8 \\
0\, & 0 & \,0\, & 1\, & 0\, & -\bt_6\ap_6 & -\bt_{67}\ap_7 & -\bt_8 \\
0\, & 0 & \,0\, & 0\, & 1\, & \ap_6 & \ap_7 & 0
\end{array}\),
\ee
with explicitly
\be
C_2=\de_2\gm_2\,C_1+\gm_2\,C_3,~~C_7=\frac{\ap_7}{\ap_6}\,C_6+\frac{\bt_7\ap_7}{\bt_8}\,C_8,
\ee
where $C_i$ is the $i$-th column. Note that there must be an overall variable for each column, which encodes the
GL(1) rescaling degree of freedom. But when one column is spanned by some other columns, we must associate each
valency with a new variable to replace the original one. For example, $\ap_6$ and $\bt_8$ above are replaced by
$\ap_7$ and $\bt_7\ap_7$, so that column 7 has $\ap_7$ as its overall variable and each entry in column 7 is a
homogeneous polynomial.

To check the property that consecutive minors are monomials, we find
\be
(4568)=\gm_6\bt_6\ap_6\de_8\bt_8,
\ee
as the only nontrivial case of the reduced matrix. For the un-reduced matrix, it also holds with
\be
(2345)=\de_2\gm_2,~~(4567)=\gm_6\bt_6\ap_6\bt_7\ap_7\de_8.
\ee
In summary, the reduced Grassmannian geometry separates vanishing maximal consecutive minors from degenerate linear
dependencies. We will first construct the positive reduced matrix with the former part, which is easily realized from
its corresponding top cell, then add the latter back to it. However, there is a subtlety in the first step.
The geometric configuration \eqref{eq-24} can be also written as
\be
(3\,4\,5\,6)\,(5\,\os{7}{\os{\,\,/}{6}\,\os{\backslash\,\,}{8}}\,1)\,(8\,\os{3}{\os{\,\,/}{1}\,\os{\backslash\,\,}{2}}\,4),
\ee
with 2 and 3 switched in $(123)$, but now its reduced matrix is unclear since label 3 appears in both of the reduced
and upstair parts, even though there is nothing wrong with the representation above. From this example we
learn the lesson that, a proper way to separate the reduced and upstair parts is to make the ``boundary'' labels
stay ``unraised'' while labels in the middle of the degenerate constraint are raised.
For example, 6 and 8 are the boundary labels in $(678)$ above that are adjacent to 5 and 1 respectively,
while 7 is the middle one that is raised. When there are two or more valencies upon a vanishing maximal minor,
this rule substantially helps clarify the parameterization. While when there is only one single valency,
it is trivial to determine which label to be raised so that $n'$ of the reduced matrix is minimized.

\subsection{Application to an N$^3$MHV $n\!=\!8$ identity}

In the following, we will present more examples of positive parameterization from reduced Grassmannian geometry
representatives. As mentioned in section \ref{sec2}, the latter are more invariant and compact representations than the
corresponding Yangian invariants. However, since the kinematical data only show up in Yangian invariants rather than
Grassmannian cells, it is still desirable to see how the geometry lands on physics by numerically confirming
all the positive Grassmannian machineries we have introduced, now in terms of Yangian invariants.

A warmup exercise which will be used later, is the top cell $[8]$ of N$^3$MHV $n\!=\!8$ amplitude.
We have met its positive matrix representative and the resulting Yangian invariant in \eqref{eq-28} and \eqref{eq-25}
respectively, which are copied below for comparison:
\be
C=\(\begin{array}{ccccccc}
1\, & 0\, & 0\, & \gm_4\bt_4\ap_4 & (\gm_4\bt_4\!+\!\gm_{45}\bt_5)\ap_5
& (\gm_4\bt_4\!+\!\gm_{45}\bt_5\!+\!\gm_{456}\bt_6)\ap_6
& (\gm_4\bt_4\!+\!\gm_{45}\bt_5\!+\!\gm_{456}\bt_6\!+\!\gm_{4567}\bt_7)\ap_7 \\
0\, & 1\, & 0\, & -\bt_4\ap_4 & -\bt_{45}\ap_5 & -\bt_{456}\ap_6 & -\bt_{4567}\ap_7 \\
0\, & 0\, & 1\, & \ap_4 & \ap_5 & \ap_6 & \ap_7
\end{array}\),
\ee
and
\be
\bal
&[34567]\,[\,2\,3\,(34)\cap(567)\,(345)\cap(67)\,7\,]\\
&\times[\,1\,2\,\,(23)\cap((34)\cap(567)\,(345)\cap(67)\,7)\,\,(2\,3\,(34)\cap(567))\cap((345)\cap(67)\,7)\,\,7\,].
\eal
\ee
Again, we have solved for all positive variables row-wise starting from the bottom. While the third row is trivial,
for the second row it is easy to check the correspondence:
\be
\bt_4\sim3\cap(4567)=\mathcal{Z}_3,~~\bt_5\sim(34)\cap(567),~~\bt_5\sim(345)\cap(67),~~\bt_5\sim(3456)\cap7=\mathcal{Z}_7,
\ee
hence we get the second 5-bracket above, on the kinematical support of the first. For the first row, it is more involved
since we have to carefully keep track of each $\gm_i$. For example, the $\gm_4$ dependence
in the top row can be explicitly factored out as
\be
C=\(\begin{array}{ccccccc}
1\, & 0\, & 0\, & \gm_4\bt_4\ap_4 & (\gm_4\bt_{45}\!+\!\gm_5\bt_5)\ap_5
& (\gm_4\bt_{456}\!+\!\gm_5\bt_{56}\!+\!\gm_6\bt_6)\ap_6
& (\gm_4\bt_{4567}\!+\!\gm_5\bt_{567}\!+\!\gm_6\bt_{67}\!+\!\gm_7\bt_7)\ap_7 \\
0\, & 1\, & 0\, & -\bt_4\ap_4 & -\bt_{45}\ap_5 & -\bt_{456}\ap_6 & -\bt_{4567}\ap_7 \\
0\, & 0\, & 1\, & \ap_4 & \ap_5 & \ap_6 & \ap_7
\end{array}\),
\ee
so that we find the correspondence
\be
\gm_4\sim2\cap(3\,(34)\cap(567)\,(345)\cap(67)\,7)=\mathcal{Z}_2,
\ee
on the kinematical support of the second row, and similarly for $\gm_5,\gm_6,\gm_7$, so the lengthy expression
\eqref{eq-25} does match its parameterization \eqref{eq-28}. However, the parameterization is not unique
because of different but equivalent gauges, or more essentially, the GL$(k)$ invariance.
On the other hand, the BCFW recursion relation in the default recursion scheme gives another
expression of $[8]$ as
\be
\bal
&[12367]\,[\,(23)\cap(671)\,3\,4\,6\,(123)\cap(67)\,]\\
&\times[\,(34)\cap(6\,(123)\cap(67)\,(23)\cap(671))\,\,4\,5\,6\,\,(6\,(123)\cap(67))\cap((23)\cap(671)\,3\,4)\,],
\eal
\ee
which is in fact $Y^3_7$. We have numerically confirmed its equivalence to \eqref{eq-25}
by using ``\verb"loop amplitudes"'' in \cite{Bourjaily:2013mma}, as a highly nontrivial consistency check for
both the GL$(k)$ invariance of (reduced) Grassmannian geometry representatives and the
positive parameterization based on them.

Now let's be ambitious to include another facet into this check: the homological identities, as we will
numerically check an N$^3$MHV $n\!=\!8$ identity proved in the last section, namely \eqref{eq-21}:
\be
\bal
0=\partial(123)(345)&=-\,[3]+(123)(45)-(123)(345)(567)+(123)(345)(678)-(123)(345)(781)+(12)(345)\\
&\equiv-\,T_1+T_2-T_3+T_4-T_5+T_6,
\eal
\ee
where each boundary cell is identified with a corresponding Yangian invariant denoted by $T_i$.

First, as a top cell, $[3]$ can be easily obtained from $[8]$. Fixing labels $1,2$, we shift $3,4,5,6,7$ of $[8]$
by $+1$ to get $4,5,6,7,8$ which correspond to $[3]$. Similarly, $T_1$ is obtained from \eqref{eq-25} as
\be
\bal
T_1=\,\,&[45678]\,[\,2\,4\,(45)\cap(678)\,(456)\cap(78)\,8\,]\\
&\times[\,1\,2\,\,(24)\cap((45)\cap(678)\,(456)\cap(78)\,8)\,\,(2\,4\,(45)\cap(678))\cap((456)\cap(78)\,8)\,\,8\,].
\eal
\ee
Next, since both $(123)(45)$ and $(12)(345)$ are reduced Grassmannian geometry representatives, $T_2$ and $T_6$ are put
together as one type, while $T_3$, $T_4$ and $T_5$ are of another. For $T_2$, its matrix representative is
\be
C=\(\begin{array}{cccccccc}
1\, & 0\, & -\gm_3\bt_3 & \,0\, & 0
& \gm_{36}\bt_6\ap_6 & (\gm_{36}\bt_6\!+\!\gm_{367}\bt_7)\ap_7
& (\gm_{36}\bt_6\!+\!\gm_{367}\bt_7\!+\!\gm_{3678}\bt_8)\ap_8 \\
0\, & 1\, & \bt_3 & \,0\, & 0 & -\bt_6\ap_6 & -\bt_{67}\ap_7 & -\bt_{678}\ap_8 \\
0\, & 0\, & 0 & \,1\, & \ap_5 & \ap_6 & \ap_7 & \ap_8
\end{array}\),
\ee
where column 5 is the upstair part of the degenerate valency $(45)$, the resulting Yangian invariant is
\be
\bal
T_2=\,\,&[45678]\,[\,2\,3\,(45)\cap(678)\,(456)\cap(78)\,8\,]\\
&\times[\,1\,2\,\,(23)\cap((45)\cap(678)\,(456)\cap(78)\,8)\,\,(2\,3\,(45)\cap(678))\cap((456)\cap(78)\,8)\,\,8\,].
\eal
\ee
For $T_6$, its matrix is
\be
C=\(\begin{array}{cccccccc}
1\, & \gm_2 & 0\, & 0\, & 0
& \gm_6\bt_6\ap_6 & (\gm_6\bt_6\!+\!\gm_{67}\bt_7)\ap_7 & (\gm_6\bt_6\!+\!\gm_{67}\bt_7\!+\!\gm_{678}\bt_8)\ap_8 \\
0\, & 0 & 1\, & 0\, & -\bt_5\ap_5 & -\bt_{56}\ap_6 & -\bt_{567}\ap_7 & -\bt_{5678}\ap_8 \\
0\, & 0 & 0\, & 1\, & \ap_5 & \ap_6 & \ap_7 & \ap_8
\end{array}\),
\ee
where column 2 is the upstair part of $(12)$, the resulting Yangian invariant is
\be
\bal
T_6=\,\,&[45678]\,[\,3\,4\,(45)\cap(678)\,(456)\cap(78)\,8\,]\\
&\times[\,1\,2\,\,(34)\cap((45)\cap(678)\,(456)\cap(78)\,8)\,\,(3\,4\,(45)\cap(678))\cap((456)\cap(78)\,8)\,\,8\,].
\eal
\ee
Then, for $T_3$, $T_4$ and $T_5$, we will discuss a new feature related to the stacking positivity relation through their
parameterizations. Explicitly, let's parameterize $(123)(345)(567)$ for $T_3$ as
\be
C_{+2}=\(\begin{array}{cccccccc}
1\, & \gm_4\bt_4 & \,0\, & 0 & \,0\, & \gm_8\bt_8\ap_8 & (\gm_8\bt_8\!+\!\gm_{81}\bt_1)\ap_1
& (\gm_8\bt_8\!+\!\gm_{81}\bt_1\!+\!\gm_2)\ap_2 \\
0\, & \bt_4 & \,1\, & \bt_6\ap_6 & \,0\, & -\bt_8\ap_8 & -\bt_{81}\ap_1 & -\bt_{81}\ap_2 \\
0\, & 0 & \,0\, & \ap_6 & \,1\, & \ap_8 & \ap_1 & \ap_2
\end{array}\), \labell{eq-29}
\ee
where unusually, these columns correspond to labels $3,4,5,6,7,8,1,2$ and we have used the subscript $+2$ to
denote this cyclic shift. The matrix above in fact results from
\be
C_{+2}=\(\begin{array}{cccccccc}
1\, & \De_{457} & \,0\, & 0 & \,0\, & \De_{578} & \De_{571} & \De_{572} \\
0\, & \De_{347} & \,1\, & \De_{367} & \,0\, & -\De_{378} & -\De_{371} & -\De_{372} \\
0\, & 0 & \,0\, & \De_{356} & \,1\, & \De_{358} & \De_{351} & \De_{352}
\end{array}\),
\ee
after imposing $(781)$ and $(812)$ to be positive, note that before imposing positivity it is convenient to first
take care of all vanishing constraints, namely $(312)\!=\!-\De_{371}\De_{352}\!+\!\De_{372}\De_{351}\!=\!0$, while
$(345)\!=\!(567)\!=\!0$ is already manifest. Then after a trivial replacement of variables, we get
\be
C_{+2}=\(\begin{array}{cccccccc}
1\, & \gm_4\bt_4 & \,0\, & 0 & \,0\, & \gm_8\bt_8\ap_8 & \De_{571} & \De_{572} \\
0\, & \bt_4 & \,1\, & \bt_6\ap_6 & \,0\, & -\bt_8\ap_8 & -\bt_{81}\ap_1 & -\bt_{81}\ap_2 \\
0\, & 0 & \,0\, & \ap_6 & \,1\, & \ap_8 & \ap_1 & \ap_2
\end{array}\).
\ee
The positivity of $(781)$ and $(812)$ unambiguously fixes $\De_{571}$ and $\De_{572}$ as those in \eqref{eq-29}.
Now, $\De_{572}$ is not a homogeneous polynomial due to the same reason for $\De_{3458}$ in \eqref{eq-30}.
We see that it is not absolute for an entry to be a homogeneous polynomial, instead it depends on the details
of parameterization. Once we have fixed the gauge and the order to parameterize as above, there is no ambiguity of
whether an entry is homogeneous or not, as it is also because the total degrees of freedom is fixed.

Switching back to the standard ordering of columns, \eqref{eq-29} becomes
\be
C=\(\begin{array}{cccccccc}
(\gm_8\bt_8\!+\!\gm_{81}\bt_1)\ap_1 & (\gm_8\bt_8\!+\!\gm_{81}\bt_1\!+\!\gm_2)\ap_2 & \,1\, & \gm_4\bt_4
& \,0\, & 0 & \,0\, & \gm_8\bt_8\ap_8 \\
-\bt_{81}\ap_1 & -\bt_{81}\ap_2 & \,0\, & \bt_4 & \,1\, & \bt_6\ap_6 & \,0\, & -\bt_8\ap_8 \\
\ap_1 & \ap_2 & \,0\, & 0 & \,0\, & \ap_6 & \,1\, & \ap_8
\end{array}\), \labell{eq-31}
\ee
where we have taken the twisted cyclicity into account, while the additional sign factor for columns $1,2$
is trivially $(-)^{3-1}\!=\!1$. The new feature of positive parameterization we want to discuss now shows up:
the sub-matrix made of the second and third rows is not positive, contradicting with the reasoning of
stacking positivity relation! This superficial failure is dissolved by the fact that,
the original positive construction in \eqref{eq-29} is legitimate, but after we cyclicly shift the order of columns
in \eqref{eq-31} the sub-matrix's positivity is not longer preserved,
while the entire matrix definitely remains positive.

The twisted cyclicity of positive parameterization provides a hint for determining the order of super momentum twistors
in each 5-bracket from the positive variables of each corresponding row, which helps clarify the sign ambiguity:
they follow the default cyclic order of subscripts for positive variables.

In \eqref{eq-31}, the $\ap_i$'s in the bottom row follow such an order, hence the resulting 5-bracket is
$[67812]\!=\![12678]$. Similarly, in other two rows their orders are
$\bt_8,\bt_1,\bt_4,1,\bt_6$ and $\gm_8,\gm_1,\gm_2,1,\gm_4$ respectively,
then the corresponding Yangian invariant is
\be
\bal
T_3=\,\,&[67812]\,[\,(67)\cap(812)\,(678)\cap(12)\,4\,5\,6\,]\\
&\times[\,((67)\cap(812)\,(678)\cap(12))\cap(456)\,\,(678)\cap(12)\,2\,3\,4\,].
\eal
\ee
Following almost the same procedure, it is straightforward to obtain
\be
C=\(\begin{array}{cccccccc}
(\gm_8\bt_8\!+\!\gm_{81}\bt_1)\ap_1 & (\gm_8\bt_8\!+\!\gm_{81}\bt_1\!+\!\gm_2)\ap_2 & \,1\,
& \gm_4\bt_4 & \,0\, & -\gm_8\bt_6\ap_6 & \,0\, & \gm_8\bt_8\ap_8 \\
-\bt_{81}\ap_1 & -\bt_{81}\ap_2 & \,0\, & \bt_4 & \,1\, & \bt_6\ap_6 & \,0\, & -\bt_8\ap_8 \\
\ap_1 & \ap_2 & \,0\, & 0 & \,0\, & \ap_6 & \,1\, & \ap_8
\end{array}\)
\ee
for $(123)(345)(678)$, and
\be
\bal
T_4=\,\,&[67812]\,[\,(67)\cap(812)\,(678)\cap(12)\,4\,5\,6\,]\\
&\times[\,(6\,(67)\cap(812)\,(678)\cap(12))\cap(45)\,\,(678)\cap(12)\,2\,3\,4\,]
\eal
\ee
respectively, as well as
\be
C=\(\begin{array}{cccccccc}
\gm_{68}\bt_{81}\ap_1 & (\gm_{68}\bt_{81}\!+\!\gm_2)\ap_2 & \,1\,
& \gm_4\bt_4 & \,0\, & -\gm_6\bt_6\ap_6 & \,0\, & \gm_{68}\bt_8\ap_8 \\
-\bt_{81}\ap_1 & -\bt_{81}\ap_2 & \,0\, & \bt_4 & \,1\, & \bt_6\ap_6 & \,0\, & -\bt_8\ap_8 \\
\ap_1 & \ap_2 & \,0\, & 0 & \,0\, & \ap_6 & \,1\, & \ap_8
\end{array}\)
\ee
for $(123)(345)(781)$, and
\be
\bal
T_5=\,\,&[67812]\,[\,(67)\cap(812)\,(678)\cap(12)\,4\,5\,6\,]\\
&\times[\,(6\,(67)\cap(812)\,(678)\cap(12))\cap(45)\,\,((67)\cap(812)\,(678)\cap(12))\cap(456)\,\,2\,3\,4\,]
\eal
\ee
respectively. Using ``\verb"loop amplitudes"'',
we have numerically confirmed $-T_1\!+\!T_2\!-\!T_3\!+\!T_4\!-\!T_5\!+\!T_6\!=\!0$.

\newpage
\section{Two-fold Simplex-like Structures of Amplitudes}
\label{sec6}

This section refines the tree BCFW recursion relation, to now reveal its
two-fold simplex-like structures: the triangle-like dissection of BCFW contour
and the simplex-like growing patterns of fully-spanning cells. The growth of full cells essentially terminates
at $n\!=\!4k\!+\!1$, then the amplitudes of a given $k$ are known for any $n$.
Two classes of amplitudes are elaborated for demonstration:
the N$^2$MHV family which terminates at $n\!=\!9$, and the N$^3$MHV family which terminates at $n\!=\!13$.

\subsection{NMHV and anti-NMHV triangles}

Before exploring the beautiful global structure of tree BCFW contour, it is instructive to first study
the simplest nontrivial cases: NMHV and anti-NMHV amplitudes. It is trivial to know that, MHV and anti-MHV amplitudes
are 1 and top cells respectively. But we will also denote top cells by `1'
for convenience in the following, with certain clarification.

Recall the NMHV $n\!=\!8$ amplitude \eqref{eq-5} which reflects a simple pattern:
\be
Y^1_8=
\(\begin{array}{cccc}
{} & {} & {} & [234] \\
{} & {} & [238] & [236] \\
{} & [278] & [258] & [256] \\
{[678]} & [478] & [458] & [456]
\end{array}\)
=\(\begin{array}{c|cccc}
[234]~ & {} & {} & {} & ~~1 \\
{[23]}~~ & {} & {} & ~1 & ~~[6] \\
{[2]}~~~ & {} & 1 & ~[5] & ~[56] \\
1~~~ & ~1 & [4] & [45] & [456] \\
\hline
{} & ~[678] & ~[78] & ~[8] & ~~1
\end{array}\),
\ee
now its pattern is even more transparent after we maximally factor out the empty slots,
where each entry in the ``NMHV triangle'' above is multiplied by its corresponding vertical and horizontal factors,
in order to match that in the un-factored form \eqref{eq-5}. In the sense of null objects, here 1 represents a top cell
since the superposition of any vanishing constraint and a top cell becomes the constraint itself, and we will use this
notation for all N$^k$MHV sectors. We can immediately generalize it for a generic $n$ as
\be
Y^1_n=
\(\begin{array}{c|cccccc}
[234\ldots n\!-\!4]~ & {} & {} & {} & {} & {} & 1 \\
\vdots~~ & {} & {} & {} & {} & \,\iddots & \vdots \\
{[234]}~ & {} & {} & {} & ~~~1 & ~\cdots & ~~~~[\ldots n\!-\!2] \\
{[23]}~~ & {} & {} & ~1 & ~~~[6] & ~\cdots & ~~~[6\ldots n\!-\!2] \\
{[\textbf{2}]}~~~ & {} & 1 & ~[5] & ~~[56] & ~\cdots & ~~[56\ldots n\!-\!2] \\
1~~~ & ~1 & [\textbf{4}] & [45] & ~[456] & ~\cdots & ~[456\ldots n\!-\!2] \\
\hline
{} & ~[\textbf{6}78\ldots n] & ~[78\ldots n] & ~[8\ldots n] & ~[\ldots n] & ~\cdots & 1
\end{array}\), \labell{eq-34}
\ee
where the bold labels $6,4,2$ uniquely determine this pattern, in fact they are the growing parameters of a quadratic
growing mode, as will be later defined. One may check it against the generic NMHV amplitude in terms of 5-brackets, namely
\be
Y^1_n=\sum_{i=4}^{n-1}\sum_{j=2}^{i-2}\,[1\,j\,j\!+\!1\,i\,i\!+\!1],
\ee
which is a special case of the tree BCFW recursion relation (in the default recursion scheme)
\be
\bal
Y^k_n(1,\ldots,n\!-\!1,n)=\,&\,Y^k_{n-1}(1,\ldots,n\!-\!1)\\
&+\sum_{j=2}^{n-3}\,\sum_{k_\textrm{L}+k_\textrm{R}=k-1}\,[1\,j\,j\!+\!1\,n\!-\!1\,n]\,
Y_\textrm{L}(1,\ldots,j,I)\,Y_\textrm{R}(I,j\!+\!1,\ldots,n\!-\!1,\widehat{n}) \labell{eq-33}
\eal
\ee
with $\mathcal{Z}_I\!=\!(\,j\,j\!+\!1)\cap(n\!-\!1\,n\,1)$
and $\widehat{\mathcal{Z}}_n\!=\!(n\!-\!1\,n)\cap(1\,j\,j\!+\!1)$.
Its matrix version, as given in figure \ref{fig-2}, is the backbone of all formal refinement in this section.

As their parity conjugate, the anti-NMHV amplitudes also have a triangle-shape pattern, though it is more intricate.
Recall \eqref{eq-1}, \eqref{eq-32} and \eqref{eq-11}, anti-NMHV amplitudes of $k\!=\!1,2,3$ with $n\!=\!k\!+\!5$
are copied below for comparison:
\be
Y^1_6=
\(\begin{array}{cc}
[\textbf{2}]~ &
\!\left\{\begin{array}{c}
\,~~[\textbf{6}] \\
\!{[\textbf{4}]}~~~
\end{array} \right.
\end{array}\!\),
\ee
\be
Y^2_7=
\(\begin{array}{ccc}
[\textbf{2}]~ &
(23)\left\{\begin{array}{c}
\!(67) \\
\!(45)~~~~~~
\end{array} \right. &
\!\!\!\!\!\!\left\{\begin{array}{c}
\,~~~[\textbf{7}] \\
\!(45)(71) \\
\!{[\textbf{5}]}~~~~
\end{array} \right.
\end{array}\!\!\),
\ee
\be
Y^3_8=
\(\begin{array}{cccc}
[\textbf{2}]~ &
(23)\left\{\begin{array}{c}
\!(678) \\
\!(456)~~~~~~
\end{array} \right. &
\!\!\!\!\!\!(234)\left\{\begin{array}{c}
~~~(78) \\
\!(456)(781)~~ \\
\!\!(56)~~~~~~~
\end{array} \right. &
\!\!\!\!\left\{\begin{array}{c}
~~~[\textbf{8}] \\
\!(456)(81)~~ \\
\,(56)(812) \\
\!\!{[\textbf{6}]}~~~~
\end{array} \right.
\end{array}\!\!\!\),
\ee
as well as a further extension of this suggestive pattern, for which $k\!=\!4$:
\be
Y^4_9=
\(\begin{array}{ccccc}
[\textbf{2}]~~ &
(23)\left\{\begin{array}{c}
\!(6789) \\
\!(4567)~~~~~~
\end{array} \right. &
\!\!\!\!(234)\left\{\begin{array}{c}
\,~~~~(789) \\
\!(4567)(7891)~~ \\
\!(567)~~~~~~~~~
\end{array} \right. &
\!(2345)\left\{\begin{array}{c}
\!(89) \\
\!(4567)(891)~~~~~~~ \\
\!(567)(8912)~~~~ \\
\!(67)~~~~~~~~~~~
\end{array} \right. &
\!\!\!\!\!\!\!\left\{\begin{array}{c}
\,~~~[\textbf{9}] \\
\!(4567)(91)~~~ \\
\!(567)(912) \\
\,~~(67)(9123) \\
\!{[\textbf{7}]}~~~~
\end{array} \right.
\end{array}\!\!\),
\ee
where the bold labels $2,k\!+\!3,k\!+\!5$ are the ``vertices'' spanning the triangles. One can prove the general form of
anti-NMHV amplitudes by induction. First, the top right cell $[n]$ in the triangle is clearly the top cell from $Y^k_{k+4}$.
Then, the top row except $[n]$ comes from the characteristic 5-brackets in \eqref{eq-33}, which are
\be
[2567\ldots k\!+\!3]~~[2367\ldots k\!+\!3]~~[2347\ldots k\!+\!3]~~\ldots~~[2345\ldots k\!+\!1],
\ee
here we have used empty slots to denote 5-brackets, as the latter's complement. Finally, the sub-triangle below
the top row comes from the characteristic 5-bracket $[4567\ldots k\!+\!3]$ and its associated sub-amplitude
$Y^{k-1}_{k+4}(2,3,4,\ldots,k\!+\!5)$. Summing all three parts, we get the desired triangle of $Y^k_{k+5}$.

An explicit example is to obtain $Y^3_8$ from $Y^2_7$. Immediately, $[8]$ is the top cell from $Y^3_7$. Then, for
the top row except $[8]$, BCFW cells
\be
[2]~~(23)(678)~~(234)(78)
\ee
correspond to the terms associated with 5-brackets
\be
[256]~~[236]~~[234]
\ee
in terms of empty slots (or $[13478]$, $[14578]$ and $[15678]$ traditionally).
Finally, the sub-triangle below the top row
\be
\(\begin{array}{ccc}
(23)(456) & ~(234)(456)(781) & \!(456)(81) \\
{} & ~(234)(56)~~~~~\, & \,~~(56)(812) \\
{} & {} & \!\![6]~~
\end{array}\!\)
\ee
corresponds to 5-bracket $[456]$ along with its associated sub-amplitude
\be
Y^2_7(2,3,4,5,6,7,8)=
\(\begin{array}{ccc}
[3]~ &
(34)\left\{\begin{array}{c}
\!(78) \\
\!(56)~~~~~~
\end{array} \right. &
\!\!\!\!\!\!\left\{\begin{array}{c}
\,~~~[8] \\
\!(56)(82) \\
\!{[6]}~~~~
\end{array} \right.
\end{array}\!\!\).
\ee
Summing all three parts, we now reproduce the triangle of $Y^3_8$.

\subsection{Triangle-like dissection of general N$^k$MHV amplitudes}

The triangle-like dissection \eqref{eq-34} of NMHV amplitudes can be extended to a general BCFW contour
with more intriguing features. Let's recall the N$^2$MHV $n\!=\!9$ amplitude \eqref{eq-3}:
\be
Y^2_9=
\(\begin{array}{cccc}
{} & {} & {} & \!\!\!\!\!\![234]~~~~~~~~ \\
{} & {} & \!\!\!\!\![239]~~~~~~ & \!\!\!\!\!\![23]I_{7,3}~~~ \\
{} & \!\!\!\![289]~~~~~ & \!\!\!\!\![29]I_{7,2}~ & \!\!\!\!\!\![2]I_{8,2}~~ \\
\,{[789]}~~~\, & \!\!\!\![89]I_{7,1} & \!\!\!\!\![9]I_{8,1} & \!\!\!\!\!I_{9,1}
\end{array}\!\!\!\!\!\!\!\!\!\)
=\(\begin{array}{c|cccc}
[234]~ & {} & {} & {} & 1 \\
{[23]}~~ & {} & {} & 1 & I_{7,3} \\
{[2]}~~~ & {} & 1 & I_{7,2} & I_{8,2} \\
1~~~ & ~1 & I_{7,1} & I_{8,1} & I_{9,1} \\
\hline
{} & ~[789] & ~[89] & ~[9] & 1
\end{array}\), \labell{eq-39}
\ee
which is a specific example of the generic N$^k$MHV $n$-particle dissection, namely
\be
Y^k_n=
\(\begin{array}{c|cccccc}
[234\ldots n\!-\!k\!-\!3]~ & {} & {} & {} & {} & {} & 1 \\
\vdots~ & {} & {} & {} & {} & ~\iddots~ & \vdots \\
{[234]}~ & {} & {} & {} & ~1 & ~\cdots~ & ~I_{n-3,4} \\
{[23]}~~ & {} & {} & 1 & ~I_{k+5,3} & ~\cdots~ & ~I_{n-2,3} \\
{[2]}~~~ & {} & 1 & I_{k+5,2} & ~I_{k+6,2} & ~\cdots~ & ~I_{n-1,2} \\
1~~~ & 1 & I_{k+5,1} & I_{k+6,1} & ~I_{k+7,1} & ~\cdots~ & I_{n,1}~~ \\
\hline
{} & ~[k\!+\!5~k\!+\!6~k\!+\!7\ldots n] & ~[k\!+\!6~k\!+\!7\ldots n] & ~[k\!+\!7\ldots n] & ~[\ldots n] & ~\cdots~ & 1
\end{array}\), \labell{eq-37}
\ee
where again, we have maximally factored out the empty slots both vertically and horizontally,
in order to isolate the $I_{i,j}$. However, only $I_{i,1}$ in the bottom row needs to be identified.
Once $I_{i,1}$ is known, $I_{i,1+j}$ can be obtained by performing a partial cyclic shift $i\!\to\!i\!+\!j$
except that label 1 is fixed, for all BCFW cells within it. Note the vertical and horizontal factors
with respect to each $I_{i,j}$ force it to be free of the labels in those empty slots,
as the partial cyclic shift also helps ensure this.

In this dissection, the vertical ``slicing'' is easy to understand. Adding an imaginary vertical line in
\be
Y^k_n=
\(\begin{array}{c|ccccc|c}
[234\ldots n\!-\!k\!-\!3]~ & {} & {} & {} & {} & {} & 1 \\
\vdots~ & {} & {} & {} & {} & ~\iddots~ & \vdots \\
{[234]}~ & {} & {} & {} & ~1 & ~\cdots~ & ~I_{n-3,4} \\
{[23]}~~ & {} & {} & 1 & ~I_{k+5,3} & ~\cdots~ & ~I_{n-2,3} \\
{[2]}~~~ & {} & 1 & I_{k+5,2} & ~I_{k+6,2} & ~\cdots~ & ~I_{n-1,2} \\
1~~~ & 1 & I_{k+5,1} & I_{k+6,1} & ~I_{k+7,1} & ~\cdots~ & I_{n,1}~~ \\
\hline
{} & ~[k\!+\!5~k\!+\!6~k\!+\!7\ldots n] & ~[k\!+\!6~k\!+\!7\ldots n] & ~[k\!+\!7\ldots n]
& ~[\ldots\textbf{\textit{n}}] & ~\cdots~ & 1
\end{array}\), \labell{eq-35}
\ee
we can clearly see its inductive structure when increasing $n$, since from \eqref{eq-33} we have
\be
Y^k_n(1,\ldots,n\!-\!1,n)=Y^k_{n-1}(1,\ldots,n\!-\!1)+\ldots
\ee
which implies the triangle of $Y^k_n$ must contain the sub-triangle of $Y^k_{n-1}$. On the other hand,
to understand the horizontal slicing needs a different perspective. We can also add an imaginary horizontal line in
\be
Y^k_n=
\(\begin{array}{c|cccccc}
[234\ldots n\!-\!k\!-\!3]~ & {} & {} & {} & {} & {} & 1 \\
\vdots~ & {} & {} & {} & {} & ~\iddots~ & \vdots \\
{[234]}~ & {} & {} & {} & ~1 & ~\cdots~ & ~I_{n-3,4} \\
{[23]}~~ & {} & {} & 1 & ~I_{k+5,3} & ~\cdots~ & ~I_{n-2,3} \\
{[\textbf{2}]}~~~ & {} & 1 & I_{k+5,2} & ~I_{k+6,2} & ~\cdots~ & ~I_{n-1,2} \\
\hline
1~~~ & 1 & I_{k+5,1} & I_{k+6,1} & ~I_{k+7,1} & ~\cdots~ & I_{n,1}~~ \\
\hline
{} & ~[k\!+\!5~k\!+\!6~k\!+\!7\ldots n] & ~[k\!+\!6~k\!+\!7\ldots n] & ~[k\!+\!7\ldots n] & ~[\ldots n] & ~\cdots~ & 1
\end{array}\),
\ee
so that it exhibits a structure suggestively analogous to \eqref{eq-35}. Here, the bold label 2 is analogous to $n$
in \eqref{eq-35}. If we rotate the triangle above clockwise by 90 degrees, it looks like the mirror image
of \eqref{eq-35}, as label 2 now becomes the mirror image of $n$!

To quantitatively justify the horizontal slicing, let's adopt the \textit{sub-reflection} of $(n\!-\!1)$ labels as
\be
(1,2,3,4,\ldots,n)\to(1,n,n\!-\!1,n\!-\!2,\ldots,2),
\ee
which has used cyclicity to fix the position of label 1, after color reflection of all $n$ labels.
If we ignore the contents of $I_{i,j}$, this sub-reflection exactly realizes the manipulation above.

Therefore, though it is less straightforward to see, $Y^k_n$ has an inductive structure from the horizontal slicing,
identical to that of the vertical one. This is natural from the experience with NMHV amplitudes, and in fact
surprising because of the more delicate structure concealed in $I_{i,j}$. The triangle-like dissection also
helps us to further refine the counting of BCFW terms. Recall \eqref{eq-36},
the number of BCFW terms in tree amplitudes is given by
\be
N^k_n=\frac{1}{n\!-\!3}\binom{n\!-\!3}{k}\binom{n\!-\!3}{k\!+\!1}.
\ee
The vertical slicing gives its first order difference as
\be
\De N^k_n=N^k_n-N^k_{n-1},
\ee
and the horizontal slicing gives its \textit{second order difference} as
\be
\De^2 N^k_n=\De N^k_n-\De N^k_{n-1}=N^k_n-2N^k_{n-1}+N^k_{n-2},
\ee
which counts the BCFW terms in $I_{n,1}$, as one can easily check in the dissection \eqref{eq-37}.
This dissection is in fact a clever rearrangement of \eqref{eq-38}, such that we only need to identify all BCFW cells
within each $I_{i,1}$ for $i\!=\!k\!+\!5,\ldots,n$, which saves a large amount of repetitive calculation.
For $k\!=\!2,3$, the second order differences of $N^k_n$ are
\be
\De^2 N^2_n=(n-5)^2, \labell{eq-40}
\ee
and
\be
\De^2 N^3_n=\frac{1}{24}(n-5)(n-6)(5n(n-11)+152). \labell{eq-41}
\ee
These two formulas will be useful in the next two subsections, in which N$^2$MHV and N$^3$MHV amplitudes are elaborated,
as highly nontrivial consistency checks of their two-fold simplex-like structures.

\subsection{N$^2$MHV amplitudes and their simplex-like growing patterns}

To further investigate the delicate structure concealed in $I_{i,j}$ we will again take N$^2$MHV $n\!=\!9$ amplitude
as an example, from which the general pattern of all N$^2$MHV amplitudes is extracted.
In its maximally factored-out form \eqref{eq-39}, namely
\be
Y^2_9=
\(\begin{array}{c|cccc}
[234]~ & {} & {} & {} & 1 \\
{[23]}~~ & {} & {} & 1 & I_{7,3} \\
{[2]}~~~ & {} & 1 & I_{7,2} & I_{8,2} \\
1~~~ & ~1 & I_{7,1} & I_{8,1} & I_{9,1} \\
\hline
{} & ~[789] & ~[89] & ~[9] & 1
\end{array}\),
\ee
we have already known $I_{7,1}$ from the anti-NMHV triangle of $k\!=\!2$ as
\be
I_{7,1}=
\(\begin{array}{cc}
\left\{\begin{array}{c}
\!(45)(71) \\
\!{[5]}~~~~
\end{array} \right. &
(23)\left\{\begin{array}{c}
\!(67) \\
\!(45)~~~~~~
\end{array} \right.
\end{array}\!\!\!\!\!\!\!\).
\ee
The partial cyclic shift immediately gives
\be
I_{7,2}=
\(\begin{array}{cc}
\left\{\begin{array}{c}
\!(56)(81) \\
\!{[6]}~~~~
\end{array} \right. &
(34)\left\{\begin{array}{c}
\!(78) \\
\!(56)~~~~~~
\end{array} \right.
\end{array}\!\!\!\!\!\!\!\),
\ee
as well as
\be
I_{7,3}=
\(\begin{array}{cc}
\left\{\begin{array}{c}
\!(67)(91) \\
\!{[7]}~~~~
\end{array} \right. &
(45)\left\{\begin{array}{c}
\!(89) \\
\!(67)~~~~~~
\end{array} \right.
\end{array}\!\!\!\!\!\!\!\),
\ee
and there is a similar simple relation between $I_{8,1}$ and $I_{8,2}$. From the honest BCFW recursion relation,
we also know that
\be
I_{8,1}=
\(\begin{array}{ccc}
[\textbf{5}]\left\{\begin{array}{c}
\!(46)(81) \\
\!{[6]}~~~~
\end{array} \right. &
[\textbf{6}](23)\left\{\begin{array}{c}
\!(78) \\
\!(45)~~~~~~
\end{array} \right. &
\!\!\!\!\!\!\left\{\begin{array}{c}
\!(234)_2(678)_2~~~ \\
\!(456)_2(781)_2~~~ \\
\!\!(23)(456)_2(81)
\end{array} \right. \\[+2.0em]
{} &
[\textbf{4}](23)\left\{\begin{array}{c}
\!(78) \\
\!(56)~~~~~~
\end{array} \right.
& {}
\end{array}\!\!\!\),
\ee
and
\be
I_{9,1}=
\(\begin{array}{cccc}
[56]\left\{\begin{array}{c}
\!(47)(91) \\
\!{[7]}~~~~
\end{array} \right. &
[67](23)\left\{\begin{array}{c}
\!(89) \\
\!(45)~~~~~~
\end{array} \right. &
\!\!\!\!\!\!\!\!\!\!\!\left\{\begin{array}{c}
\![\textbf{7}](234)_2(689)_2 \\
\!{[\textbf{4}]}(235)_2(789)_2
\end{array} \right. &
\!\!\!\!\!\left\{\begin{array}{c}
\!(2345)_2(6789)_2~~~~ \\
\!(23)(4567)_2(891)_2
\end{array} \right. \\[+1.5em]
{} &
[47](23)\left\{\begin{array}{c}
\!(89) \\
\!(56)~~~~~~
\end{array} \right. &
\!\!\!\!\!\!\!\!\!\!\!\left\{\begin{array}{c}
\![\textbf{7}](456)_2(891)_2 \\
\!{[\textbf{5}]}(467)_2(891)_2
\end{array} \right.
& {} \\[+1.5em]
{} &
[45](23)\left\{\begin{array}{c}
\!(89) \\
\!(67)~~~~~~
\end{array} \right. &
\!\!\!\!\!\left\{\begin{array}{c}
\![\textbf{6}](23)(457)_2(91) \\
\!{[\textbf{4}]}(23)(567)_2(91)
\end{array} \right.
& {}
\end{array}\!\!\!\),
\ee
where the bold labels are the growing parameters. To see how the BCFW cells grow in $I_{7,1}$, $I_{8,1}$ and $I_{9,1}$,
let's focus on
\be
\left\{\begin{array}{c}
\!(45)(71) \\
\!{[5]}~~~~
\end{array} \right.\!\!\to
[5]\left\{\begin{array}{c}
\!(46)(81) \\
\!{[6]}~~~~
\end{array} \right.\!\!\to
[56]\left\{\begin{array}{c}
\!(47)(91) \\
\!{[7]}~~~~
\end{array} \right.
\ee
respectively, where the growing parameter 5 uniquely determines the pattern
\be
[5]\to[56]\to[567]\to\ldots,
\ee
and it induces a partial cyclic shift for each associated subset of cells, maintaining the old ``topology'' of
the original subset of cells. For example, the topology of $(45)(71)$ upon labels $1,2,3,4,5,6,7$ is identical to
that of $(46)(81)$ upon $1,2,3,4,6,7,8$, as well as $(47)(91)$ upon $1,2,3,4,7,8,9$.

Again in $I_{7,1}$, $I_{8,1}$ and $I_{9,1}$, we find
\be
\begin{array}{ccc}
(23)\left\{\begin{array}{c}
\!(67) \\
\!(45)~~~~~~
\end{array} \right. &
\!\!\!\!\!\!\!\!\to[6](23)\left\{\begin{array}{c}
\!(78) \\
\!(45)~~~~~~
\end{array} \right. &
\!\!\!\!\!\!\!\!\to[67](23)\left\{\begin{array}{c}
\!(89) \\
\!(45)~~~~~~
\end{array} \right. \\[+1.5em]
{} &
\!\!\!\!\!\!\!\!\to[4](23)\left\{\begin{array}{c}
\!(78) \\
\!(56)~~~~~~
\end{array} \right. &
\!\!\!\!\!\!\!\!\to[47](23)\left\{\begin{array}{c}
\!(89) \\
\!(56)~~~~~~
\end{array} \right. \\[+1.5em]
{} & {} &
\!\!\!\!\!\!\!\!\to[45](23)\left\{\begin{array}{c}
\!(89) \\
\!(67)~~~~~~
\end{array} \right.
\end{array}
\ee
respectively, where the growing parameters $6,4$ uniquely determine the pattern
\be
[6]+[4]\to[67]+[47]+[45]\to[678]+[478]+[458]+[456]\to\ldots,
\ee
and similarly, this induces a partial cyclic shift for each associated subset of cells.
While a single growing parameter leads to a sequence of empty slots where each term contains one more label
than its preceding one following the default increasing order, two growing parameters give a sequence of
sums of empty slots where each sum exhibits an interesting pattern: the first term is gradually ``eaten'' by
the last term, as we move forward term by term, labels in the former are replaced by those in the latter one by one.
These two terms alone follow the pattern of a single growing parameter respectively, in the meanwhile.

The same pattern applies to $(234)_2(678)_2$, $(456)_2(781)_2$ and $(23)(456)_2(81)$ in $I_{8,1}$ of which
the corresponding growing parameters are $(7,4)$, $(7,5)$ and $(6,4)$. For $(2345)_2(6789)_2$ and $(23)(4567)_2(891)_2$
in $I_{9,1}$ we will not explicitly present $I_{10,1}$, $I_{11,1}$ and so on, instead, by focusing on the relevant terms
in $I_{9,1}$, $I_{10,1}$ and $I_{11,1}$, we find
\be
\begin{array}{ccc}
\left\{\begin{array}{c}
\!(2345)_2(6789)_2~~~~ \\
\!(23)(4567)_2(891)_2
\end{array} \right. &
\!\to[8]\left\{\begin{array}{c}
\!\!(2345)_2(679\,10)_2~~~~ \\
\!(23)(4567)_2(9\,10\,1)_2
\end{array} \right. &
\!\to[89]\left\{\begin{array}{c}
\!(2345)_2(67\,10\,\,11)_2~~~~ \\
\!(23)(4567)_2(10\,\,11\,1)_2
\end{array} \right. \\[+1.5em]
{} &
\!\to[6]\left\{\begin{array}{c}
\!\!(2345)_2(789\,10)_2~~~~ \\
\!(23)(4578)_2(9\,10\,1)_2
\end{array} \right. &
\!\to[69]\left\{\begin{array}{c}
\!(2345)_2(78\,10\,\,11)_2~~~~ \\
\!(23)(4578)_2(10\,\,11\,1)_2
\end{array} \right. \\[+1.5em]
{} &
\!\to[4]\left\{\begin{array}{c}
\!\!(2356)_2(789\,10)_2~~~~ \\
\!(23)(5678)_2(9\,10\,1)_2
\end{array} \right. &
\!\to[67]\left\{\begin{array}{c}
\!(2345)_2(89\,10\,\,11)_2~~~~ \\
\!(23)(4589)_2(10\,\,11\,1)_2
\end{array} \right. \\[+1.5em]
{} & {} &
\!\to[49]\left\{\begin{array}{c}
\!(2356)_2(78\,10\,\,11)_2~~~~ \\
\!(23)(5678)_2(10\,\,11\,1)_2
\end{array} \right. \\[+1.5em]
{} & {} &
\!\to[47]\left\{\begin{array}{c}
\!(2356)_2(89\,10\,\,11)_2~~~~ \\
\!(23)(5689)_2(10\,\,11\,1)_2
\end{array} \right. \\[+1.5em]
{} & {} &
\!\to[45]\left\{\begin{array}{c}
\!(2367)_2(89\,10\,\,11)_2~~~~ \\
\!(23)(6789)_2(10\,\,11\,1)_2
\end{array} \right. \\[+1.5em]
\end{array}
\ee
respectively, where the growing parameters $8,6,4$ uniquely determine the pattern
\be
\(\begin{array}{cc}
[8] & {} \\
{[6]} & [4]
\end{array}\)\to
\(\,\begin{array}{ccc}
[89]~ & {} & {} \\
{[69]}~ & [49]~ & {} \\
{[67]}~ & [47]~ & [45]
\end{array}\,\)\to
\(\,\begin{array}{cccc}
[89\,10]~ & {} & {} & {} \\
{[69\,10]}~ & [49\,10]~ & {} & {} \\
{[67\,10]}~ & [47\,10]~ & [45\,10] & {} \\
{[678]}~~~ & [478]~~~ & [458]~~ & ~[456]
\end{array}\,\)\to\ldots, \labell{eq-43}
\ee
now each ``edge'' of the ``solid triangle'' above follows the pattern of two growing parameters respectively.
This is identical to the growing pattern of NMHV amplitudes, if $6,4,2$ became the growing parameters!
Incidentally, we find that beyond $n\!=\!9$ there is no more BCFW cell of new topology in $I_{i,1}$ since all of its cells
descend from the characteristic objects above, which are named as the fully-spanning cells due to the fact that
none of the $n$ columns are removed from them when they first show up in $I_{n,1}$.
But of course, $[5]$ in $I_{7,1}$ is the only exception, still,
for convenience it is put together with $(45)(71)$, as they share the same growing parameter.
One may ask why the new full cells stop showing up beyond $n\!=\!9$, its reason is quite physical:
minimally at $n\!=\!10$, the only possible Yangian invariant is the trivial product of two 5-brackets without overlapped
labels, namely $[12345]\,[6789\,10]$, there is no place for any BCFW deformation,
or more formally, there is no place for positivity to play its role in forging amplitude-like cells.
Strictly speaking, as we will recursively prove later, there exists no more new full cell beyond $n\!=\!4k\!+\!1$.
But naively, if we ``collapse'' $k$ 5-brackets without overlapped labels so that each adjacent pair of them
has one overlapped slot, we also get the correct critical relation between $n$ and $k$. For example, in
\be
C=\(\begin{array}{ccccccccccccc}
\cdot\, & \cdot\, & \cdot\, & \cdot\, & \cdot\, & 0\, & 0\, & 0\, & 0\, & 0\, & 0\, & 0\, & 0\, \\
0\, & 0\, & 0\, & 0\, & \cdot\, & \cdot\, & \cdot\, & \cdot\, & \cdot\, & 0\, & 0\, & 0\, & 0\, \\
0\, & 0\, & 0\, & 0\, & 0\, & 0\, & 0\, & 0\, & \cdot\, & \cdot\, & \cdot\, & \cdot\, & \cdot\,
\end{array}\)
\ee
for $k\!=\!3$, where the unspecified dots are the only nonzero entries which form a descending ``staircase'',
we find that $n\!=\!5\!\cdot\!3\!-\!(3\!-\!1)\!=\!13$.
In general, this is obviously $n\!=\!5k\!-\!(k\!-\!1)\!=\!4k\!+\!1$.

\begin{figure}
\begin{center}
\includegraphics[width=0.55\textwidth]{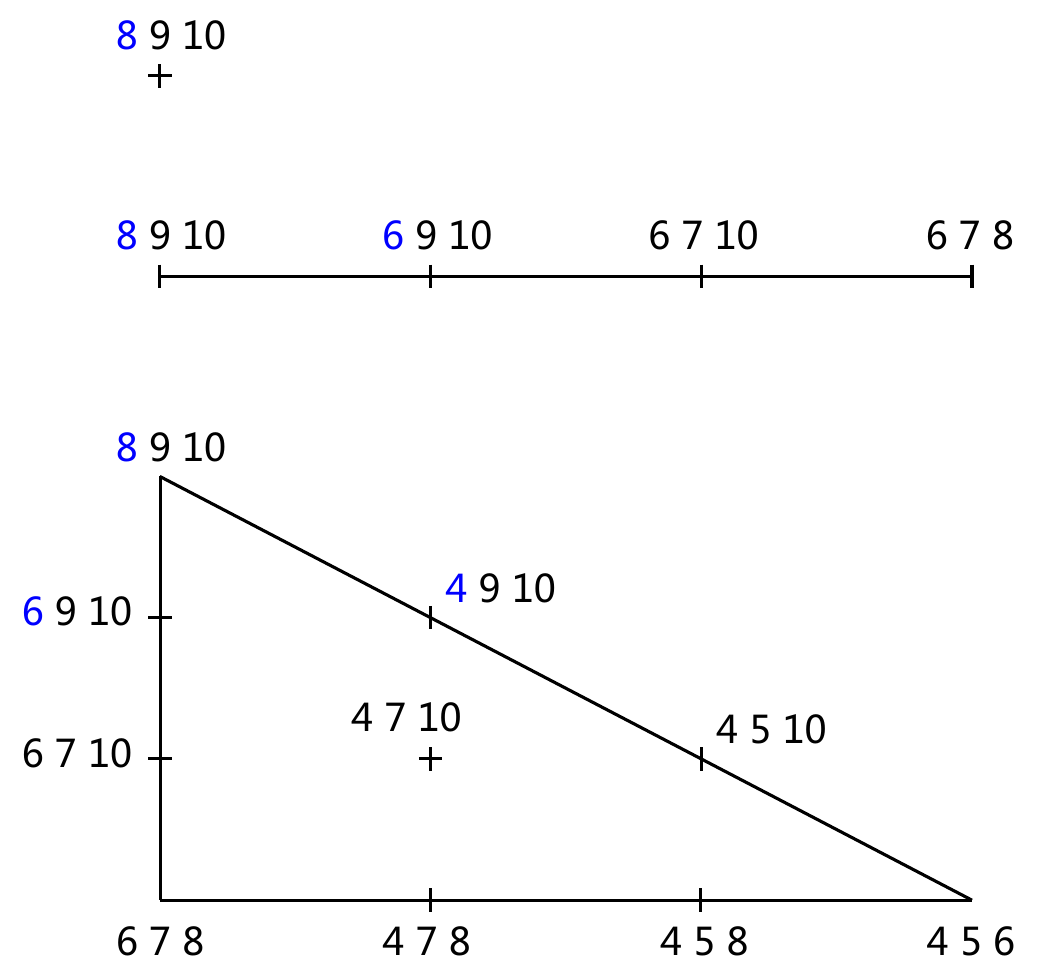}
\caption{Solid 0-, 1-, 2-simplices of level 3 with growing parameters $(8)$, $(8,6)$, $(8,6,4)$ respectively.}
\label{fig-17}
\end{center}
\end{figure}

It is favorable to characterize various growing patterns by introducing the concept of \textit{solid} $m$-simplex.
Such a simple geometric object has three ingredients: growing mode, growing parameter(s) and \textit{level},
and now let's describe them via three examples in figure \ref{fig-17}. There, three typical objects that
we have encountered in the N$^2$MHV amplitudes are shown: solid 0-, 1-, 2-simplices of level 3, with
growing parameters $(8)$, $(8,6)$, $(8,6,4)$. The 0-mode is represented by a 0-simplex or a point, the 1-mode
a line segment and the 2-mode a triangle. The growing parameters are highlighted as the starting points of a solid simplex.
It is solid because ``inside'' it there are also a number of points, where this number is its level minus one
for 1-mode, for instance.
The level is the number of labels at each point, but for compactness this ingredient is often not specified.

To see how this formalism helps simplify amplitudes, we reformulate a general N$^2$MHV amplitude by
a few lines as
\be
G_{7,0}=\left\{\begin{array}{c}
\!(45)(71) \\
\!{[5]}~~~~
\end{array} \right.~~~~~~~~~~~~~~~~~~~~~(5)~~~~~~ \labell{eq-45}
\ee
\be
G_{7,1}=(23)\left\{\begin{array}{c}
\!(67) \\
\!(45)~~~~~~
\end{array} \right.~~~~~~~~~~~~~~~(6,4)~~~ \labell{eq-46}
\ee
\be
G_{8,1}=\left\{\begin{array}{c}
\!(234)_2(678)_2~~~~~~~~~~~~~~~~~(7,4)~~ \\
\!(456)_2(781)_2~~~~~~~~~~~~~~~~~(7,5)~~ \\
\!(23)(456)_2(81)~~~~~~~~~~~~~~\,(6,4)~~
\end{array} \right. \labell{eq-47}
\ee
\be
G_{9,2}=\left\{\begin{array}{c}
\!(2345)_2(6789)_2~~~~ \\
\!(23)(4567)_2(891)_2
\end{array} \right.~~~~~~~~~(8,6,4) \labell{eq-48}
\ee
where $G_{i,m}$ is the part purely made of full cells in $I_{i,1}$ and $m$ is its corresponding growing mode,
followed by their growing parameters. In this way, $I_{i,1}$ can be expressed in terms of $G_{i,m}$ as
\be
\bal
I_{7,1}&=G_{7,0}+G_{7,1},\\
I_{8,1}&=G_{8,1}+(G_{7,0,2}+G_{7,1,2}),\\
I_{9,1}&=G_{9,2}+G_{8,1,2}+(G_{7,0,3}+G_{7,1,3}),\\
I_{10,1}&=G_{9,2,2}+G_{8,1,3}+(G_{7,0,4}+G_{7,1,4}),\\
I_{11,1}&=G_{9,2,3}+G_{8,1,4}+(G_{7,0,5}+G_{7,1,5}),
\eal
\ee
and so on, where the additional label $l$ in $G_{i,m,l}$ represents its level during the simplex-like growth.
Now, according to the growing modes and levels of full cells, for any $n$, it is easy to count the terms in $I_{n,1}$ as
\be
2+2(n-6)+3(n-7)+2\cdot\frac{(n-7)(n-8)}{2}=(n-5)^2,
\ee
which nicely matches $\De^2 N^2_n$ in \eqref{eq-40} as expected.

So far we have understood the entire profile of \textit{all} N$^2$MHV amplitudes by capturing the full cells
and their growing patterns. ``Experimentally'', all these results can be confirmed by ``\verb"positroids"'' up to any $n$,
even though beyond $n\!=\!11$ or so it inevitably becomes boring. In practice, it is useful to first take the advantage of
the triangle-like dissection \eqref{eq-37} of the full amplitude honestly produced by ``\verb"positroids"''.
The required combinatoric knowledge is given by \eqref{eq-42}, but we will use
\be
\sigma(a\!-\!1)=\sigma_\textrm{MT}(a)+1,
\ee
instead, for identifying the corresponding permutations in massless spinor space shown in ``\verb"positroids"''.
From \cite{ArkaniHamed:2012nw} we know that, $\sigma_\textrm{MT}(a)\!=\!a$ if label $a$ is absent, so that in this case
$\sigma(a\!-\!1)\!=\!a\!+\!1$. Therefore, $\sigma(n\!-\!1)\!=\!n\!+\!1$ if label $n$ is absent and $\sigma(1)\!=\!3$
if label 2 is absent. Neglecting the permutations
satisfying $\sigma(n\!-\!1)\!=\!n\!+\!1$ or $\sigma(1)\!=\!3$, we easily avoid the redundant cells
with the aid of the triangle-like dissection, and the surviving part is simply $I_{n,1}$.

In {\S 12} of \cite{ArkaniHamed:2012nw} all 14 distinct N$^2$MHV Yangian invariants are listed. Among them,
one is the top cell at $n\!=\!6$, one is a trivial product at $n\!=\!10$, one is quadratic at $n\!=\!8$
and one is composite-linear at $n\!=\!9$.
Still, not all of the rest ten legitimate candidates of full cells are present in a general N$^2$MHV amplitude,
partly due to the default recursion scheme we have adopted (although the present full cells can be traded for other
candidates in this list via the homological identities).

Note that, we have used different orientations for the triangle-like dissection and a solid 2-simplex to distinguish them:
the former is growing ``rightward'' while the latter is growing ``downward'', as exhibited in \eqref{eq-37}
and \eqref{eq-43}, but of course they are conceptually identical. What's more, the former is in the form which maximally
factors out the empty slots both vertically and horizontally, while the latter is not since we find it more convenient
to generalize a 2-simplex to an $m$-simplex in the un-factored form.

Finally, we must remark that all the unproved facts: why the full cells are so, as well as their growing modes and
parameters, and why their simplex-like growing patterns are valid up to any $n$, will be further investigated along with
the more lengthy and complicated sector: the N$^3$MHV amplitudes, in the following subsections.
After the N$^3$MHV case is understood, the N$^2$MHV one would look quite trivial.

\subsection{N$^3$MHV amplitudes and their simplex-like growing patterns}

As we have known, the nontrivial N$^3$MHV amplitudes start with anti-NMHV amplitude $Y^3_8$ in \eqref{eq-44},
as copied below
\be
Y^3_8=
\(\begin{array}{cc}
{} & [2] \\
{[8]} & J_{8,1}
\end{array}\),
\ee
where
\be
J_{8,1}=
\(\begin{array}{ccc}
(23)\left\{\begin{array}{c}
\!(678) \\
\!(456)~~~~~~
\end{array} \right. &
\!\!\!\!\!\!(234)\left\{\begin{array}{c}
~~~(78) \\
\!(456)(781)~~ \\
\!\!(56)~~~~~~~
\end{array} \right. &
\!\!\!\!\left\{\begin{array}{c}
\!(456)(81)~~ \\
\,(56)(812) \\
\!\!{[6]}~~~~
\end{array} \right.
\end{array}\!\!\!\),
\ee
which is rearranged to fit the triangle-like dissection \eqref{eq-37}. Immediately, at $n\!=\!9$ we find the full cells
in $J_{8,1}$ grow according to
\be
H_{8,0}=\left\{\begin{array}{c}
\!(456)(81)~~ \\
\,(56)(812) \\
\!\!{[6]}~~~~
\end{array} \right.~~~~(6)
\ee
as well as
\be
H_{8,1}=\(\begin{array}{c}
~\,(23)\left\{\begin{array}{c}
\!(678) \\
\!(456)~~~~~~
\end{array} \right.~~~~~(7,4) \\[+1.5em]
(234)\left\{\begin{array}{c}
~~~(78) \\
\!(456)(781)~~ \\
\!\!(56)~~~~~~~
\end{array} \right.~~(7,5)
\end{array}\)
\ee
where $H_{i,m}$ is the part purely made of full cells in $J_{i,1}$ and $m$ is its corresponding growing mode,
followed by their growing parameters. Similarly, $[6]$ in $H_{8,0}$ is the only exception that is not a full cell.

These objects are reminiscent of $G_{7,0}$ and $G_{7,1}$ in $I_{7,1}$,
but as $n$ increases, the complexity of N$^3$MHV amplitudes soon
becomes overwhelming. For example, $G_{8,1}$ has three terms while $(H_{9,0}\!+\!H_{9,1}\!+\!H_{9,2})$ in total has
18 terms of various topologies. This is in fact not surprising since its complexity gets enhanced by the
Grassmannian geometry of $k\!=\!3$.
Now, let's list all the full cells in $H_{9,0}$, $H_{9,1}$ and $H_{9,2}$ respectively in the following:
\be
~H_{9,0}=(4\,\os{6}{\os{|}{5}}\,7)\,(8\,\os{1}{\os{|}{9}}\,2)~~~~~~~~~~~~~~~~~~~~~(6)~~~~
\ee
as well as
\be
H_{9,1}=\left\{\begin{array}{c}
(2345)_3(789)_2~~~~~~~~~~~~~~~~~(8,5) \\[+0.5em]
(2\,3\,\os{5}{\os{|}{4}})\,(7\,\os{9}{\os{|}{8}}\,1)~~~~~~~~~~~~~~~~~~\,(8,5) \\[+0.5em]
(234)_2(6789)_3~~~~~~~~~~~~~~~~~(8,4) \\[+0.5em]
(\os{6}{\os{|}{7}}\,8\,9)\,(9\,1\,\os{3}{\os{|}{2}})~~~~~~~~~~~~~~~~~~\,(7,4) \\[+0.5em]
(4567)_3(891)_2~~~~~~~~~~~~~~~~~(8,6) \\[+0.5em]
(234)(4567)_3\,(7\,8\,\os{1}{\os{|}{9}})~~~~~~~~~~(7,5) \\[+0.5em]
(23)(4567)_3(91)~~~~~~~~~~~~~~\,(7,4) \\[+0.5em]
(234)\,(4\,\os{6}{\os{|}{5}}\,7)\,(91)~~~~~~~~~~~~~~(7,5) \\[+0.5em]
(2\,3\,\os{5}{\os{|}{4}})\,(467)(7891)_3~~~~~~~~~~(8,5) \\[+0.5em]
(4\,5\,\os{7}{\os{|}{6}})\,(9\,1\,\os{3}{\os{|}{2}})~~~~~~~~~~~~~~~~~~\,(7,4) \\[+0.5em]
(567)_2(8912)_3~~~~~~~~~~~~~~~~~(8,6) \\[+0.5em]
(234)(567)_2(912)~~~~~~~~~~~~~(7,5)
\end{array} \right.
\ee
and the terms of 2-mode
\be
H_{9,2}=\left\{\begin{array}{c}
(\os{2}{\os{|}{3}}\,4\,5)\,(6\,7\,\os{9}{\os{|}{8}})~~~~~ \\[+0.5em]
(23)(45)(89)~~~~~ \\[+0.5em]
(\os{2}{\os{|}{3}}\,4\,5)\,(\os{6}{\os{|}{7}}\,8\,9)~~~~~ \\[+0.5em]
(23)\,(\os{4}{\os{|}{5}}\,6\,7)\,(891) \\[+0.5em]
(23)(45)(67)~~~~~
\end{array} \right.~~~~~~(8,6,4)
\ee
which can be found in appendix \ref{app3} without the further classification above.
Of course, the growing modes and parameters of full cells must be identified from $J_{10,1}$ by explicit recursion,
but soon we will introduce a more clever way to achieve this from their sub-cells.

\begin{figure}
\begin{center}
\includegraphics[width=0.9\textwidth]{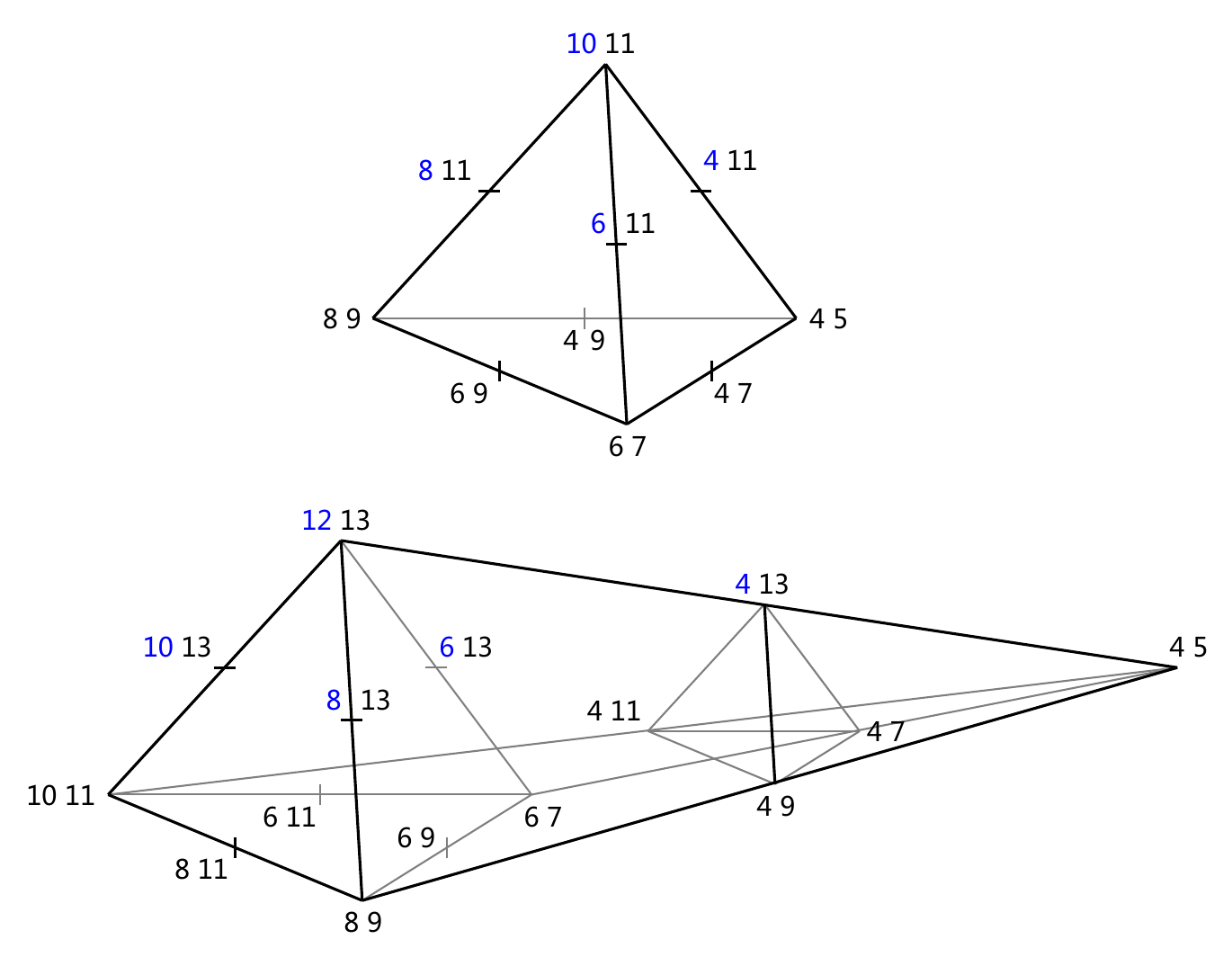}
\caption{Solid 3- and 4-simplices of level 2 with growing parameters $(10,8,6,4)$ and $(12,10,8,6,4)$ respectively.}
\label{fig-18}
\end{center}
\end{figure}

The numbers of cells in $H_{9,0},H_{9,1},H_{9,2}$ are $1,12,5$ respectively. Then, pushing ahead, we find
27 full cells in $J_{10,1}$, among which 7 belong to $H_{10,1}$ and 20 belong to $H_{10,2}$,
and the lengthy details are relegated to appendix \ref{app3}. At $n\!=\!11$ something novel happens:
we encounter the full cells of 3-mode, among which one example is
\be
(\os{2}{\os{|}{3}}\,4\,5\,\os{7}{\os{|}{6}})_3\,(89\,10\,\,11)_2~~~~~~(10,8,6,4)
\ee
where the growing parameters $10,8,6,4$ uniquely determine the pattern
\be
\(\begin{array}{c}
[10]~~ \\[+0.5em]
\left\{\begin{array}{cc}
[8] & {} \\
{[6]} & ~[4]
\end{array} \right.
\end{array}\)\to
\(\begin{array}{c}
[10\,\,11]~~~~~~~ \\[+0.5em]
\left\{\begin{array}{cc}
[8\,11] & {} \\
{[6\,11]} & ~[4\,11]
\end{array} \right.~~ \\[+1.5em]
\left\{\begin{array}{ccc}
[89]~ & {} & {} \\
{[69]}~ & [49]~ & {} \\
{[67]}~ & [47]~ & [45]
\end{array} \right.
\end{array}\)\to\ldots,
\ee
now each ``face'' of the ``solid tetrahedron'' above follows the pattern of three growing parameters respectively.
This can be better visualized in the upper part of figure \ref{fig-18} and there, the growing parameters are highlighted as
the starting points of a solid simplex.
In $J_{11,1}$ we find 26 full cells, among which 16 belong to $H_{11,2}$ and 10 belong to $H_{11,3}$.
Similar to those five full cells of 2-mode in $J_{9,1}$ which share the growing parameters $8,6,4$,
the ten full cells of 3-mode in $J_{11,1}$ share the growing parameters $10,8,6,4$.

Next, in $J_{12,1}$ we find 15 full cells and all of them are of 3-mode with various quadruples of growing parameters.
Finally, in $J_{13,1}$ we find 5 full cells of 4-mode and all of them share the growing parameters $12,10,8,6,4$.
Among them one example is
\be
(2345)_2(6789)_2(10\,\,11\,\,12\,\,13)_2~~~~~~(12,10,8,6,4)
\ee
where the growing parameters $12,10,8,6,4$ uniquely determine the pattern
\be
\(\begin{array}{cc}
[12]~~ & {} \\[+0.5em]
\left\{\begin{array}{cc}
[10] & {} \\
\!{[8]}~ & [6]
\end{array} \right. &
\begin{array}{c}
{} \\
{[4]}
\end{array}
\end{array}\)\to
\(\begin{array}{ccc}
[12\,\,13]~~~~~~~~~~~~ & {} & {} \\[+0.5em]
\left\{\begin{array}{cc}
[10\,\,13] & {} \\
{[8\,13]}~~ & [6\,13]
\end{array} \right.~~~~~~ &
\begin{array}{c}
{} \\
\!{[4\,13]}~~~~~~
\end{array} & {} \\[+1.5em]
\left\{\begin{array}{ccc}
[10\,\,11] & {} & {} \\
{[8\,11]}~~ & [6\,11]~ & {} \\
{[89]}~~~~ & [69]~~~ & [67]
\end{array} \right. &
\begin{array}{c}
{} \\
\,\left\{\begin{array}{cc}
[4\,11] & {} \\
{[49]}~~ & \,[47]
\end{array} \right.~~
\end{array} &
\begin{array}{c}
{} \\
{} \\
\!\!\![45]
\end{array}
\end{array}\)\to\ldots,
\ee
now each codimension-1 ``face'' or ``solid tetrahedron'', of the solid 4-simplex above follows
the pattern of four growing parameters respectively.
This can be better visualized in the lower part of figure \ref{fig-18}, and the growing parameters
are highlighted as the starting points of a solid simplex.

For convenience, we summarize the numbers of cells in $G_{i,m}$ with respect to $I_{i,1}$
and those of $H_{i,m}$ with respect to $J_{i,1}$ below
\be
\begin{array}{cc|c|c}
~G_{7,0}~ & ~G_{7,1}~ & ~G_{8,1}~ & ~G_{9,2}~ \\
\hline
1+1 & 2 & 3 & 2
\end{array}
\ee
\be
\begin{array}{cc|ccc|cc|cc|c|c}
~H_{8,0}~ & ~H_{8,1}~ & ~H_{9,0}~ & ~H_{9,1}~ & ~H_{9,2}~ & ~H_{10,1}~ & ~H_{10,2}~
& ~H_{11,2}~ & ~H_{11,3}~ & ~H_{12,3}~ & ~H_{13,4}~ \\
\hline
2+1 & 5 & 1 & 12 & 5 & 7 & 20 & 16 & 10 & 15 & 5
\end{array}
\ee
where the `+1' denotes the only fake full cell for each case.
In this way, $J_{i,1}$ can be expressed in terms of $H_{i,m}$ as
\be
\bal
J_{8,1}&=H_{8,0}+H_{8,1},\\
J_{9,1}&=(H_{9,0}+H_{9,1}+H_{9,2})+(H_{8,0,2}+H_{8,1,2}),\\
J_{10,1}&=(H_{10,1}+H_{10,2})+(H_{9,0,2}+H_{9,1,2}+H_{9,2,2})+(H_{8,0,3}+H_{8,1,3}),\\
J_{11,1}&=(H_{11,2}+H_{11,3})+(H_{10,1,2}+H_{10,2,2})+(H_{9,0,3}+H_{9,1,3}+H_{9,2,3})+(H_{8,0,4}+H_{8,1,4}),\\
J_{12,1}&=H_{12,3}+(H_{11,2,2}+H_{11,3,2})+(H_{10,1,3}+H_{10,2,3})+(H_{9,0,4}+H_{9,1,4}+H_{9,2,4})+(H_{8,0,5}+H_{8,1,5}),\\
J_{13,1}&=H_{13,4}+H_{12,3,2}+(H_{11,2,3}+H_{11,3,3})+(H_{10,1,4}+H_{10,2,4})+(H_{9,0,5}+H_{9,1,5}+H_{9,2,5})\\
&~~~+(H_{8,0,6}+H_{8,1,6}),\\
J_{14,1}&=H_{13,4,2}+H_{12,3,3}+(H_{11,2,4}+H_{11,3,4})+(H_{10,1,5}+H_{10,2,5})+(H_{9,0,6}+H_{9,1,6}+H_{9,2,6})\\
&~~~+(H_{8,0,7}+H_{8,1,7}),
\eal
\ee
and so on, where the additional label $l$ in $H_{i,m,l}$ represents its level during the simplex-like growth.
Now, according to the growing modes and levels of full cells, for any $n$, it is easy to count the terms in $J_{n,1}$ as
\be
\bal
&\,(3+5(n-7))+\(1+12(n-8)+5\cdot\frac{(n-7)(n-8)}{2}\)+\(7(n-9)+20\cdot\frac{(n-8)(n-9)}{2}\)\\
&+\(16\cdot\frac{(n-9)(n-10)}{2}+10\cdot\frac{(n-8)(n-9)(n-10)}{6}\)\\
&+15\cdot\frac{(n-9)(n-10)(n-11)}{6}+5\cdot\frac{(n-9)(n-10)(n-11)(n-12)}{24}\\
=\,&\frac{1}{24}(n-5)(n-6)(5n(n-11)+152),
\eal
\ee
which nicely matches $\De^2 N^3_n$ in \eqref{eq-41} as expected.

So far we have understood the entire profile of \textit{all} N$^3$MHV amplitudes by capturing the full cells
and their growing patterns. In appendix \ref{app3}, we list all N$^3$MHV full cells for $n\!=\!8,9,10,11,12,13$
of which the growing modes vary from 0 to 4. The growth of full cells essentially terminates at $n\!=\!13$,
which justifies the critical relation $n\!=\!4k\!+\!1$.

\subsection{Determination of growing modes and parameters}

Having gathered enough ``experimental data'', it is time for us to finish deriving
all the unproved facts in the previous two subsections.
For this purpose, we will make use of the tree BCFW recursion relation in its matrix form
given in figure \ref{fig-2}, along with the triangle-like dissection \eqref{eq-37}, which can greatly
enhance the former's efficiency by providing more structural guidance.

Let's completely clarify the N$^2$MHV full cells as warmup exercises.
Explicitly, we first derive the full cells, then identify their growing parameters,
with which we justify their simplex-like growing patterns of various modes up to any $n$ finally.
The essential termination of N$^2$MHV as well as N$^3$MHV full cells will be proved
in the next subsection, as indicated by a general formula for any $k$.

The first example is $(45)(71)$ of 0-mode at $n\!=\!7$, as recursively constructed from the matrix
\be
C=\(\begin{array}{ccccccc}
c_1~ & c_2~ & c_3 & 0~ & 0~ & -c_6 & -c_7 \\[+0.5em]
0~ & r_2~ & \(r_3\!+\!\dfrac{c_3}{c_2}r_2\) & r_4~ & r_5~ & r_6 & 0
\end{array}\),
\ee
where the first row is its characteristic 5-bracket $[12367]$. If $n$ is increased by one,
we can push the $c_2,c_3$ pair forward by one slot. However, doing so will produce the
empty slot $[2]$ and the resulting cell is not an essential object lying in the bottom row of \eqref{eq-37},
hence we must stretch the matrix without pushing them forward as
\be
C=\(\begin{array}{cccccccc}
c_1~ & c_2~ & c_3 & 0~ & 0~ & 0~ & -c_7 & -c_8 \\[+0.5em]
0~ & r_2~ & \(\und{r_3}\!+\!\dfrac{c_3}{c_2}r_2\) & r_4~ & \und{r_5}~ & r_6~ & \und{r_7} & 0
\end{array}\).
\ee
This leads to three cells with respect to the three underlined entries above
to be removed individually for maintaining an NMHV sub-amplitude. As the removal of $r_3$ or $r_7$
does not lead to a cell with one empty slot, but a full cell of $n\!=\!8$ instead, therefore, to obtain an identical
configuration with one extra empty slot, we can only choose to remove $r_5$ and get
\be
C=\(\begin{array}{cccccccc}
c_1~ & c_2 & c_3 & 0~ & 0~ & 0~ & -c_7 & -c_8 \\[+0.5em]
0~ & r_2 & \(r_3\!+\!\dfrac{c_3}{c_2}r_2\) & r_4~ & 0~ & r_6~ & r_7 & 0
\end{array}\),
\ee
which is $[5](46)(81)$. For $n\!=\!9$, we need to similarly remove two entries as above, and the same reasoning
gives $[56](47)(91)$. Such a pattern continues up to any $n$, which clarifies the behavior of a 0-mode
with $(5)$ as its growing parameter.

Next, we consider $(23)(67)$ of 1-mode at $n\!=\!7$, as constructed from
\be
C=\(\begin{array}{ccccccc}
c_1~ & 0~ & 0 & -c_4 & -c_5 & -c_6 & -c_7 \\[+0.5em]
l_1~ & l_2~ & l_3 & \(l_4\!+\!\dfrac{c_4}{c_5}l_5\) & l_5 & 0 & 0
\end{array}\).
\ee
Again, if $n$ is increased by one, we can push $c_4,c_5$ forward by one slot, or not. In this case, both choices are
legitimate as we choose to remove $l_6$ in
\be
C=\(\begin{array}{cccccccc}
c_1~ & 0~ & 0 & -c_4 & -c_5 & ~0~ & -c_7 & -c_8\\[+0.5em]
l_1~ & \und{l_2}~ & l_3 & \(\und{l_4}\!+\!\dfrac{c_4}{c_5}l_5\) & l_5 & ~\und{l_6}~ & 0 & 0
\end{array}\),
\ee
and $l_4$ in
\be
C=\(\begin{array}{cccccccc}
c_1~ & 0~ & 0~ & 0 & -c_5 & -c_6 & -c_7 & -c_8 \\[+0.5em]
l_1~ & \und{l_2}~ & l_3~ & \und{l_4} & \(l_5\!+\!\dfrac{c_5}{c_6}l_6\) & \und{l_6} & 0 & 0
\end{array}\),
\ee
hence the resulting cells are $[6](23)(78)$ and $[4](23)(78)$, of which the
characteristic 5-brackets are $[14578]$ and $[15678]$ respectively.
For $n\!=\!9$, the characteristic 5-brackets are now $[14589]$, $[15689]$ and $[16789]$,
so we obtain $[67](23)(89)$, $[47](23)(89)$ and $[45](23)(89)$.
Such a pattern continues up to any $n$, which clarifies the behavior of a 1-mode
with $(6,4)$ as its growing parameters.

The full cell $(23)(45)$ of 1-mode at $n\!=\!7$ shares the same growing parameters with $(23)(67)$,
though their origin is slightly different from that of $(23)(67)$. It is constructed from
\be
C=\(\begin{array}{ccccccc}
c_1~ & c_2 & c_3 & 0~ & 0 & -c_6 & -c_7 \\[+0.5em]
0~ & r_2 & ~\dfrac{c_3}{c_2}r_2~ & r_4~ & r_5 & \(r_6\!+\!\dfrac{c_6}{c_7}r_7\) & r_7
\end{array}\),
\ee
and if $n$ is increased by one, we should get
\be
C=\(\begin{array}{cccccccc}
c_1~ & c_2 & c_3 & 0 & ~0 & ~0 & -c_7 & -c_8 \\[+0.5em]
0~ & r_2 & ~\dfrac{c_3}{c_2}r_2~ & \und{r_4} & ~r_5 & ~\und{r_6} & \(r_7\!+\!\dfrac{c_7}{c_8}r_8\) & \und{r_8}
\end{array}\),
\ee
where $r_4$ and $r_6$ can be removed. For $n\!=\!9$, it is easy to obtain $[67](23)(45)$, $[47](23)(56)$ and $[45](23)(67)$
and such a pattern continues up to any $n$. From the three examples above, we learn that for the removal of an entry
to be valid, it must not belong to column 2 or be ``covered'' by any entry in the characteristic 5-bracket.

Let's see how this works in the following, first for $(234)_2(678)_2$ of 1-mode at $n\!=\!8$ with the matrix
\be
C=\(\begin{array}{cccccccc}
c_1~ & 0~ & 0~ & 0 & -c_5 & -c_6 & -c_7 & -c_8 \\
l_1~ & l_2~ & l_3~ & l_4 & l_5 & 0 & 0 & 0
\end{array}\).
\ee
If $n$ is increased by one, it can be stretched as
\be
C=\(\begin{array}{ccccccccc}
c_1~ & 0~ & 0~ & 0 & -c_5 & -c_6 & ~0~ & -c_8 & -c_9 \\
l_1~ & l_2~ & l_3~ & l_4 & l_5 & 0 & ~0~ & 0 & 0
\end{array}\),
\ee
which immediately gives $[7](234)_2(689)_2$, and the other choice is
\be
C=\(\begin{array}{ccccccccc}
c_1~ & 0~ & 0~ & 0~ & 0 & -c_6 & -c_7 & -c_8 & -c_9 \\
l_1~ & \und{l_2}~ & l_3~ & \und{l_4}~ & l_5 & \und{l_6} & 0 & 0 & 0
\end{array}\),
\ee
where $\l_2$ belongs to column 2, and $l_6$ is covered by $-c_6$, so the resulting cell is $[4](235)_2(789)_2$.
Next, for $(456)_2(781)_2$ of 1-mode at $n\!=\!8$ with the matrix
\be
C=\(\begin{array}{cccccccc}
c_1~ & c_2 & c_3 & 0 & 0~ & 0~ & -c_7 & -c_8 \\[+0.5em]
0~ & r_2 & \(r_3\!+\!\dfrac{c_3}{c_2}r_2\) & r_4~ & r_5~ & r_6 & 0 & 0
\end{array}\),
\ee
if $n$ is increased by one, it can be stretched as
\be
C=\(\begin{array}{ccccccccc}
c_1~ & c_2 & c_3 & 0 & 0~ & 0~ & 0 & -c_8 & -c_9 \\[+0.5em]
0~ & r_2 & \(\und{r_3}\!+\!\dfrac{c_3}{c_2}r_2\) & r_4 & \und{r_5}~ & r_6~ & \und{r_7} & 0 & 0
\end{array}\),
\ee
from which we obtain $[7](456)_2(891)_2$ and $[5](467)_2(891)_2$. Similarly for $(23)(456)_2(81)$ of 1-mode at $n\!=\!8$
with the matrix
\be
C=\(\begin{array}{cccccccc}
c_1~ & c_2 & c_3 & 0~ & 0~ & 0 & -c_7 & -c_8 \\[+0.5em]
0~ & r_2 & ~\dfrac{c_3}{c_2}r_2~ & r_4~ & r_5~ & r_6 & r_7 & 0
\end{array}\),
\ee
if $n$ is increased by one, it can be stretched as
\be
C=\(\begin{array}{ccccccccc}
c_1~ & c_2 & c_3 & 0~ & 0~ & 0~ & 0 & -c_8 & -c_9 \\[+0.5em]
0~ & r_2 & ~\dfrac{c_3}{c_2}r_2~ & \und{r_4}~ & r_5~ & \und{r_6}~ & r_7 & \und{r_8} & 0
\end{array}\),
\ee
from which we obtain $[6](23)(457)_2(91)$ and $[4](23)(567)_2(91)$.
From these examples, we know the 1-mode (of level 1) either has two entries to be removed in one
stretched matrix, or has a bifurcation of stretched matrices, each of both has an entry to be removed
or directly results in an empty column. For a 2-mode, such a mechanism will lead to three empty slots.

For example, we consider $(2345)_2(6789)_2$ of 2-mode at $n\!=\!9$ with the matrix
\be
C=\(\begin{array}{ccccccccc}
c_1~ & 0~ & 0~ & 0~ & 0 & -c_6 & -c_7 & -c_8 & -c_9 \\
l_1~ & l_2~ & l_3~ & l_4~ & l_5 & 0 & 0 & 0 & 0
\end{array}\),
\ee
if $n$ is increased by one, it can be stretched as
\be
C=\(\begin{array}{cccccccccc}
c_1~ & 0~ & 0~ & 0~ & 0 & -c_6 & -c_7 & ~0~ & -c_9 & -c_{10} \\
l_1~ & l_2~ & l_3~ & l_4~ & l_5 & 0 & 0 & ~0~ & 0 & 0
\end{array}\),
\ee
which gives $[8](2345)_2(679\,10)_2$, and the other choice is
\be
C=\(\begin{array}{cccccccccc}
c_1~ & 0~ & 0~ & 0~ & 0~ & 0 & -c_7 & -c_8 & -c_9 & -c_{10} \\
l_1~ & \und{l_2}~ & l_3~ & \und{l_4}~ & l_5~ & \und{l_6} & 0 & 0 & 0 & 0
\end{array}\),
\ee
which gives $[6](2345)_2(789\,10)_2$ and $[4](2356)_2(789\,10)_2$.
Again, as one can check, such a pattern continues up to any $n$, which clarifies the behavior of a 2-mode
with $(8,6,4)$ as its growing parameters.

Another different example is $(23)(4567)_2(891)_2$ of 2-mode at $n\!=\!9$ with the matrix
\be
C=\(\begin{array}{ccccccccc}
c_1~ & c_2 & c_3 & 0~ & 0~ & 0~ & 0 & -c_8 & -c_9 \\[+0.5em]
0~ & r_2 & ~\dfrac{c_3}{c_2}r_2~ & r_4~ & r_5~ & r_6~ & r_7 & 0 & 0
\end{array}\),
\ee
if $n$ is increased by one, it can be stretched as
\be
C=\(\begin{array}{ccccccccccc}
c_1~ & c_2 & c_3 & 0~ & 0~ & 0~ & 0~ & 0 & -c_9 & -c_{10} \\[+0.5em]
0~ & r_2 & ~\dfrac{c_3}{c_2}r_2~ & \und{r_4}~ & r_5~ & \und{r_6}~ & r_7~ & \und{r_8} & 0 & 0
\end{array}\),
\ee
which directly results in $[8](23)(4567)_2(9\,10\,1)_2$, $[6](23)(4578)_2(9\,10\,1)_2$ and $[4](23)(5678)_2(9\,10\,1)_2$.

Now we have finished proving \eqref{eq-45}, \eqref{eq-46}, \eqref{eq-47} and \eqref{eq-48}.
But one may note that, we have not mentioned the fake full cell $[5]$. This issue is dissolved if we also include
the top cells as full cells, since a top cell is in fact a trivial full cell without any vanishing constraint,
hence $[5]$ is a descendent of `1'. Such a unified perspective will be necessary for the
general formula in the next subsection, which is ``from full cells to full cells'' without
involving any redundant ingredient.

Next, we will only pick three N$^3$MHV full cells for demonstration, since the number of these objects is too large
for a complete proof. The first one is
\be
(2\,3\!\os{5\,\,\,\,6}{\os{\backslash\,/}{4}}\!\!)\,(789\,10)_2(11\,\,12\,\,1)_2~~~~~~(11,9,7,5) \labell{eq-49}
\ee
of 3-mode at $n\!=\!12$, characterized by the matrix consisting of 2-vectors $R_i$ as
\be
C=\(\begin{array}{cccccccccccc}
c_1 & \,c_2 & c_3 & 0 & \,0 & \,0 & 0 & \,0 & \,0 & \,0 & c_{11} & \,c_{12} \\[+0.5em]
0 & \,R_2 & \(R_3\!+\!\dfrac{c_3}{c_2}R_2\) & \ap_4R_3 & \,\ap_5R_3 & \,\ap_6R_3
& R_7 & \,\ap_8R_7 & \,\ap_9R_7 & \,\ap_{10}R_7 & 0 & \,0
\end{array}\),
\ee
where the $\ap_i$'s help manifest linear dependencies $(3456)_2(789\,10)_2$ of the sub-cell.
If $n$ is increased by one, it can be stretched as
\be
C=\(\begin{array}{ccccccccccccc}
c_1 & \,c_2 & c_3 & 0 & \,0 & \,0 & \,0 & \,0 & \,0 & \,0 & \,0 & c_{12} & \,c_{13} \\[+0.5em]
0 & \,R_2 & \(\und{R_3}\!+\!\dfrac{c_3}{c_2}R_2\) & R_4 & \,\und{R_5} & \,R_6 & \,\und{R_7}
& \,R_8 & \,\und{R_9} & \,R_{10} & \,\und{R_{11}} & 0 & \,0
\end{array}\),
\ee
now for compactness, we have hidden the vanishing constraint $(3456)_2(789\,10)_2$, but keep in mind that it is still there.
As the growing parameters of $(2345)_2(6789)_2$ at $n\!=\!9$ are $(8,6,4)$, they are now $(9,7,5)$ for $(3456)_2(789\,10)_2$
simply because the N$^2$MHV sub-amplitude starts at column 2. In fact, it is the sub-cell that is stretched, then we only
need to recall its growing pattern for obtaining the new parameters of the N$^3$MHV full cell.
Besides $(9,7,5)$, the triangle-like dissection \eqref{eq-37} also brings in empty slots $[2]$ and $[10]$
with respect to $(2345)_2(6789)_2$, which now become $[3]$ and $[11]$ for the sub-cell.
But as $R_3$ is covered by $c_3$, the growing parameters are $(11,9,7,5)$ without 3.
This reasoning can be clearly described as
\be
\(\begin{array}{c|cc}
[2]~ & {} & \!\!\cdots \\[+0.5em]
\begin{array}{c}
{} \\
1~
\end{array} &
\begin{array}{c}
{} \\
~1
\end{array} &
\,\left\{\begin{array}{cc}
[8] & {} \\
{[6]} & [4]
\end{array} \right. \\[+1.3em]
\hline
{} & ~[10] & \!1~
\end{array}\)\Rightarrow
\(\begin{array}{cc}
\begin{array}{c}
{} \\
{[11]}
\end{array} &
\,\left\{\begin{array}{cc}
[9] & {} \\
{[7]} & [5]
\end{array} \right.
\end{array}\!\),
\ee
where $\cdots$ denotes the irrelevant contribution that is ``covered'', and $\Rightarrow$ denotes
the label shift $i\!\to\!i\!+\!1$. For $n\!=\!14$, it is trivial to find
\be
\(\begin{array}{c|ccc}
[23]~ & {} & {} & \!\!\!\cdots~~~~~~~~~ \\[+0.5em]
[2]~~ & {} & \!\!\cdots & \!\!\!\cdots~~~~~~~~~ \\[+0.5em]
\begin{array}{c}
{} \\
{} \\
1~~
\end{array} &
\begin{array}{c}
{} \\
{} \\
~1
\end{array} &
\begin{array}{c}
{} \\
\left\{\begin{array}{cc}
[8] & {} \\
{[6]} & [4]
\end{array} \right.
\end{array} &
\,\left\{\begin{array}{ccc}
[89]~ & {} & {} \\
{[69]}~ & [49]~ & {} \\
{[67]}~ & [47]~ & [45]
\end{array} \right. \\[+2.0em]
\hline
{} & ~[10\,\,11] & \![11] & \!\!1~~~~~~~~~~~
\end{array}\)\Rightarrow
\(\begin{array}{ccc}
\begin{array}{c}
{} \\
{} \\
{[11\,\,12]}
\end{array} &
\begin{array}{c}
{} \\
\,\left\{\begin{array}{cc}
[9\,12] & {} \\
{[7\,12]} & ~[5\,12]
\end{array} \right.
\end{array} &
\left\{\begin{array}{ccc}
[9\,10] & {} & {} \\
{[7\,10]} & ~[5\,10] & {} \\
{[78]}~~ & [58]~ & ~[56]
\end{array} \right.
\end{array}\).
\ee
Therefore, the growing parameters of an N$^3$MHV full cell in fact rely on not only their counterpart of the N$^2$MHV
sub-cell, but also the triangle-like dissection! It is quite desirable to see the two-fold simplex-like structures
of amplitudes have such a close interconnection. As one can check, this pattern continues up to any $n$,
which clarifies the behavior of a 3-mode.

The second example is
\be
(\os{7}{\os{|}{6}}{\os{\,\,9\,\,10\,\,11}{\os{\backslash\,|\,/}{8}}}\,\os{13}{\os{|}{12}})\,
(12\,\,1\!\!{\os{3\,\,4\,\,5}{\os{\backslash\,|\,/}{2}}}\!\!)~~~~~~(12,10,8,6,4)
\ee
of 4-mode at $n\!=\!13$, characterized by the matrix
\be
C=\(\begin{array}{ccccccccccccc}
c_1~ & 0~ & 0~ & 0~ & 0 & -c_6 & -c_7 & 0~ & 0~ & 0~ & 0~ & c_{12}~ & c_{13} \\
l_1~ & l_2~ & l_3~ & l_4~ & l_5 & 0 & 0 & 0~ & 0~ & 0~ & 0~ & 0~ & 0 \\[+0.5em]
0~ & 0~ & 0~ & 0~ & 0~ & r_6 & ~\dfrac{c_7}{c_6}r_6~ & r_8~ & r_9~ & r_{10}~ & r_{11}~ & 0~ & 0
\end{array}\).
\ee
If $n$ is increased by one, it can be stretched as
\be
C=\(\begin{array}{cccccccccccccc}
c_1~ & 0~ & 0~ & 0~ & 0 & -c_6 & -c_7 & 0~ & 0~ & 0~ & 0~ & 0~ & c_{13}~ & c_{14} \\
l_1~ & l_2~ & l_3~ & l_4~ & l_5 & 0 & 0 & 0~ & 0~ & 0~ & 0~ & 0~ & 0~ & 0 \\[+0.5em]
0~ & 0~ & 0~ & 0~ & 0 & r_6 & ~\dfrac{c_7}{c_6}r_6~ & \und{r_8}~ & r_9~ & \und{r_{10}}~
& r_{11}~ & \und{r_{12}}~ & 0~ & 0
\end{array}\),
\ee
which gives $[12]$, $[10]$ and $[8]$, and the other choice is
\be
C=\(\begin{array}{cccccccccccccc}
c_1~ & 0~ & 0~ & 0~ & 0~ & 0 & -c_7 & -c_8 & 0~ & 0~ & 0~ & 0~ & c_{13}~ & c_{14} \\
l_1~ & \und{l_2}~ & l_3~ & \und{l_4}~ & l_5~ & \und{l_6} & 0 & 0 & 0~ & 0~ & 0~ & 0~ & 0~ & 0 \\[+0.5em]
0~ & 0~ & 0~ & 0~ & 0~ & 0 & r_7 & ~\dfrac{c_8}{c_7}r_7~ & r_9~ & r_{10}~ & r_{11}~ & r_{12}~ & 0~ & 0
\end{array}\),
\ee
which gives $[6]$ and $[4]$. The last example shares the same growing parameters with the one above, it is
\be
(23)(4567)_2(89\,10\,\,11)_2(12\,\,13\,\,1)_2~~~~~~(12,10,8,6,4)
\ee
of 4-mode at $n\!=\!13$, characterized by the matrix
\be
C=\(\begin{array}{ccccccccccccc}
c_1 & \,c_2 & c_3 & 0 & \,0 & \,0 & \,0 & 0 & \,0 & \,0 & \,0 & c_{12} & \,c_{13} \\[+0.5em]
0 & \,R_2 & ~\dfrac{c_3}{c_2}R_2~ & R_4 & \,\ap_5R_4 & \,\ap_6R_4 & \,\ap_7R_4
& R_8 & \,\ap_9R_8 & \,\ap_{10}R_8 & \,\ap_{11}R_8 & 0 & \,0
\end{array}\),
\ee
where the linear dependencies $(4567)_2(89\,10\,\,11)_2$ of the sub-cell are manifest, which will be again hidden
in the following. If $n$ is increased by one, it can be stretched as
\be
C=\(\begin{array}{cccccccccccccc}
c_1 & \,c_2 & c_3 & 0 & \,0 & \,0 & \,0 & \,0 & \,0 & \,0 & \,0 & \,0 & c_{13} & \,c_{14} \\[+0.5em]
0 & \,R_2 & ~\dfrac{c_3}{c_2}R_2~ & \und{R_4} & \,R_5 & \,\und{R_6} & \,R_7 & \,\und{R_8} & \,R_9 & \,\und{R_{10}}
& \,R_{11} & \,\und{R_{12}} & 0 & \,0
\end{array}\),
\ee
which gives $[12]$, $[10]$, $[8]$, $[6]$ and $[4]$. Among these five empty slots, $(10,8,6)$ are the growing parameters
of $(4567)_2(89\,10\,\,11)_2$ simply because the N$^2$MHV sub-amplitude starts at column 3,
and $(12,4)$ are from the triangle-like dissection of this sub-amplitude of $n\!=\!9$. This can be clearly described as
\be
\(\begin{array}{c|cc}
[2]~ & {} & \!1~ \\[+0.5em]
\begin{array}{c}
{} \\
1~
\end{array} &
\begin{array}{c}
{} \\
~1
\end{array} &
\,\left\{\begin{array}{cc}
[8] & {} \\
{[6]} & [4]
\end{array} \right. \\[+1.3em]
\hline
{} & ~[10] & \!1~
\end{array}\)\Rightarrow
\(\begin{array}{cc}
{} & \!\![4]~~ \\[+0.5em]
\begin{array}{c}
{} \\
{[12]}
\end{array} &
\,\left\{\begin{array}{cc}
[10] & {} \\
{[8]}~\, & [6]
\end{array} \right.
\end{array}\!\)
\ee
where $\Rightarrow$ similarly denotes the label shift $i\!\to\!i\!+\!2$. For $n\!=\!15$, it is trivial to find
\be
\(\begin{array}{c|ccc}
[23]~ & {} & {} & \!\!\!1~~~~~~~~~~~ \\[+1.0em]
\begin{array}{c}
{} \\
{[2]}~~
\end{array} & {} &
\begin{array}{c}
{} \\
\!\!\!1
\end{array} &
\!\!\!\left\{\begin{array}{cc}
[9] & {} \\
{[7]} & [5]
\end{array} \right.~~~~~~~~ \\[+1.5em]
\begin{array}{c}
{} \\
{} \\
1~~
\end{array} &
\begin{array}{c}
{} \\
{} \\
~1
\end{array} &
\begin{array}{c}
{} \\
\,\left\{\begin{array}{cc}
[8] & {} \\
{[6]} & [4]
\end{array} \right.
\end{array} &
\,\left\{\begin{array}{ccc}
[89]~ & {} & {} \\
{[69]}~ & [49]~ & {} \\
{[67]}~ & [47]~ & [45]
\end{array} \right. \\[+2.0em]
\hline
{} & ~[10\,\,11] & [11] & \!\!1~~~~~~~~~~~
\end{array}\)\Rightarrow
\(\begin{array}{ccc}
{} & {} & \!\!\![45]~~~~~~~~~~~~~~~ \\[+1.0em]
{} &
\begin{array}{c}
{} \\
\!\![4\,13]~~~~~~
\end{array} &
\!\!\!\left\{\begin{array}{cc}
[4\,11] & {} \\
{[49]}~~ & [47]
\end{array} \right.~~~~~~~~~ \\[+1.5em]
\begin{array}{c}
{} \\
{} \\
{[12\,\,13]}
\end{array} &
\begin{array}{c}
{} \\
\,\left\{\begin{array}{cc}
[10\,\,13] & {} \\
{[8\,13]}~~ & [6\,13]
\end{array} \right.
\end{array} &
\,\left\{\begin{array}{ccc}
[10\,\,11]~ & {} & {} \\
{[8\,11}]~~~ & [6\,11]~ & {} \\
{[89]}~~~~~ & [69]~~~ & [67]
\end{array} \right.
\end{array}\),
\ee
and as one can check, this pattern continues up to any $n$, which clarifies the behavior of a 4-mode.
Note the 4-mode comes from a 2-mode upgraded by the triangle-like dissection, which schematically
looks like ``(2-mode)$\,\otimes\,$(2-mode)$\,=\,$(4-mode)'', if there is no additional truncation resulting from
the removal of an entry that belongs to column 2 or is ``covered''. In contrast, such a truncation happens for
the previous example \eqref{eq-49}, hence it is degraded to a 3-mode.

\subsection{Refined BCFW recursion relation of fully-spanning cells and its termination at $n\!=\!4k\!+\!1$}

Finally, we will derive the general formula which counts the full cells through the refined BCFW
recursion relation as promised. This formula provides a clear guidance without either missing or overcounting terms
for us to construct full cells solely from full cells. We also know the ``experimental'' fact that the
essential termination of full cells for $k\!=\!1,2,3$ happens at $n\!=\!4k\!+\!1$, and now it can be generalized for any $k$
in the meanwhile.

First, let's summarize the numbers of full cells for $k\!=\!1,2,3$ below
\be
\begin{array}{c|cccccccccc}
\,k\backslash n\, & ~~5~ & ~6~ & ~7~ & ~8~ & ~9~ & ~10~ & ~11~ & ~12~ & ~13~ & ~14~ \\
\hline
1~ & ~1 & {} & {} & {} & {} & {} & {} & {} & {} & {} \\
2~ & {} & 1 & 3 & 3 & 2 & {} & {} & {} & {} & {} \\
3~ & {} & {} & 1 & 7 & 18 & 27 & 26 & 15 & 5
\end{array} \labell{eq-50}
\ee
where we have included top cells as full cells to make the aimed formula work,
and all the blank areas are implicitly filled with zeros.
We may denote each number above as ($F$ stands for fully-spanning)
\be
F^k_n~~\textrm{for}~~k+4\leq n\leq 4k+1,
\ee
then for the top cells, we trivially have $F^k_{k+4}\!=\!1$.
Note the fake full cell $[k\!+\!3]$ has been excluded in $F^k_{k+5}$
for $k\!\geq\!2$, because it is in fact a descendent of the corresponding top cell denoted by `1'.
If we rearrange the anti-NMHV triangles according to the triangle-like dissection, the vertex $[k\!+\!3]$
spanning this class of triangles clearly descends from the top cell. It is of 0-mode with the growing parameter $(k\!+\!3)$,
which is definitely consistent with the NMHV triangle \eqref{eq-34}. In this way amplitudes of all $k$'s are nicely
unified, as we now separate $[k\!+\!3]$ from $I_{k+5,1}$, and this peculiar exception turns out to be
a natural ingredient of the two-fold simplex-like structures.

\begin{figure}
\begin{center}
\includegraphics[width=0.6\textwidth]{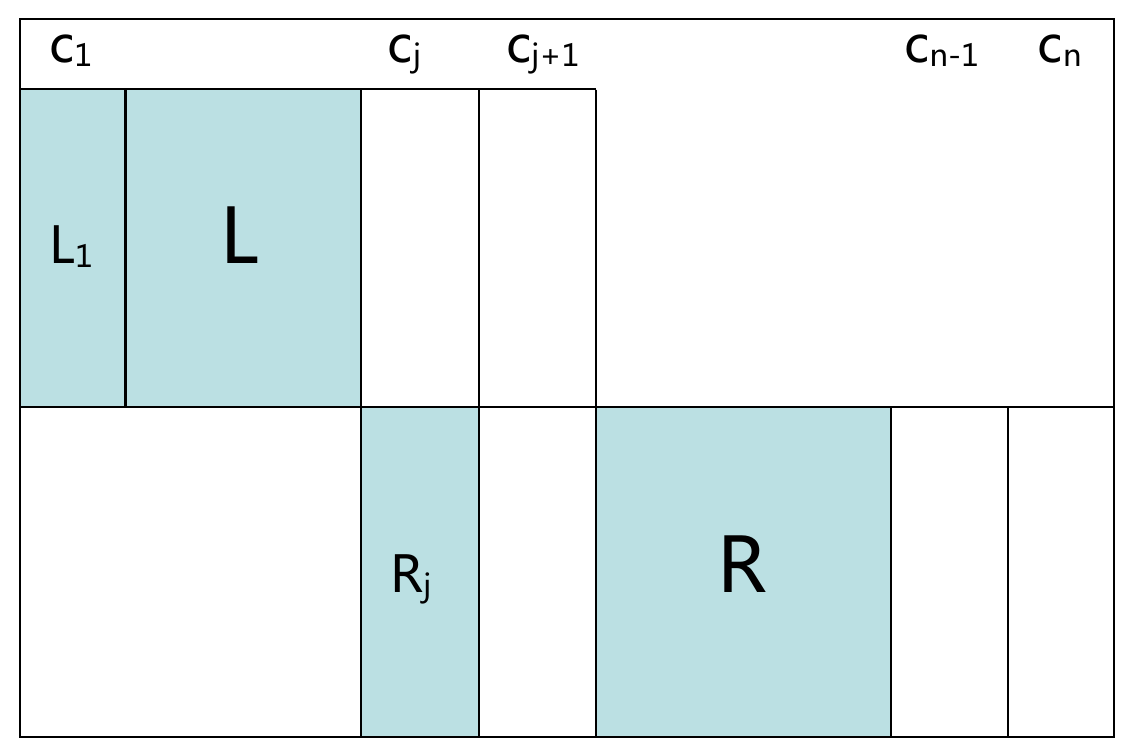}
\caption{Maximally constructing full cells from full cells through the refined BCFW recursion relation.} \label{fig-19}
\end{center}
\end{figure}

Next, we present how to construct full cells solely from full cells through the refined BCFW recursion relation,
as sketched in figure \ref{fig-19}. There, the left and right sub-matrices are maximally moved away from
the ``shadows'' of the $c_i$'s above their last two columns, while the resulting cell is
maintained to be fully-spanning. Furthermore,
the second column of the right sub-matrix is removed since it is covered by $c_{j+1}$, as its removal will not render
column $j\!+\!1$ empty. But this is not available for the left one, as there is no $c_2$ above its second column.

Therefore, the minimal full cells in the left and right sub-matrices are of widths $(n_\textrm{L}\!-\!2)$ and
$(n_\textrm{R}\!-\!3)$ respectively, while the maximal ones are trivially of $n_\textrm{L}$ and $n_\textrm{R}$.
From this, we can count the numbers of contributing full cells in the left and right sub-matrices as
\be
\(F^{k_\textrm{L}}_{n_\textrm{L}-2}+F^{k_\textrm{L}}_{n_\textrm{L}-1}+F^{k_\textrm{L}}_{n_\textrm{L}}\)~~\textrm{for}~~
k_\textrm{L}+4\leq n_\textrm{L}\leq(4k_\textrm{L}+1)+2,
\ee
from the triangle-like dissection of $Y_\textrm{L}$ (only the contributing fraction is presented)
\be
\(\begin{array}{c|cccc}
\vdots~ & {} & \vdots & ~\vdots & ~\vdots \\
1~\, & ~\cdots
& F^{k_\textrm{L}}_{n_\textrm{L}-2} & ~F^{k_\textrm{L}}_{n_\textrm{L}-1} & ~F^{k_\textrm{L}}_{n_\textrm{L}} \\[+0.25em]
\hline
{} & ~\cdots & ~[n_\textrm{L}\!-\!1~n_\textrm{L}] & ~~[n_\textrm{L}] & ~~1~
\end{array}\),
\ee
as well as
\be
\(F^{k_\textrm{R}}_{n_\textrm{R}-3}
+2\(F^{k_\textrm{R}}_{n_\textrm{R}-2}+F^{k_\textrm{R}}_{n_\textrm{R}-1}\)+F^{k_\textrm{R}}_{n_\textrm{R}}\)
~~\textrm{for}~~k_\textrm{R}+4\leq n_\textrm{R}\leq(4k_\textrm{R}+1)+3,
\ee
from the triangle-like dissection of $Y_\textrm{R}$
\be
\(\begin{array}{c|cccc}
\,\vdots~ & {} & \vdots & ~\vdots & ~\vdots \\
{[n_\textrm{L}]}~ & ~\cdots
& F^{k_\textrm{R}}_{n_\textrm{R}-3} & ~F^{k_\textrm{R}}_{n_\textrm{R}-2} & ~F^{k_\textrm{R}}_{n_\textrm{R}-1} \\[+0.25em]
1~ & ~\cdots
& F^{k_\textrm{R}}_{n_\textrm{R}-2} & ~F^{k_\textrm{R}}_{n_\textrm{R}-1} & ~F^{k_\textrm{R}}_{n_\textrm{R}} \\[+0.25em]
\hline
{} & ~\cdots & ~[n\!-\!1~n] & ~~[n] & ~~1~
\end{array}\).
\ee
Note that we have assumed the critical relation $n\!=\!4k\!+\!1$ for both sub-amplitudes,
but the upper bounds of $n_\textrm{L}$ and $n_\textrm{R}$ are increased by $+2$ and $+3$ respectively,
due to the construction above.

\begin{figure}
\begin{center}
\includegraphics[width=0.45\textwidth]{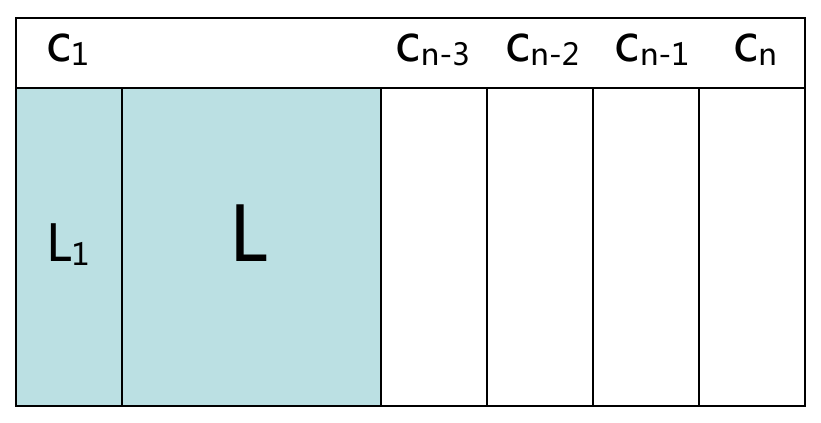}
\caption{Special case $k_\textrm{R}\!=\!0$ of maximally constructing full cells from full cells.} \label{fig-20}
\end{center}
\end{figure}

\begin{figure}
\begin{center}
\includegraphics[width=0.45\textwidth]{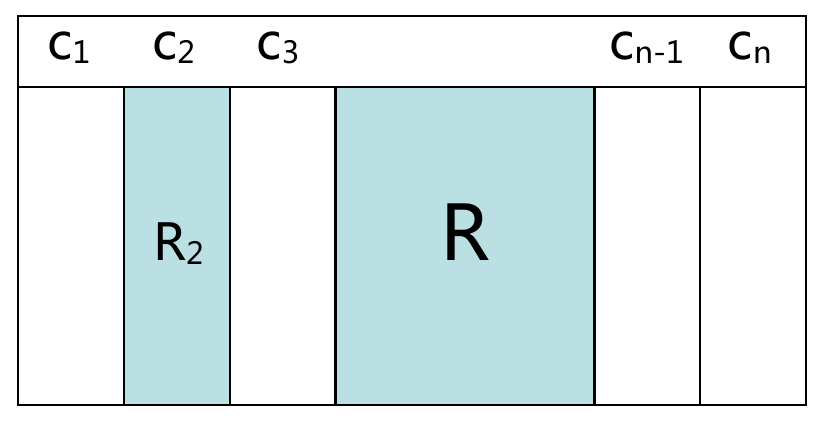}
\caption{Special case $k_\textrm{L}\!=\!0$ of maximally constructing full cells from full cells.} \label{fig-21}
\end{center}
\end{figure}

However, there are two special cases we need to consider separately:
$k_\textrm{R}\!=\!0$ and $k_\textrm{L}\!=\!0$, as shown in figures \ref{fig-20} and \ref{fig-21}.
Similarly, the numbers of contributing full cells can be counted from their triangle-like dissections given by
\be
\(\begin{array}{c|cccc}
\,\vdots~ & {} & \vdots & ~\vdots & ~\vdots \\
1~ & ~\cdots & F^{k-1}_{n-4} & ~F^{k-1}_{n-3} & ~F^{k-1}_{n-2} \\[+0.25em]
\hline
{} & ~\cdots & ~[n\!-\!3~n\!-\!2] & ~~[n\!-\!2] & ~~1~
\end{array}\),
\ee
as well as
\be
\(\begin{array}{c|cccc}
\,\vdots~ & {} & \vdots & ~\vdots & ~\vdots \\
{[3]}~ & ~\cdots & F^{k-1}_{n-4} & ~F^{k-1}_{n-3} & ~F^{k-1}_{n-2} \\[+0.25em]
1~ & ~\cdots & F^{k-1}_{n-3} & ~F^{k-1}_{n-2} & ~F^{k-1}_{n-1} \\[+0.25em]
\hline
{} & ~\cdots & ~[n\!-\!1~n] & ~~[n] & ~~1~
\end{array}\),
\ee
which are $\(F^{k-1}_{n-4}\!+\!F^{k-1}_{n-3}\!+\!F^{k-1}_{n-2}\)$ and
$\(F^{k-1}_{n-4}\!+\!2\(F^{k-1}_{n-3}\!+\!F^{k-1}_{n-2}\)\!+\!F^{k-1}_{n-1}\)$ respectively. \vskip 0.2 cm%

Now, we abuse the same symbol in \eqref{eq-36}, following the recursion
\be
N^k_n=(N_\textrm{L})^{k-1}_{n-2}+(N_\textrm{R})^{k-1}_{n-1}
+\sum_{k_\textrm{L},\,k_\textrm{R}\geq1}N^{k_\textrm{L}}_{n_\textrm{L}}N^{k_\textrm{R}}_{n_\textrm{R}},
\ee
to count the full cells \textit{only}, hence this is ``from full cells to full cells''. Note the two
special cases $k_\textrm{R}\!=\!0$ and $k_\textrm{L}\!=\!0$ have been separated from the rest contributions,
as $(N_\textrm{L})^{k-1}_{n-2}$ and $(N_\textrm{R})^{k-1}_{n-1}$ count the full cells with respect to figures
\ref{fig-20} and \ref{fig-21}. Summing the three parts according to the discussion above,
we get the general inductive formula:
\be
\bal
F^k_n=&\(2F^{k-1}_{n-4}+3\(F^{k-1}_{n-3}+F^{k-1}_{n-2}\)+F^{k-1}_{n-1}\)\\
&+\sum^{k_\textrm{L}+k_\textrm{R}=k-1}_{k_\textrm{L},\,k_\textrm{R}\geq1}
\sum^{n_\textrm{L}+n_\textrm{R}=n+2}_{\substack{k_\textrm{L}+4\,\leq\,n_\textrm{L}\,\leq\,4k_\textrm{L}+3\\{}\\
k_\textrm{R}+4\,\leq\,n_\textrm{R}\,\leq\,4k_\textrm{R}+4}}
\(F^{k_\textrm{L}}_{n_\textrm{L}-2}+F^{k_\textrm{L}}_{n_\textrm{L}-1}+F^{k_\textrm{L}}_{n_\textrm{L}}\)
\(F^{k_\textrm{R}}_{n_\textrm{R}-3}
+2\(F^{k_\textrm{R}}_{n_\textrm{R}-2}+F^{k_\textrm{R}}_{n_\textrm{R}-1}\)+F^{k_\textrm{R}}_{n_\textrm{R}}\), \labell{eq-51}
\eal
\ee
which can recursively reproduce \eqref{eq-50} as a nontrivial consistency check.

At last, we are able to prove the critical relation $n\!=\!4k\!+\!1$ for the essential termination of full cells,
which is straightforwardly confirmed by the formula above, as we find $F^k_n\!=\!0$ for $n\!>\!4k\!+\!1$.
Actually, this can be seen much more intuitively in figures \ref{fig-19}, \ref{fig-20} and \ref{fig-21}.
Imagine the critical relation $n\!=\!4k\!+\!1$ holds for full cells of smaller $k$ and $n$,
from figures \ref{fig-20} and \ref{fig-21} we know the maximal $n$ is given by
\be
(4(k-1)+1)+4=4k+1,
\ee
where $+4$ comes from four empty sub-columns below the $c_i$'s, for both special cases.
For the general case of which $k_\textrm{L},k_\textrm{R}\!\geq\!1$, from figure \ref{fig-19} the maximal $n$ is given by
\be
(4k_\textrm{L}+1)+(4k_\textrm{R}+1)+3=4(k_\textrm{L}+k_\textrm{R}+1)+1=4k+1,
\ee
where $+3$ comes from three empty sub-columns. Therefore, we have proved the critical relation by simple induction.
This relation has a nontrivial consistency check, as given in the following.

From the N$^2$MHV and N$^3$MHV amplitudes after dissection,
we have found an interesting property of the growing parameters: regardless of $k$,
an $m$-mode \textit{first} shows up at $n\!=\!2m\!+\!5$, of which the
growing parameters are $(2m\!+\!4,\ldots,6,4)$. Explicitly, we know the relevant ``experimental data'' up to $k\!=\!3$ as
\be
\bal
&n=5:~~~~~~~~~~~~~~~~~~~~~~~(4)\\
&n=7:~~~~~~~~~~~~~~~~~~~~(6,4)\\
&n=9:~~~~~~~~~~~~~~~~~(8,6,4)\\
&n=11:~~~~~~~~~~~(10,8,6,4)\\
&n=13:~~~~~~\,(12,10,8,6,4)
\eal
\ee
for $m\!=\!0,1,2,3,4$, where all the parameters are even. When we push the $c_j,c_{j+1}$ pair forward in figure \ref{fig-2},
the only way to expose all $(m\!+\!1)$ empty slots demands the growing parameters to be $(2m\!+\!4,\ldots,6,4)$. The distance
between two closest parameters is 2 due to the moving $c_j,c_{j+1}$ pair. There are two left even numbers, however,
namely 2 and $(2m\!+\!6)$, as the truncated ``head and tail'' of growing parameters. It is not surprising to find that
they are nothing but $[2]$ and $[n\!+\!1]$ associated with $I_{n,2}$ and $I_{n,1}$ respectively
in the triangle-like dissection \eqref{eq-37}, but the growing parameters are associated with $I_{n+1,1}$.
Therefore, they cannot be part of the growing parameters.

Based on the observation above, now we know that for a given $k$, the highest mode shows up only at $n\!=\!4k\!+\!1$,
which uses all possible growing parameters. The resulting $m$-simplex of which $m\!=\!(n\!-\!5)/2\!=\!2(k\!-\!1)$
implies for $\De^2 N^k_n$, the highest order in $n$ is $2(k\!-\!1)$.
On the other hand, in \eqref{eq-36} the number of BCFW terms in tree amplitudes is given by
\be
N^k_n=\frac{1}{n\!-\!3}\binom{n\!-\!3}{k}\binom{n\!-\!3}{k\!+\!1}
=\frac{(n-4)(n-5)\ldots(n-k-2)}{k!}\times\frac{(n-3)(n-4)\ldots(n-k-3)}{(k+1)!},
\ee
then the second order difference of $N^k_n$ is analogous to its second order derivative with respect to $n$,
which gives the highest order in $n$ as expected:
\be
(k-1)+(k+1)-2=2(k-1).
\ee
While the general formula \eqref{eq-51} guides us to enumerate full cells solely from full cells,
with the correct counting, the termination of this procedure at $n\!=\!4k\!+\!1$ tells us to stop
when the positivity of full cells is maximally ``diluted''.

\newpage
\section{Discussions}

This ending section investigates some further technical aspects around the four major facets
of positive Grassmannian geometry as potential future directions. They may help perfect the present understanding
and calculational capability towards a wider range of applicability.
\\ \\
1. \textit{2-loop BCFW recursion relation from positivity. Alternative positive construction.}\\ \\
It is immediately desirable to reformulate the 2-loop BCFW recursion relation, by manipulating positive components
introduced in section \ref{sec3}. In the 2-loop case, we will see how two $D$-matrices interact on top of the
tree counterpart (namely the $C$-matrix), besides the known interaction between $D$ and $C$ enforced by positivity
in the 1-loop case. Its logic of ``positive interaction'' is analogous to that of the amplituhedron formulation for
4-particle MHV integrands to all loop orders, while the latter is far simpler, although its idea of
summing all the projective volumes of non-overlapping positive regions lacks an elegant extension to more generic sectors,
probably due to the specialty of 4 particles (a loop integration involves 4 degrees of freedom, a momentum twistor has
4 components, $G(2,4)$ has 4 degrees of freedom). Nevertheless, there might exist an eclectic way, which combines both
the wide-range applicability of brute-force recursion and the extreme simplicity of 4-particle all-loop amplituhedron,
if we dive deep enough into the essential role of positivity for understanding amplitudes and integrands.
To study purely the interaction between two $D$-matrices, one may first consider the 2-loop MHV integrand for a
generic $n$. Of course, this is already much more nontrivial than its 1-loop counterpart, as it is quite different from
the amplituhedron proposal which stacks one $D$ on top of the other and then imposes positivity of the entire matrix.
In fact, the latter approach also encounters significant difficulties, as positivity of both the momentum twistor data
and the matrix made of two $D$'s is more subtle to be implemented, unlike the simple case of a $4\!\times\!4$
square matrix which has only one positive determinant.

Therefore, in this intriguing mist of positivity it is hard to tell what would be the ultimate paradigm
for calculating amplitudes and integrands. More new structures may emerge from more complex results, which may be
considered as laborious, repetitive, yet lacking in merits in the old approach, since without enough complexity
new global insights can be hardly possible. We are ready for switching perspectives to be adapted to
alternatives which can generate physical quantities more efficiently from positivity.
\\ \\
2. \textit{Signs of non-BCFW-like cells. Cyclicity of amplitudes.}
\\ \\
We have not confronted the sign problems of two special N$^2$MHV cells, namely $(12)(34)(56)(78)$ at $n\!=\!8$ and
$(123)_2(456)_2(789)_2$ at $n\!=\!9$, in the related homological identities. As such cells cannot be connected with
BCFW-like cells which do not have extra non-unity factors, other types of residue theorems beyond the simple
6-term NMHV identity must be explored for understanding their signs. We expect the tedious calculation involving
solving quadratic and linear system of equations could be circumvented as much as possible, after
their essence is better captured in a purely geometric way.

There is a fascinating interconnection between the cyclicity of amplitudes, and definitions of physical as well as
spurious poles, while the former relies on various homological identities as we have known. The physical poles
are defined as momentum twistor contractions of the form $\<i\,i\!+\!1\,j\,j\!+\!1\>$, but in terms of Grassmannian
geometry representatives they may be more complicated. And the rest poles are spurious,
which must cancel pair-wise as internal faces of an amplituhedron. Curiously, we find that under
a cyclic shift (by $+1$) only the physical poles of a generic NMHV amplitude will survive,
while all of its spurious poles disappear prior to cancelation.
The cyclicity of amplitudes serves as a filter for separating physical and spurious poles in this sense,
and we conjecture it holds for any $k$. It can perhaps even become a novel distinction between
them, without making reference to momentum twistor contractions.

A natural yet tricky question about cyclicity is: can it be manifested without any knowledge of non-BCFW-like cells
since they in fact never show up in amplitudes constructed by recursion? We guess they cancel in the
related homological identities due to some nontrivial reasons, or identities containing these cells
are simply irrelevant to cyclicity.
\\ \\
3. \textit{How the two-fold simplex-like structures of amplitudes help manifest cyclicity?}
\\ \\
Since the two-fold simplex-like structures greatly refine amplitudes, by capturing the full cells along with
their growing modes and parameters, it is appealing to expect they can help manifest cyclicity. At least,
they should save considerable amount of repetitive manipulation of homological identities. This is nicely realized
in the case of NMHV amplitudes in section \ref{sec4}, as we derive an inductive relation of cyclicity.
But we should not be so wishful, when trying to prove the cyclicity of N$^2$MHV $n\!=\!8$ amplitude based on that of
the $n\!=\!7$ amplitude, for example, due to BCFW cells of new topologies and new identities connecting them,
while there is only one full cell and one distinct identity in the NMHV case.
Nevertheless, there are limited numbers of full cells and distinct identities, so after we manage to prove
the cyclicity of N$^2$MHV $n\!=\!9$ amplitude in which all N$^2$MHV full cells have shown up,
cyclicity of a larger $n$ should be available in a simpler way.
\\ \\
4. \textit{1-loop invariant Grassmannian characterization.}
\\ \\
It is yet unknown how to naturally incorporate the 1-loop kermit function with its corresponding BCFW
forward-limit product of Yangian invariants. We demand more than just positivity as done in section \ref{sec3},
since we need to clarify how to characterize 1-loop BCFW cells invariantly, while this invariance calls for
clarification as well. Such a characterization should ultimately be geometric like its tree counterpart,
then we can reclaim all the positive Grassmannian machineries as previously introduced and upgrade them for 1-loop
integrands. First, it is necessary to establish the bijection between 1-loop Grassmannian geometric configurations
and positive matrix representatives. This bijection is indispensable if we want a numerical check, whenever
feeling unsure about the formal relation. After that, we can use the compact geometric objects to explore
the analogous aspects of 1-loop integrands, such as the homological identities and the
two-fold simplex-like structures.

The 1-loop Grassmannian geometry representatives have a notable difference: a $D$-matrix is unlike a 5-bracket
since when it is forged with a $C$-matrix by positivity, it is still a separable part, but a 5-bracket
can be totally fused into the $C$-matrix and become an ``organic'' part of it.
It is then inevitable to use a composite representation that separates the kermit and tree parts, though as a whole
it is positive.
\\ \\
5. \textit{1-loop homological identities. Residue theorems connecting FAC and FWD terms.}
\\ \\
Among the potential 1-loop homological identities, we are especially interested in those residue theorems
which can connect FAC and FWD terms. The reason is straightforward: we would like to have a unified description for
both FAC and FWD terms as mentioned in section \ref{sec3}, otherwise the somehow ``manual'' separation between them
obstructs a global formalism similar to the two-fold simplex-like structures.
As both of them originate from the BCFW deformation, and are identified as residues of physical poles,
it is appealing to expect they are not essentially different.

But of course, all 1-loop identities deserve attention because they may be relevant to the cyclicity of
1-loop integrands. In fact, the latter becomes an arena for checking the identities, similar to that of tree amplitudes.
Furthermore, these identities are ideal objects for checking the positive parameterization of
1-loop Grassmannian geometry representatives via their actual integrands, since it is highly nontrivial to get a zero
from the cancelation of many nasty numerical terms.
\\ \\
6. \textit{Botany of full cells.}
\\ \\
Although the full cells are generated by the refined BCFW recursion relation as done in section \ref{sec6},
these geometric objects are obviously independent of any specific matrix configuration. Is there a geometric or
combinatoric approach which can generate full cells without the aid of recursion in its matrix form, along with
their growing modes and parameters? In other words, we are looking for a completely intrinsic way independent of
physical factorization limits to define amplitudes, such that in its formalism, the two-fold simplex-like structures
are manifest! However, a hidden obstacle is: we seem to give some privilege to the default recursion scheme.
To be less biased, we have to recall the homological identities that account for the cyclicity of amplitudes,
so it is hard to overestimate their role even in this ``botany'' of full cells.

\section*{Acknowledgments}

The author would like to thank Bo Feng and Song He for reading the manuscript and giving valuable\\ comments.
This work is supported by Qiu-Shi Funding and Chinese NSF funding under contracts\\
No.11135006, No.11125523 and No.11575156.

\newpage
\appendix
\section{Matrix Representatives of 1-loop NMHV $n\!=\!6$ Integrand}
\label{app1}

Below we present the 16 matrix representatives of 1-loop NMHV $n\!=\!6$ integrand,
from BCFW recursion relation in the positive matrix form.
For FAC terms, $d_i$'s and $b_i$'s are the kermit variables, while $c_i$'s are ordinary Grassmannian variables.
But for FWD terms, $d_i$'s and $c_i$'s are the kermit variables, as $l_i$'s and $r_i$'s are ordinary
Grassmannian variables with respect to $Y_\textrm{L}$ and $Y_\textrm{R}$.
It is straightforward to rewrite them as familiar 1-loop integrands, as we can read off the deformed
momentum twistors and reconstruct their corresponding kermit functions and Yangian invariants.
Note the GL(1) redundancies of kermit variables, and the GL($k$) redundancy (with $k\!=\!1$) of Grassmannian variables,
are all unfixed.

\subsection{$n\!=\!5$ FAC}

\be
C=\(\begin{array}{ccccc}
c_1 & ~c_2~ & c_3 & c_4 & c_5 \\[+0.5em]
0 & ~d_2~ & \dfrac{c_3}{c_2}d_2 & -\!\(d_4\!+\!\dfrac{c_4}{c_5}d_5\) & -d_5 \\[+0.5em]
0 & ~b_2~ & \(b_3\!+\!\dfrac{c_3}{c_2}b_2\) & b_4 & 0
\end{array}\)
\ee

\subsection{$n\!=\!5$ FWD}

\be
C=\(\begin{array}{ccccc}
d_1 & ~0~ & 0 & d_4 & d_5 \\[+0.5em]
c_1 & ~c_2~ & c_3 & 0 & 0 \\[+0.5em]
r_1 & ~r_2~ & \(r_3\!+\!\dfrac{c_3}{c_2}r_2\)
& \(r_4\!+\!\dfrac{d_4}{d_5}r_5\!+\!\dfrac{d_4}{d_1}r_1\) & \(r_5\!+\!\dfrac{d_5}{d_1}r_1\)
\end{array}\)
\ee
\be
C=\(\begin{array}{ccccc}
d_1 & ~0~ & 0 & d_4 & ~d_5 \\[+0.5em]
c_1 & ~0~ & -c_3 & -c_4 & ~0 \\[+0.5em]
\(l_1\!+\!\dfrac{d_1}{d_5}l_5\) & ~l_2~ & \(l_3\!+\!\dfrac{c_3}{c_4}l_4\) & \(l_4\!+\!\dfrac{d_4}{d_5}l_5\) & ~l_5
\end{array}\)
\ee

\subsection{$n\!=\!6$ FAC}

\be
C=\(\begin{array}{cccccc}
c_1 & ~c_2~ & c_3 & 0 & ~c_5 & ~c_6 \\[+0.5em]
0 & ~d_2~ & \dfrac{c_3}{c_2}d_2 & -d_4 & ~-\!d_5 & ~0 \\[+0.5em]
0 & ~b_2~ & \(b_3\!+\!\dfrac{c_3}{c_2}b_2\) & b_4 & ~0 & ~0
\end{array}\)
\ee
\be
C=\(\begin{array}{cccccc}
c_1 & ~c_2~ & c_3 & ~0~ & c_5 & c_6 \\[+0.5em]
0 & ~d_2~ & \dfrac{c_3}{c_2}d_2 & ~0~ & -\!\(d_5\!+\!\dfrac{c_5}{c_6}d_6\) & -d_6 \\[+0.5em]
0 & ~b_2~ & \(b_3\!+\!\dfrac{c_3}{c_2}b_2\) & ~b_4~ & 0 & 0
\end{array}\)
\ee
\newpage
\be
C=\(\begin{array}{cccccc}
c_1 & ~c_2 & c_3 & 0~ & c_5 & ~c_6 \\[+0.5em]
0 & ~d_2 & ~\dfrac{c_3}{c_2}d_2~ & 0~ & -\!\(d_5\!+\!\dfrac{c_5}{c_6}d_6\) & ~-\!d_6 \\[+1.0em]
0 & ~b_2 & ~\dfrac{c_3}{c_2}b_2~ & b_4~ & b_5 & ~0
\end{array}\)
\ee
\be
C=\(\begin{array}{cccccc}
c_1 & ~0~ & c_3 & ~c_4 & ~c_5 & ~c_6 \\[+0.5em]
d_1 & ~0~ & -\!\(d_3\!+\!\dfrac{c_3}{c_4}d_4\) & ~-\!d_4 & ~0 & ~0 \\[+0.5em]
b_1 & ~b_2~ & b_3 & ~0 & ~0 & ~0
\end{array}\)
\ee
\be
C=\(\begin{array}{cccccc}
c_1 & ~0 & ~c_3~ & c_4 & c_5 & ~c_6 \\[+0.5em]
0 & ~0 & ~d_3~ & \dfrac{c_4}{c_3}d_3 & -\!\(d_5\!+\!\dfrac{c_5}{c_6}d_6\) & ~-\!d_6 \\[+0.5em]
0 & ~0 & ~b_3~ & \(b_4\!+\!\dfrac{c_4}{c_3}b_3\) & b_5 & ~0
\end{array}\)
\ee

\subsection{$n\!=\!6$ FWD}

\be
C=\(\begin{array}{cccccc}
d_1 & ~0~ & 0 & ~0~ & d_5 & ~d_6 \\[+0.5em]
c_1 & ~c_2~ & c_3 & ~0~ & 0 & ~0 \\[+0.5em]
0 & ~r_2~ & \(r_3\!+\!\dfrac{c_3}{c_2}r_2\) & ~r_4~ & \(r_5\!+\!\dfrac{d_5}{d_6}r_6\) & ~r_6
\end{array}\)
\ee
\be
C=\(\begin{array}{cccccc}
d_1 & ~0~ & 0 & ~0~ & d_5 & d_6 \\[+0.5em]
c_1 & ~c_2~ & c_3 & ~0~ & 0 & 0 \\[+0.5em]
r_1 & ~r_2~ & \(r_3\!+\!\dfrac{c_3}{c_2}r_2\) & ~r_4~
& \(\dfrac{d_5}{d_6}r_6\!+\!\dfrac{d_5}{d_1}r_1\) & \(r_6\!+\!\dfrac{d_6}{d_1}r_1\)
\end{array}\)
\ee
\be
C=\(\begin{array}{cccccc}
d_1 & ~0 & 0 & 0~ & d_5 & d_6 \\[+0.5em]
c_1 & ~c_2 & c_3 & 0~ & 0 & 0 \\[+0.5em]
r_1 & ~r_2 & ~\dfrac{c_3}{c_2}r_2~ & r_4~
& \(r_5\!+\!\dfrac{d_5}{d_6}r_6\!+\!\dfrac{d_5}{d_1}r_1\) & \(r_6\!+\!\dfrac{d_6}{d_1}r_1\)
\end{array}\)
\ee
\be
C=\(\begin{array}{cccccc}
d_1 & ~0~ & 0 & 0 & d_5 & d_6 \\[+0.5em]
c_1 & ~0~ & -c_3 & -c_4 & 0 & 0 \\[+0.5em]
\(l_1\!+\!\dfrac{d_1}{d_6}l_6\) & ~l_2~ & \(l_3\!+\!\dfrac{c_3}{c_4}l_4\) & l_4 & ~\dfrac{d_5}{d_6}l_6~ & l_6
\end{array}\)
\ee
\newpage
\be
C=\(\begin{array}{cccccc}
d_1 & ~0 & ~0~ & 0 & d_5 & d_6 \\[+0.5em]
c_1 & ~0 & ~c_3~ & c_4 & 0 & 0 \\[+0.5em]
r_1 & ~0 & ~r_3~ & \(r_4\!+\!\dfrac{c_4}{c_3}r_3\)
& \(r_5+\dfrac{d_5}{d_6}r_6\!+\!\dfrac{d_5}{d_1}r_1\) & \(r_6\!+\!\dfrac{d_6}{d_1}r_1\)
\end{array}\)
\ee
\be
C=\(\begin{array}{cccccc}
d_1 & ~0 & ~0 & ~0 & d_5 & d_6 \\[+0.5em]
c_1 & ~0 & ~0 & ~-\!c_4 & -c_5 & 0 \\[+0.5em]
\(l_1\!+\!\dfrac{d_1}{d_6}l_6\) & ~l_2 & ~l_3 & ~l_4 & ~\dfrac{d_5}{d_6}l_6~ & l_6
\end{array}\)
\ee
\be
C=\(\begin{array}{cccccc}
d_1 & ~0 & ~0~ & 0 & d_5 & ~d_6 \\[+0.5em]
c_1 & ~0 & ~0~ & -c_4 & -c_5 & ~0 \\[+0.5em]
\(l_1\!+\!\dfrac{d_1}{d_6}l_6\) & ~l_2 & ~0~ & \(l_4\!+\!\dfrac{c_4}{c_5}l_5\) & \(l_5\!+\!\dfrac{d_5}{d_6}l_6\) & ~l_6
\end{array}\)
\ee
\be
C=\(\begin{array}{cccccc}
d_1 & ~0 & ~0~ & 0 & d_5 & ~d_6 \\[+0.5em]
c_1 & ~0 & ~0~ & -c_4 & -c_5 & ~0 \\[+0.5em]
~\dfrac{d_1}{d_6}l_6~ & ~l_2 & ~l_3~ & \(l_4\!+\!\dfrac{c_4}{c_5}l_5\) & \(l_5\!+\!\dfrac{d_5}{d_6}l_6\) & ~l_6
\end{array}\)
\ee

\newpage
\section{All Distinct N$^2$MHV Homological Identities}
\label{app2}

Below we list all distinct N$^2$MHV homological identities and underline those non-BCFW-like cells,
which include composite-linear cells such as $(123)_2(456)_2(789)_2$
and quadratic cells such as $(12)(34)(56)(78)$.
All of them are generated by the \textsc{Mathematica} package ``\verb"positroids"'',
and translated to the form in terms of Grassmannian geometry representatives.
For clarity, we have separated those vanishing boundary cells from each identity
and denote them by `VBC'. Each part of the relative signs with respect to a fixed gauge
has been simplified to highlight the key vanishing constraints.

Although we have declared that, the signs of simple BCFW-like cells can be identified by incarnating
the NMHV $n\!=\!6$ identity, incidentally, there is still an exceptional case:
in the $n\!=\!9$ identity generated by $(123)_2(45)(67)(89)$,
(partly underlined) boundary cells $(123)_2(4567)_2(89)$ and $(123)_2(45)(6789)_2$
cannot be connected with the rest six BCFW-like cells.
It is an interesting future problem to explore the residue theorem for this tricky identity.

\subsection{$n\!=\!7$}

$(12)$
\be
\textrm{VBC:}~~(123)_2,~(712)_2,
\ee
\be
0=-\,[1]+[2]-(12)(34)+(12)(45)-(12)(56)+(12)(67).
\ee
{}
\be
C=\(\begin{array}{ccccccc}
\os{-}{1}\, & \os{+}{\De_{24}} & \os{-}{\De_{34}} & \,0\, & -\os{+}{\De_{45}} & -\os{-}{\De_{46}} & -\os{+}{\De_{47}} \\
0\, & 0 & \De_{13} & \,1\, & \De_{15} & \De_{16} & \De_{17}
\end{array}\)\Longrightarrow-\,[1]+[2]-(34)+(45),
\ee
\be
C=\(\begin{array}{ccccccc}
\os{-}{1}\, & \os{+}{\De_{26}} & \os{-}{\De_{36}} & \os{+}{\De_{46}} & \os{-}{\De_{56}} & \,0\, & -\os{+}{\De_{67}} \\
0\, & 0 & \De_{13} & \De_{14} & \De_{15} & \,1\, & \De_{17}
\end{array}\)\Longrightarrow-\,[1]+[2]-(56)+(67).
\ee

\subsection{$n\!=\!8$}

$(123)_2(45)$
\be
\textrm{VBC:}~~(123)_2[4],~(123)_2[5],~(12345)_2,~(8123)_2(45),
\ee
\be
0=-\,[1](23)(45)+(13)[2](45)-(12)[3](45)+(123)_2(456)_2-(123)_2(45)(67)+(123)_2(45)(78).
\ee
{}
\be
C=\(\begin{array}{cccccccc}
\os{-}{1}\, & \os{+}{\De_{26}} & \os{-}{\De_{36}} & ~\os{+}{\De_{56}}\dfrac{\De_{14}}{\De_{15}}~ & \os{+}{\De_{56}}
& \,0\, & -\os{-}{\De_{67}} & -\os{+}{\De_{68}} \\[+0.5em]
0\, & 0 & 0 & \De_{14} & \De_{15} & \,1\, & \De_{17} & \De_{18}
\end{array}\)\Longrightarrow-\,[1]+[2]-[3]+(56)-(67),
\ee
\be
C=\(\begin{array}{cccccccc}
\os{-}{1}\, & \os{+}{\De_{27}} & \os{-}{\De_{37}} & ~\os{+}{\De_{57}}\dfrac{\De_{14}}{\De_{15}}~ & \os{+}{\De_{57}}
& \os{-}{\De_{67}} & \,0\, & -\os{+}{\De_{78}} \\[+0.5em]
0\, & 0 & 0 & \De_{14} & \De_{15} & \De_{16} & \,1\, & \De_{18}
\end{array}\)\Longrightarrow-\,[1]+[2]-[3]-(67)+(78).
\ee
\\
$(123)_2(56)$
\be
\textrm{VBC:}~~(123)_2[5],~(123)_2[6],~(1234)_2(56),~(8123)_2(56),
\ee
\be
0=-\,[1](23)(56)+(13)[2](56)-(12)[3](56)+(123)_2(456)_2-(123)_2(567)_2+(123)_2(56)(78).
\ee
{}
\be
C=\(\begin{array}{cccccccc}
\os{-}{1}\, & \os{+}{\De_{26}} & \os{-}{\De_{36}} & \os{+}{\De_{46}} & 0 & \,0\, & -\os{-}{\De_{67}} & -\os{+}{\De_{68}} \\
0\, & 0 & 0 & \De_{14} & \De_{15} & \,1\, & \De_{17} & \De_{18}
\end{array}\)\Longrightarrow-\,[1]+[2]-[3]+(45)-(67),
\ee
\be
C=\(\begin{array}{cccccccc}
\os{-}{1}\, & \os{+}{\De_{27}} & \os{-}{\De_{37}} & \os{+}{\De_{47}} & ~\os{-}{\De_{67}}\dfrac{\De_{15}}{\De_{16}}~
& \os{-}{\De_{67}} & \,0\, & -\os{+}{\De_{78}} \\[+0.5em]
0\, & 0 & 0 & \De_{14} & \De_{15} & \De_{16} & \,1\, & \De_{18}
\end{array}\)\Longrightarrow-\,[1]+[2]-[3]-(67)+(78).
\ee
\\
$(123)_2(67)$
\be
\textrm{VBC:}~~(123)_2[6],~(123)_2[7],~(1234)_2(67),~(8123)_2(67),
\ee
\be
0=-\,[1](23)(67)+(13)[2](67)-(12)[3](67)+(123)_2(45)(67)-(123)_2(567)_2+(123)_2(678)_2.
\ee
{}
\be
C=\(\begin{array}{cccccccc}
\os{-}{1}\, & \os{+}{\De_{24}} & \os{-}{\De_{34}} & \,0\, & -\os{+}{\De_{45}}
& -\os{-}{\De_{46}} & -\os{-}{\De_{46}}\dfrac{\De_{17}}{\De_{16}} & -\os{+}{\De_{48}} \\[+0.5em]
0\, & 0 & 0 & \,1\, & \De_{15} & \De_{16} & \De_{17} & \De_{18}
\end{array}\)\Longrightarrow-\,[1]+[2]-[3]+(45),
\ee
\be
C=\(\begin{array}{cccccccc}
\os{-}{1}\, & \os{+}{\De_{27}} & \os{-}{\De_{37}} & \os{+}{\De_{47}} & \os{-}{\De_{57}} & 0 & \,0\, & -\os{+}{\De_{78}} \\
0\, & 0 & 0 & \De_{14} & \De_{15} & \De_{16} & \,1\, & \De_{18}
\end{array}\)\Longrightarrow-\,[1]+[2]-[3]-(56)+(78).
\ee
\\
$(123)_2(78)$
\be
\textrm{VBC:}~~(123)_2[7],~(123)_2[8],~(1234)_2(78),~(78123)_2,
\ee
\be
0=-\,[1](23)(78)+(13)[2](78)-(12)[3](78)+(123)_2(45)(78)-(123)_2(56)(78)+(123)_2(678)_2.
\ee
{}
\be
C=\(\begin{array}{cccccccc}
\os{-}{1}\, & \os{+}{\De_{24}} & \os{-}{\De_{34}} & \,0\, & -\os{+}{\De_{45}}
& -\os{-}{\De_{46}} & -\os{+}{\De_{47}} & -\os{+}{\De_{47}}\dfrac{\De_{18}}{\De_{17}} \\[+0.5em]
0\, & 0 & 0 & \,1\, & \De_{15} & \De_{16} & \De_{17} & \De_{18}
\end{array}\)\Longrightarrow-\,[1]+[2]-[3]+(45),
\ee
\be
C=\(\begin{array}{cccccccc}
\os{-}{1}\, & \os{+}{\De_{26}} & \os{-}{\De_{36}} & \os{+}{\De_{46}} & \os{-}{\De_{56}} & \,0\,
& -\os{+}{\De_{67}} & -\os{+}{\De_{67}}\dfrac{\De_{18}}{\De_{17}} \\[+0.5em]
0\, & 0 & 0 & \De_{14} & \De_{15} & \,1\, & \De_{17} & \De_{18}
\end{array}\)\Longrightarrow-\,[1]+[2]-[3]-(56)+(67).
\ee
\\
$(12)(34)(56)$
\be
\textrm{VBC:}~~(12)(3456)_2,~(1234)_2(56),
\ee
\be
\bal
0=\,&-[1](34)(56)+[2](34)(56)-(12)[3](56)+(12)[4](56)-(12)(34)[5]+(12)(34)[6]\\
&-(12)(34)(567)_2+(812)_2(34)(56)-\und{(12)(34)(56)(78)}.
\eal
\ee
{}
\be
C=\(\begin{array}{cccccccc}
\os{-}{1}\, & \os{+}{\De_{25}} & \os{-}{\De_{35}} & \os{+}{\De_{45}} & \,0\, & 0
& -\os{-}{\De_{57}} & -\os{+}{\De_{58}} \\[+0.5em]
0\, & 0 & \De_{13} & ~\De_{13}\dfrac{\De_{45}}{\De_{35}}~ & \,1\, & \De_{16} & \De_{17} & \De_{18}
\end{array}\)\Longrightarrow-\,[1]+[2]-[3]+[4]-(67),
\ee
\be
C=\(\begin{array}{cccccccc}
1\, & \De_{25} & \De_{35} & ~\De_{35}\dfrac{\De_{14}}{\De_{13}}~ & \,0\, & 0 & -\De_{57} & -\De_{58} \\[+0.5em]
0\, & 0 & \os{-}{\De_{13}} & \os{+}{\De_{14}} & \,\os{-}{1}\, & \os{+}{\De_{16}} & \os{-}{\De_{17}} & \os{+}{\De_{18}}
\end{array}\)\Longrightarrow-\,[3]+[4]-[5]+[6]+(81).
\ee
\\
$(12)(34)(67)$
\be
\textrm{VBC:}~~(1234)_2(67),
\ee
\be
\bal
0=\,&-[1](34)(67)+[2](34)(67)-(12)[3](67)+(12)[4](67)+(12)(34)[6]-(12)(34)[7]\\
&-(12)(345)_2(67)-(12)(34)(567)_2+(12)(34)(678)_2+(812)_2(34)(67).
\eal
\ee
{}
\be
C=\(\begin{array}{cccccccc}
\os{-}{1}\, & \os{+}{\De_{23}} & \,0\, & 0 & -\os{-}{\De_{35}} & -\os{+}{\De_{36}}
& -\os{-}{\De_{37}} & -\os{+}{\De_{38}} \\[+0.5em]
0\, & 0 & \,1\, & \De_{14} & \De_{15} & \De_{16} & \De_{16}\dfrac{\De_{37}}{\De_{36}} & \De_{18}
\end{array}\)\Longrightarrow-\,[1]+[2]-(45)+[6]-[7],
\ee
\be
C=\(\begin{array}{cccccccc}
\os{-}{1}\, & \os{+}{\De_{26}} & \os{-}{\De_{36}} & \os{+}{\De_{46}} & \os{-}{\De_{56}}
& \,0\, & 0 & -\os{+}{\De_{68}} \\[+0.5em]
0\, & 0 & \De_{13} & ~\De_{13}\dfrac{\De_{46}}{\De_{36}}~ & \De_{15} & \,1\, & \De_{17} & \De_{18}
\end{array}\)\Longrightarrow-\,[1]+[2]-[3]+[4]-(56)+(78),
\ee
\be
C=\(\begin{array}{cccccccc}
1\, & \De_{26} & \De_{36} & ~\De_{36}\dfrac{\De_{14}}{\De_{13}}~ & \De_{56} & \,0\, & 0 & -\De_{68} \\[+0.5em]
0\, & 0 & \os{-}{\De_{13}} & \os{+}{\De_{14}} & \os{-}{\De_{15}} & \,\os{+}{1}\, & \os{-}{\De_{17}} & \os{+}{\De_{18}}
\end{array}\)\Longrightarrow-\,[3]+[4]+[6]-[7]+(81).
\ee

\subsection{$n\!=\!9$}

$(1234)_2(567)_2$
\be
\textrm{VBC:}~~(1234)_2[5](67),~(1234)_2(57)[6],~(1234)_2(56)[7],~(1234567)_2,~(91234)_2(567)_2,
\ee
\be
\bal
0=&-[1](234)_2(567)_2+(134)_2[2](567)_2-(124)_2[3](567)_2+(123)_2[4](567)_2\\
&-(1234)_2(5678)_2+(1234)_2(567)_2(89).
\eal
\ee
{}
\be
C=\(\begin{array}{ccccccccc}
\os{-}{1}\, & \os{+}{\De_{28}} & \os{-}{\De_{38}} & \os{+}{\De_{48}}
& ~\os{-}{\De_{78}}\dfrac{\De_{15}}{\De_{17}}~ & \os{-}{\De_{78}}\dfrac{\De_{16}}{\De_{17}}~
& \os{-}{\De_{78}} & \,0\, & -\os{+}{\De_{89}} \\[+0.5em]
0\, & 0 & 0 & 0 & \De_{15} & \De_{16} & \De_{17} & \,1\, & \De_{19}
\end{array}\)\Longrightarrow-\,[1]+[2]-[3]+[4]-(78)+(89).
\ee
\\
$(1234)_2(678)_2$
\be
\textrm{VBC:}~~(1234)_2[6](78),~(1234)_2(68)[7],~(1234)_2(67)[8],~(12345)_2(678)_2,~(91234)_2(678)_2,
\ee
\be
\bal
0=&-[1](234)_2(678)_2+(134)_2[2](678)_2-(124)_2[3](678)_2+(123)_2[4](678)_2\\
&-(1234)_2(5678)_2+(1234)_2(6789)_2.
\eal
\ee
{}
\be
C=\(\begin{array}{ccccccccc}
\os{-}{1}\, & \os{+}{\De_{28}} & \os{-}{\De_{38}} & \os{+}{\De_{48}} & \os{-}{\De_{58}}
& 0 & 0 & \,0\, & -\os{+}{\De_{89}} \\
0\, & 0 & 0 & 0 & \De_{15} & \De_{16} & \De_{17} & \,1\, & \De_{19}
\end{array}\)\Longrightarrow-\,[1]+[2]-[3]+[4]-(56)+(89).
\ee
\\
$(1234)_2(789)_2$
\be
\textrm{VBC:}~~(1234)_2[7](89),~(1234)_2(79)[8],~(1234)_2(78)[9],~(12345)_2(789)_2,~(7891234)_2,
\ee
\be
\bal
0=&-[1](234)_2(789)_2+(134)_2[2](789)_2-(124)_2[3](789)_2+(123)_2[4](789)_2\\
&-(1234)_2(56)(789)_2+(1234)_2(6789)_2.
\eal
\ee
{}
\be
C=\(\begin{array}{ccccccccc}
\os{-}{1}\, & \os{+}{\De_{26}} & \os{-}{\De_{36}} & \os{+}{\De_{46}} & \os{-}{\De_{56}} & \,0\,
& -\os{+}{\De_{67}} & -\os{+}{\De_{67}}\dfrac{\De_{18}}{\De_{17}} & -\os{+}{\De_{67}}\dfrac{\De_{19}}{\De_{17}} \\[+0.5em]
0\, & 0 & 0 & 0 & \De_{15} & \,1\, & \De_{17} & \De_{18} & \De_{19}
\end{array}\)\Longrightarrow-\,[1]+[2]-[3]+[4]-(56)+(67).
\ee
\\
$(1234)_2(56)(78)$
\be
\bal
\textrm{VBC:}~~&(1234)_2[5](78),~(1234)_2[6](78),~(1234)_2(56)[7],~(1234)_2(56)[8],\\
&(123456)_2(78),~(91234)_2(56)(78),
\eal
\ee
\be
\bal
0=\,&-[1](234)_2(56)(78)+(134)_2[2](56)(78)-(124)_2[3](56)(78)+(123)_2[4](56)(78)\\
&-(1234)_2(5678)_2+(1234)_2(56)(789)_2.
\eal
\ee
{}
\be
C=\(\begin{array}{ccccccccc}
\os{-}{1}\, & \os{+}{\De_{28}} & \os{-}{\De_{38}} & \os{+}{\De_{48}}
& ~\os{-}{\De_{68}}\dfrac{\De_{15}}{\De_{16}}~ & \os{-}{\De_{68}} & 0 & \,0\, & -\os{+}{\De_{89}} \\[+0.5em]
0\, & 0 & 0 & 0 & \De_{15} & \De_{16} & \De_{17} & \,1\, & \De_{19}
\end{array}\)\Longrightarrow-\,[1]+[2]-[3]+[4]-(67)+(89).
\ee
\\
$(1234)_2(56)(89)$
\be
\bal
\textrm{VBC:}~~&(1234)_2[5](89),~(1234)_2[6](89),~(1234)_2(56)[8],~(1234)_2(56)[9],\\
&(123456)_2(89),~(891234)_2(56),
\eal
\ee
\be
\bal
0=\,&-[1](234)_2(56)(89)+(134)_2[2](56)(89)-(124)_2[3](56)(89)+(123)_2[4](56)(89)\\
&-(1234)_2(567)_2(89)+(1234)_2(56)(789)_2.
\eal
\ee
{}
\be
C=\(\begin{array}{ccccccccc}
\os{-}{1}\, & \os{+}{\De_{27}} & \os{-}{\De_{37}} & \os{+}{\De_{47}}
& ~\os{-}{\De_{67}}\dfrac{\De_{15}}{\De_{16}}~ & \os{-}{\De_{67}}
& \,0\, & -\os{+}{\De_{78}} & -\os{+}{\De_{78}}\dfrac{\De_{19}}{\De_{18}} \\[+0.5em]
0\, & 0 & 0 & 0 & \De_{15} & \De_{16} & \,1\, & \De_{18} & \De_{19}
\end{array}\)\Longrightarrow-\,[1]+[2]-[3]+[4]-(67)+(78).
\ee
\\
$(1234)_2(67)(89)$
\be
\bal
\textrm{VBC:}~~&(1234)_2[6](89),~(1234)_2[7](89),~(1234)_2(67)[8],~(1234)_2(67)[9],\\
&(12345)_2(67)(89),~(891234)_2(67),
\eal
\ee
\be
\bal
0=\,&-[1](234)_2(67)(89)+(134)_2[2](67)(89)-(124)_2[3](67)(89)+(123)_2[4](67)(89)\\
&-(1234)_2(567)_2(89)+(1234)_2(6789)_2.
\eal
\ee
{}
\be
C=\(\begin{array}{ccccccccc}
\os{-}{1}\, & \os{+}{\De_{27}} & \os{-}{\De_{37}} & \os{+}{\De_{47}}
& \os{-}{\De_{57}} & 0 & \,0\, & -\os{+}{\De_{78}} & -\os{+}{\De_{78}}\dfrac{\De_{19}}{\De_{18}} \\
0\, & 0 & 0 & 0 & \De_{15} & \De_{16} & \,1\, & \De_{18} & \De_{19}
\end{array}\)\Longrightarrow-\,[1]+[2]-[3]+[4]-(56)+(78).
\ee
\\
$(123)_2(456)_2(78)$
\be
\textrm{VBC:}~~(123456)_2(78),~(123)_2(45678)_2,
\ee
\be
\bal
0=\,&-[1](23)(456)_2(78)+(13)[2](456)_2(78)-(12)[3](456)_2(78)\\
&-(123)_2[4](56)(78)+(123)_2(46)[5](78)-(123)_2(45)[6](78)\\
&+(123)_2(456)_2[7]-(123)_2(456)_2[8]\\
&+(9123)_2(456)_2(78)-\und{(123)_2(456)_2(789)_2}.
\eal
\ee
{}
\be
C=\(\begin{array}{ccccccccc}
\os{-}{1}\, & \os{+}{\De_{24}} & \os{-}{\De_{34}} & \,0\, & 0 & 0
& -\os{+}{\De_{47}} & -\os{-}{\De_{48}} & -\os{+}{\De_{49}} \\[+0.5em]
0\, & 0 & 0 & \,1\, & \De_{15} & \De_{16} & \De_{17} & \De_{17}\dfrac{\De_{48}}{\De_{47}} & \De_{19}
\end{array}\)\Longrightarrow-\,[1]+[2]-[3]+[7]-[8],
\ee
\be
C=\(\begin{array}{ccccccccc}
1\, & \De_{24} & \De_{34} & \,0\, & 0 & 0 & -\De_{47} & -\De_{47}\dfrac{\De_{18}}{\De_{17}} & -\De_{49} \\[+0.5em]
0\, & 0 & 0 & \,\os{-}{1}\, & \os{+}{\De_{15}} & \os{-}{\De_{16}} & \os{+}{\De_{17}} & \os{-}{\De_{18}} & \os{+}{\De_{19}}
\end{array}\)\Longrightarrow-\,[4]+[5]-[6]+[7]-[8]+(91).
\ee
\\
$(123)_2(456)_2(89)$
\be
\textrm{VBC:}~~(123456)_2(89),~(89123)_2(456)_2,
\ee
\be
\bal
0=\,&-[1](23)(456)_2(89)+(13)[2](456)_2(89)-(12)[3](456)_2(89)\\
&-(123)_2[4](56)(89)+(123)_2(46)[5](89)-(123)_2(45)[6](89)\\
&-(123)_2(456)_2[8]+(123)_2(456)_2[9]\\
&+(123)_2(4567)_2(89)-\und{(123)_2(456)_2(789)_2}.
\eal
\ee
{}
\be
C=\(\begin{array}{ccccccccc}
\os{-}{1}\, & \os{+}{\De_{24}} & \os{-}{\De_{34}} & \,0\, & 0 & 0
& -\os{+}{\De_{47}} & -\os{-}{\De_{48}} & -\os{+}{\De_{49}} \\[+0.5em]
0\, & 0 & 0 & \,1\, & \De_{15} & \De_{16} & \De_{17} & \De_{18} & \De_{18}\dfrac{\De_{49}}{\De_{48}}
\end{array}\)\Longrightarrow-\,[1]+[2]-[3]+(67)-[8]+[9],
\ee
\be
C=\(\begin{array}{ccccccccc}
1\, & \De_{24} & \De_{34} & \,0\, & 0 & 0 & -\De_{47} & -\De_{48} & -\De_{48}\dfrac{\De_{19}}{\De_{18}} \\[+0.5em]
0\, & 0 & 0 & \,\os{-}{1}\, & \os{+}{\De_{15}} & \os{-}{\De_{16}} & \os{+}{\De_{17}} & \os{-}{\De_{18}} & \os{+}{\De_{19}}
\end{array}\)\Longrightarrow-\,[4]+[5]-[6]-[8]+[9].
\ee
\\
$(123)_2(45)(678)_2$
\be
\textrm{VBC:}~~(12345)_2(678)_2,~(123)_2(45678)_2,
\ee
\be
\bal
0=\,&-[1](23)(45)(678)_2+(13)[2](45)(678)_2-(12)[3](45)(678)_2\\
&+(123)_2[4](678)_2-(123)_2[5](678)_2+(123)_2(45)[6](78)\\
&-(123)_2(45)(68)[7]+(123)_2(45)(67)[8]\\
&+(123)_2(45)(6789)_2-(9123)_2(45)(678)_2.
\eal
\ee
{}
\be
C=\(\begin{array}{ccccccccc}
\os{-}{1}\, & \os{+}{\De_{26}} & \os{-}{\De_{36}} & \os{+}{\De_{46}} & \os{-}{\De_{56}}
& \,0\, & 0 & 0 & -\os{+}{\De_{69}} \\[+0.5em]
0\, & 0 & 0 & \De_{14} & ~\De_{14}\dfrac{\De_{56}}{\De_{46}}~ & \,1\, & \De_{17} & \De_{18} & \De_{19}
\end{array}\)\Longrightarrow-\,[1]+[2]-[3]+[4]-[5]+(89),
\ee
\be
C=\(\begin{array}{ccccccccc}
1\, & \De_{26} & \De_{36} & \De_{46} & ~\De_{46}\dfrac{\De_{15}}{\De_{14}}~ & \,0\, & 0 & 0 & -\De_{69} \\[+0.5em]
0\, & 0 & 0 & \os{+}{\De_{14}} & \os{-}{\De_{15}} & \,\os{+}{1}\, & \os{-}{\De_{17}} & \os{+}{\De_{18}} & \os{-}{\De_{19}}
\end{array}\)\Longrightarrow+\,[4]-[5]+[6]-[7]+[8]-(91).
\ee
\\
$(123)_2(45)(67)(89)$
\be
\textrm{VBC:}~~(12345)_2(67)(89),~(89123)_2(45)(67),
\ee
\be
\bal
0=\,&-\und{[1](23)(45)(67)(89)}
+\und{(13)[2](45)(67)(89)}-\und{(12)[3](45)(67)(89)}\\
&-(123)_2[4](67)(89)+(123)_2[5](67)(89)-(123)_2(45)[6](89)\\
&+(123)_2(45)[7](89)-(123)_2(45)(67)[8]+(123)_2(45)(67)[9]\\
&+(123)_2\und{(4567)_2}(89)-(123)_2(45)\und{(6789)_2}.
\eal
\ee
{}
\be
C=\(\begin{array}{ccccccccc}
1\, & \De_{24} & \De_{34} & \,0\, & 0 & -\De_{46} & -\De_{46}\dfrac{\De_{17}}{\De_{16}}
& -\De_{48} & -\De_{48}\dfrac{\De_{19}}{\De_{18}} \\[+0.5em]
0\, & 0 & 0 & \,\os{-}{1}\, & \os{+}{\De_{15}} & \os{-}{\De_{16}} & \os{+}{\De_{17}} & \os{-}{\De_{18}} & \os{+}{\De_{19}}
\end{array}\)\Longrightarrow-\,[4]+[5]-[6]+[7]-[8]+[9].
\ee

\subsection{$n\!=\!10$}

$(12345)_2(6789)_2$
\be
\bal
\textrm{VBC:}~~&(12345)_2[6](789)_2,~(12345)_2(689)_2[7],~(12345)_2(679)_2[8],~(12345)_2(678)_2[9],\\
&(123456789)_2,~(10\,12345)_2(6789)_2,
\eal
\ee
\be
\bal
0=\,&-[1](2345)_2(6789)_2+(1345)_2[2](6789)_2-(1245)_2[3](6789)_2\\
&+(1235)_2[4](6789)_2-(1234)_2[5](6789)_2+(12345)_2(6789\,10)_2.
\eal
\ee
{}
\be
C=\(\begin{array}{cccccccccc}
\os{-}{1}\, & \os{+}{\De_{29}} & \os{-}{\De_{39}} & \os{+}{\De_{49}} & \os{-}{\De_{59}}
& 0 & 0 & 0 & \,0\, & -\os{+}{\De_{9,10}} \\
0\, & 0 & 0 & 0 & 0 & \De_{16} & \De_{17} & \De_{18} & \,1\, & \De_{1,10}
\end{array}\)\Longrightarrow-\,[1]+[2]-[3]+[4]-[5]+(9\,10).
\ee
\\
$(12345)_2(789\,10)_2$
\be
\bal
\textrm{VBC:}~~&(12345)_2[7](89\,10)_2,~(12345)_2(79\,10)_2[8],~(12345)_2(78\,10)_2[9],~(12345)_2(789)_2[10],\\
&(123456)_2(789\,10)_2,~(789\,10\,12345)_2,
\eal
\ee
\be
\bal
0=\,&-[1](2345)_2(789\,10)_2+(1345)_2[2](789\,10)_2-(1245)_2[3](678\,10)_2\\
&+(1235)_2[4](789\,10)_2-(1234)_2[5](789\,10)_2+(12345)_2(6789\,10)_2.
\eal
\ee
{}
\be
C=\(\begin{array}{cccccccccc}
\os{-}{1}\, & \os{+}{\De_{27}} & \os{-}{\De_{37}} & \os{+}{\De_{47}} & \os{-}{\De_{57}} & \os{+}{\De_{67}}
& \,0\, & 0 & 0 & 0 \\
0\, & 0 & 0 & 0 & 0 & \De_{16} & \,1\, & \De_{18} & \De_{19} & \De_{1,10}
\end{array}\)\Longrightarrow-\,[1]+[2]-[3]+[4]-[5]+(67).
\ee
\\
$(12345)_2(678)_2(9\,10)$
\be
\bal
\textrm{VBC:}~~&(12345)_2[6](78)(9\,10),~(12345)_2(68)[7](9\,10),~(12345)_2(67)[8](9\,10),~(12345)_2(678)_2[9],\\
&(12345)_2(678)_2[10],~(12345678)_2(9\,10),~(9\,10\,12345)_2(678)_2,
\eal
\ee
\be
\bal
0=\,&-[1](2345)_2(678)_2(9\,10)+(1345)_2[2](678)_2(9\,10)-(1245)_2[3](678)_2(9\,10)\\
&+(1235)_2[4](678)_2(9\,10)-(1234)_2[5](678)_2(9\,10)+(12345)_2(6789\,10)_2.
\eal
\ee
{}
\be
C=\(\begin{array}{cccccccccc}
\os{-}{1}\, & \os{+}{\De_{28}} & \os{-}{\De_{38}} & \os{+}{\De_{48}} & \os{-}{\De_{58}} & 0 & 0 & \,0\,
& -\os{+}{\De_{89}} & -\os{+}{\De_{89}}\dfrac{\De_{1,10}}{\De_{19}} \\[+0.5em]
0\, & 0 & 0 & 0 & 0 & \De_{16} & \De_{17} & \,1\, & \De_{19} & \De_{1,10}
\end{array}\)\Longrightarrow-\,[1]+[2]-[3]+[4]-[5]+(89).
\ee
\\
$(12345)_2(67)(89\,10)_2$
\be
\bal
\textrm{VBC:}~~&(12345)_2[6](89\,10)_2,~(12345)_2[7](89\,10)_2,~(12345)_2(67)[8](9\,10)_2,~(12345)_2(67)(8\,10)[9],\\
&(12345)_2(67)(89)[10],~(1234567)_2(89\,10)_2,~(89\,10\,12345)_2(67),
\eal
\ee
\be
\bal
0=\,&-[1](2345)_2(67)(89\,10)_2+(1345)_2[2](67)(89\,10)_2-(1245)_2[3](67)(89\,10)_2\\
&+(1235)_2[4](67)(89\,10)_2-(1234)_2[5](67)(89\,10)_2+(12345)_2(6789\,10)_2.
\eal
\ee
{}
\be
C=\(\begin{array}{cccccccccc}
\os{-}{1}\, & \os{+}{\De_{28}} & \os{-}{\De_{38}} & \os{+}{\De_{48}} & \os{-}{\De_{58}}
& ~\os{+}{\De_{78}}\dfrac{\De_{16}}{\De_{17}}~ & \os{+}{\De_{78}} & \,0\, & 0 & 0 \\[+0.5em]
0\, & 0 & 0 & 0 & 0 & \De_{16} & \De_{17} & \,1\, & \De_{19} & \De_{1,10}
\end{array}\)\Longrightarrow-\,[1]+[2]-[3]+[4]-[5]+(78).
\ee
\\
$(1234)_2(5678)_2(9\,10)$
\be
\textrm{VBC:}~~(12345678)_2(9\,10),~(1234)_2(56789\,10)_2,~(9\,10\,1234)_2(5678)_2,
\ee
\be
\bal
0=\,&-[1](234)_2(5678)_2(9\,10)+(134)_2[2](5678)_2(9\,10)-(124)_2[3](5678)_2(9\,10)\\
&+(123)_2[4](5678)_2(9\,10)-(1234)_2[5](678)_2(9\,10)+(1234)_2(578)_2[6](9\,10)\\
&-(1234)_2(568)_2[7](9\,10)+(1234)_2(567)_2[8](9\,10)\\
&-(1234)_2(5678)_2[9]+(1234)_2(5678)_2[10].
\eal
\ee
{}
\be
C=\(\begin{array}{cccccccccc}
\os{-}{1}\, & \os{+}{\De_{25}} & \os{-}{\De_{35}} & \os{+}{\De_{45}} & \,0\, & 0 & 0 & 0
& -\os{-}{\De_{59}} & -\os{+}{\De_{5,10}} \\[+0.5em]
0\, & 0 & 0 & 0 & \,1\, & \De_{16} & \De_{17} & \De_{18} & \De_{19} & \De_{19}\dfrac{\De_{5,10}}{\De_{59}}
\end{array}\)\Longrightarrow-\,[1]+[2]-[3]+[4]-[9]+[10],
\ee
\be
C=\(\begin{array}{cccccccccc}
1\, & \De_{25} & \De_{35} & \De_{45} & \,0\, & 0 & 0 & 0 & -\De_{59} & -\De_{59}\dfrac{\De_{1,10}}{\De_{19}} \\[+0.5em]
0\, & 0 & 0 & 0 & \,\os{-}{1}\,
& \os{+}{\De_{16}} & \os{-}{\De_{17}} & \os{+}{\De_{18}} & \os{-}{\De_{19}} & \os{+}{\De_{1,10}}
\end{array}\)\Longrightarrow-\,[5]+[6]-[7]+[8]-[9]+[10].
\ee
\\
$(1234)_2(567)_2(89\,10)_2$
\be
\textrm{VBC:}~~(1234567)_2(89\,10)_2,~(1234)_2(56789\,10)_2,~(89\,10\,1234)_2(567)_2,
\ee
\be
\bal
0=\,&-\und{[1](234)_2(567)_2(89\,10)_2}+\und{(134)_2[2](567)_2(89\,10)_2}\\
&-\und{(124)_2[3](567)_2(89\,10)_2}+\und{(123)_2[4](567)_2(89\,10)_2}\\
&-(1234)_2[5](67)(89\,10)_2+(1234)_2(57)[6](89\,10)_2-(1234)_2(56)[7](89\,10)_2\\
&+(1234)_2(567)_2[8](9\,10)-(1234)_2(567)_2(8\,10)[9]+(1234)_2(567)_2(89)[10].
\eal
\ee
{}
\be
C=\(\begin{array}{cccccccccc}
1\, & \De_{25} & \De_{35} & \De_{45} & \,0\, & 0 & 0
& -\De_{58} & -\De_{58}\dfrac{\De_{19}}{\De_{18}} & -\De_{58}\dfrac{\De_{1,10}}{\De_{18}} \\[+0.5em]
0\, & 0 & 0 & 0 & \,\os{-}{1}\,
& \os{+}{\De_{16}} & \os{-}{\De_{17}} & \os{+}{\De_{18}} & \os{-}{\De_{19}} & \os{+}{\De_{1,10}}
\end{array}\)\Longrightarrow-\,[5]+[6]-[7]+[8]-[9]+[10].
\ee

\subsection{$n\!=\!11$}

$(123456)_2(789\,10\,\,11)_2$
\be
\bal
\textrm{VBC:}~~&(123456)_2[7](89\,10\,\,11)_2,~(123456)_2(79\,10\,\,11)_2[8],~(123456)_2(78\,10\,\,11)_2[9],\\
&(123456)_2(789\,11)_2[10],~(123456)_2(789\,10)_2[11],
\eal
\ee
\be
\bal
0=\,&-[1](23456)_2(789\,10\,\,11)_2+(13456)_2[2](789\,10\,\,11)_2-(12456)_2[3](789\,10\,\,11)_2\\
&+(12356)_2[4](789\,10\,\,11)_2-(12346)_2[5](789\,10\,\,11)_2+(12345)_2[6](789\,10\,\,11)_2.
\eal
\ee
{}
\be
C=\(\begin{array}{ccccccccccc}
\os{-}{1}\, & \os{+}{\De_{27}} & \os{-}{\De_{37}} & \os{+}{\De_{47}} & \os{-}{\De_{57}} & \os{+}{\De_{67}}
& \,0\, & 0 & 0 & 0 & 0 \\
0\, & 0 & 0 & 0 & 0 & 0 & \,1\, & \De_{18} & \De_{19} & \De_{1,10} & \De_{1,11}
\end{array}\)\Longrightarrow-\,[1]+[2]-[3]+[4]-[5]+[6].
\ee

\newpage
\section{All N$^3$MHV Fully-spanning Cells}
\label{app3}

Below we list all N$^3$MHV full cells (the only exception is $[6]$ at $n\!=\!8$) for $n\!=\!8,9,10,11,12,13$, while
at $n\!=\!14$ there is no more new full cell. Only the $n\!=\!8$ part is arranged following the anti-NMHV triangle form,
the rest are arranged individually following their default order produced by ``\verb"positroids"'', for the convenience
of cross check. The growing parameters are given along with the full cells,
as their growing modes vary from 0 to 4. For these six parts, the numbers of cells are $8,18,27,26,15,5$ respectively.

\subsection{$n\!=\!8$}

\be
~~~\,\left\{\begin{array}{c}
\!(456)(81)~~ \\
\,(56)(812) \\
\!\!{[6]}~~~~
\end{array} \right.~~~~~~~~~~(6)
\ee
\be
~\,(23)\left\{\begin{array}{c}
\!(678) \\
\!(456)~~~~~~
\end{array} \right.~~~~~~~~~~~(7,4)
\ee
\be
(234)\left\{\begin{array}{c}
~~~(78) \\
\!(456)(781)~~ \\
\!\!(56)~~~~~~~
\end{array} \right.~~~~~~~~(7,5)
\ee

\subsection{$n\!=\!9$}

\be
(4\,\os{6}{\os{|}{5}}\,7)\,(8\,\os{1}{\os{|}{9}}\,2)~~~~~~~~~~~~~~~~~~(6)~~~~~~
\ee
\be
(2345)_3(789)_2~~~~~~~~~~~~~~~~\,(8,5)~~~
\ee
\be
(\os{2}{\os{|}{3}}\,4\,5)\,(6\,7\,\os{9}{\os{|}{8}})~~~~~~~~~~~~~~~~~~(8,6,4)
\ee
\be
(2\,3\,\os{5}{\os{|}{4}})\,(7\,\os{9}{\os{|}{8}}\,1)~~~~~~~~~~~~~~~~~~(8,5)~~~
\ee
\be
(23)(45)(89)~~~~~~~~~~~~~~~~~~(8,6,4)
\ee
\be
(234)_2(6789)_3~~~~~~~~~~~~~~~~\,(8,4)~~~
\ee
\be
(\os{6}{\os{|}{7}}\,8\,9)\,(9\,1\,\os{3}{\os{|}{2}})~~~~~~~~~~~~~~~~~~(7,4)~~~
\ee
\be
(\os{2}{\os{|}{3}}\,4\,5)\,(\os{6}{\os{|}{7}}\,8\,9)~~~~~~~~~~~~~~~~~~(8,6,4)
\ee
\be
(4567)_3(891)_2~~~~~~~~~~~~~~~~\,(8,6)~~~
\ee
\be
(234)(4567)_3\,(7\,8\,\os{1}{\os{|}{9}})~~~~~~~~~\,(7,5)~~~
\ee
\be
(23)(4567)_3(91)~~~~~~~~~~~~~~(7,4)~~~
\ee
\be
(234)\,(4\,\os{6}{\os{|}{5}}\,7)\,(91)~~~~~~~~~~~~~\,(7,5)~~~
\ee
\be
(2\,3\,\os{5}{\os{|}{4}})\,(467)(7891)_3~~~~~~~~~~(8,5)~~~
\ee
\be
(23)\,(\os{4}{\os{|}{5}}\,6\,7)\,(891)~~~~~~~~~~~~~~(8,6,4)
\ee
\be
(4\,5\,\os{7}{\os{|}{6}})\,(9\,1\,\os{3}{\os{|}{2}})~~~~~~~~~~~~~~~~~~~(7,4)~~\,
\ee
\be
\,(23)(45)(67)~~~~~~~~~~~~~~~~~~~(8,6,4)
\ee
\be
(567)_2(8912)_3~~~~~~~~~~~~~~~~~\,(8,6)~~\,
\ee
\be
(234)(567)_2(912)~~~~~~~~~~~~~\,(7,5)~~\,
\ee

\subsection{$n\!=\!10$}

\be
(23456)_3(789\,10)_2~~~~~~~~~~(9,7,5)\,
\ee
\be
(\os{2}{\os{|}{3}}\,4\,5\,6)_3\,(6\,7\!\!\os{\,9\,\,\,10}{\os{\backslash\,/}{8}}\!\!\!)~~~~~~~~~~~~\,(9,6,4)
\ee
\be
(2\,3\,\os{5}{\os{|}{4}}\,6)_3\,(7\!\!\os{\,9\,\,\,10}{\os{\backslash\,/}{8}}\!\!1)~~~~~~~~~~~~\,(9,5)~~~
\ee
\be
(\os{3}{\os{|}{2}}\,\os{5}{\os{|}{4}}\,6)\,(89\,10)_2~~~~~~~~~~~~~~(9,6,4)
\ee
\be
(\!\!\os{2\,\,\,\,3}{\os{\backslash\,/}{4}}\!5\,6)\,(6\,7\,8\,\os{10}{\os{|}{9}})_3~~~~~~~~~~~~(9,7,4)
\ee
\be
(2\,3\!\os{5\,\,\,\,6}{\os{\backslash\,/}{4}}\!\!)\,(7\,8\,\os{10}{\os{|}{9}}\,1)_3~~~~~~~~~~~~(9,7,5)
\ee
\be
(23)(456)_2\,(8\,\os{10}{\os{|}{9}}\,1)~~~~~~~~~~(9,6,4)
\ee
\be
(2345)_2(6789\,10)_3~~~~~~~~~~(9,6,4)
\ee
\be
(6\,\os{8}{\os{|}{7}}\,9\,10)_3\,(10\,1\!\os{3\,\,\,\,4}{\os{\backslash\,/}{2}}\!\!)~~~~~~~~~~(8,4)~~~
\ee
\be
(\!\!\os{2\,\,\,\,3}{\os{\backslash\,/}{4}}\!5\,6)\,(6\,\os{8}{\os{|}{7}}\,9\,10)_3~~~~~~~~~~~\,(9,7,4)
\ee
\be
(678)_2\,(\os{9}{\os{|}{10}}\,1\,\os{3}{\os{|}{2}})~~~~~~~~~~~~~~~(9,7,4)
\ee
\be
(\os{2}{\os{|}{3}}\,4\,5)\,(\!\!\os{6\,\,\,\,7}{\os{\backslash\,/}{8}}\!9\,10)\,(10\,1\,3)~~~~~~~~~~~\,(8,6,4)
\ee
\be
(234)\,(4\,5\,6\,\os{8}{\os{|}{7}})_3\,(9\,10\,1)_2~~~~~~~~~~(9,7,5)
\ee
\be
(23)(45678)_3(9\,10\,1)_2~~~~~~~~~~~~\,(9,7,4)
\ee
\be
(4\,\os{6}{\os{|}{5}}\,7\,8)_3\,(8\!\!\os{10\,\,\,\,1}{\os{\backslash\,/}{9}}\!\!2)~~~~~~~~~~~~~~~~~~~~(9,6)~~~
\ee
\be
(234)\,(4\,\os{6}{\os{|}{5}}\,7\,8)_3\,(9\,10\,1)_2~~~~~~~~~~\,(9,7,5)
\ee
\be
(2\,3\,\os{5}{\os{|}{4}})\,(4\,6\,\os{8}{\os{|}{7}})\,(7\,9\,\os{1}{\os{|}{10}})~~~~~~~~~~~~~~(8,5)~~\,
\ee
\be
(23)\,(\os{4}{\os{|}{5}}\,6\,7\,8)_3\,(8\,9\,\os{1}{\os{|}{10}})~~~~~~~~~~~~~\,(8,6,4)
\ee
\be
(4\,5\,\os{7}{\os{|}{6}}\,8)_3\,(9\,\os{1}{\os{|}{10}}\,\,\os{3}{\os{|}{2}})~~~~~~~~~~~~~~~~~~~(7,4)~~~
\ee
\be
(23)\,(\os{5}{\os{|}{4}}\,\os{7}{\os{|}{6}}\,8)\,(10\,1)~~~~~~~~~~~~~~~~~~\,(8,6,4)
\ee
\be
(4\!\os{6\,\,\,\,7}{\os{\backslash\,/}{5}}\!8)\,(8\,9\,\os{1}{\os{|}{10}}\,\,2)_3~~~~~~~~~~~~~~~~~~\,(8,6)~~~
\ee
\be
(234)\,(4\!\os{6\,\,\,\,7}{\os{\backslash\,/}{5}}\!8)\,(9\,\os{1}{\os{|}{10}}\,\,2)~~~~~~~~~~~~~~\,(7,5)~~~
\ee
\be
(2\,3\!\os{5\,\,\,\,6}{\os{\backslash\,/}{4}}\!\!)\,(\os{7}{\os{|}{8}}\,9\,10\,1)_3~~~~~~~~~~~~~~~~~~~(9,7,5)
\ee
\be
(23)\,(\!\!\os{4\,\,\,\,5}{\os{\backslash\,/}{6}}\!7\,8)\,(89\,10\,1)_3~~~~~~~~~~~~~~(9,6,4)
\ee
\be
(4\,5\!\os{7\,\,\,\,8}{\os{\backslash\,/}{6}}\!\!)\,(9\,10\,1\,\os{3}{\os{|}{2}})_3~~~~~~~~~~~~~~~~~~~\,(9,7,4)
\ee
\be
(45)(678)_2\,(10\,1\,\os{3}{\os{|}{2}})~~~~~~~~~~~~~~~~~\,(8,6,4)
\ee
\be
(234)(5678)_2(9\,10\,1\,2)_3~~~~~~~~~~~~(9,7,5)
\ee

\subsection{$n\!=\!11$}

\be
(\os{2}{\os{|}{3}}\,4\,5\,\os{7}{\os{|}{6}})_3\,(89\,10\,\,11)_2~~~~~~~~~~~~~~~~~(10,8,6,4)~
\ee
\be
(2\,3\,\os{5}{\os{|}{4}}\,6\,7)_3\,(7\!\!\!\!\!{\os{\,\,9\,\,10\,\,11}{\os{\backslash\,|\,/}{8}}}\!\!\!\!\!1)
~~~~~~~~~~~~~~~~~~~~~(10,8,5)~~~~
\ee
\be
(\os{3}{\os{|}{2}}\,\os{5}{\os{|}{4}}\,6\,7)_3\,(89\,10\,\,11)_2~~~~~~~~~~~~~~~~~(10,8,6,4)~
\ee
\be
(\!\!\os{2\,\,\,\,3}{\os{\backslash\,/}{4}}\!5\,\os{7}{\os{|}{6}})\,
(6\,8\!\!\!\os{10\,\,\,11}{\os{\backslash\,/}{9}}\!\!\!\!)~~~~~~~~~~~~~~~~~~~~~~~~~~(10,7,4)~~~\,
\ee
\be
(2\,3\!\os{5\,\,\,\,6}{\os{\backslash\,/}{4}}\!7)_3\,(7\,8\!\!\!\os{10\,\,\,11}{\os{\backslash\,/}{9}}\!\!\!1)_3
~~~~~~~~~~~~~~~~~~~\,(10,7,5)~~~\,
\ee
\be
(\os{3}{\os{|}{2}}\,\os{5\,\,\,6}{\os{\backslash\,/}{4}}7)\,(8\!\!\!\os{10\,\,\,11}{\os{\backslash\,/}{9}}\!\!\!1)
~~~~~~~~~~~~~~~~~~~~~~~~\,(10,6,4)~~~
\ee
\be
(2345)_2\,(\os{6}{\os{|}{7}}\,8\,9\,\os{11}{\os{|}{10}})_3~~~~~~~~~~~~~~~~~~~~(10,8,6,4)
\ee
\be
(23)(4567)_2\,(8\,9\,\os{11}{\os{|}{10}}\,1)_3~~~~~~~~~~~~~~~(10,8,6,4)
\ee
\be
(6\,7\,\os{9}{\os{|}{8}}\,10\,\,11)_3\,(11\,1\!\!{\os{3\,\,4\,\,5}{\os{\backslash\,|\,/}{2}}}\!\!)~~~~~~~~~~~~~~~\,(9,6,4)~~~~
\ee
\be
(2345)_2\,(\os{7}{\os{|}{6}}\,\os{9}{\os{|}{8}}\,10\,\,11)_3~~~~~~~~~~~~~~~~~~\,(10,8,6,4)
\ee
\be
(6\os{8\,\,\,9}{\os{\backslash\,/}{7}}\,\os{11}{\os{|}{10}})\,(10\,1\!\os{3\,\,\,\,4}{\os{\backslash\,/}{2}}\!\!)
~~~~~~~~~~~~~~~~~~~~~~(10,8,4)~~~
\ee
\be
(\!\!\os{2\,\,\,\,3}{\os{\backslash\,/}{4}}\!5\,6)\,(6\!\os{8\,\,\,\,9}{\os{\backslash\,/}{7}}\!10\,\,11)_3\,(11\,1\,4)
~~~~~~~~~\,(9,7,4)~~~~
\ee
\be
(\os{2}{\os{|}{3}}\,4\,5)\,(6789)_2\,(\os{10}{\os{|}{11}}\,1\,3)~~~~~~~~~~~~~~~\,(10,8,6,4)
\ee
\be
(2\,3\,\os{5}{\os{|}{4}})\,(4\,6\!\os{8\,\,\,\,9}{\os{\backslash\,/}{7}}\!\!)\,(10\,\,11\,1)_2~~~~~~~~~~~~~~(10,8,5)~~~
\ee
\be
(23)\,(\os{4}{\os{|}{5}}\,6\,7\,\os{9}{\os{|}{8}})_3\,(10\,\,11\,1)_2~~~~~~~~~~~~~\,(10,8,6,4)
\ee
\be
(4\,5\,\os{7}{\os{|}{6}}\,8\,9)_3\,(9\os{11\,\,\,1\,}{\os{\backslash\,/}{10}}\,\os{3}{\os{|}{2}})
~~~~~~~~~~~~~~~~~~~(10,7,4)~~~
\ee
\be
(23)\,(\os{5}{\os{|}{4}}\,\os{7}{\os{|}{6}}\,8\,9)_3\,(10\,\,11\,1)_2~~~~~~~~~~~~~\,(10,8,6,4)
\ee
\be
(4\!\os{6\,\,\,\,7}{\os{\backslash\,/}{5}}\,\os{9}{\os{|}{8}})\,(8\os{11\,\,\,1\,}{\os{\backslash\,/}{10}}2)
~~~~~~~~~~~~~~~~~~~~(10,8,6)~~~
\ee
\be
(234)\,(4\!\os{6\,\,\,\,7}{\os{\backslash\,/}{5}}\!8\,9)_3\,(9\os{11\,\,\,1\,}{\os{\backslash\,/}{10}}2)
~~~~~~~~~~\,(10,7,5)~~~
\ee
\be
(2\,3\!\os{5\,\,\,\,6}{\os{\backslash\,/}{4}}\!\!)\,\,
(\!\!\os{7\,\,\,\,8}{\os{\backslash\,/}{9}}\!10\,\,\os{1}{\os{|}{11}})~~~~~~~~~~~~~~~~~~~(9,7,5)~~~~
\ee
\be
(23)\,(\!\!\os{4\,\,\,\,5}{\os{\backslash\,/}{6}}\!7\,\os{9}{\os{|}{8}})\,(8\,10\,\,\os{1}{\os{|}{11}})
~~~~~~~~~~~~~~(9,6,4)~~~~
\ee
\be
(4\,5\!\os{7\,\,\,\,8}{\os{\backslash\,/}{6}}\!9)_3\,(9\,10\,\,\os{1}{\os{|}{11}}\,\,\os{3}{\os{|}{2}})_3
~~~~~~~~~~~~~(9,7,4)~~~~
\ee
\be
(\os{5}{\os{|}{4}}\,\os{7\,\,\,\,8}{\os{\backslash\,/}{6}}9)_3\,(10\,\,\os{1}{\os{|}{11}}\,\,\os{3}{\os{|}{2}})_3
~~~~~~~~~~~~~~~\,(8,6,4)~~~~
\ee
\be
(234)\,(4\!\!{\os{6\,\,7\,\,8}{\os{\backslash\,|\,/}{5}}}\!\!9)\,(9\,10\,\,\os{1}{\os{|}{11}}\,\,2)_3
~~~~~~~~~(9,7,5)~~~~
\ee
\be
(23)(4567)_2\,(\os{8}{\os{|}{9}}\,10\,\,11\,1)_3~~~~~~~~~~\,(10,8,6,4)
\ee
\be
(45)(6789)_2\,(10\,\,11\,1\,\os{3}{\os{|}{2}})_3~~~~~~~~~~\,(10,8,6,4)
\ee

\subsection{$n\!=\!12$}

\be
(\!\!\os{2\,\,\,\,3}{\os{\backslash\,/}{4}}5\os{7\,\,\,\,8}{\os{\backslash\,/}{6}}\!\!)\,(9\,10\,\,11\,\,12)_2
~~~~~~~~~~~~~(11,9,7,4)
\ee
\be
(2\,3\!\os{5\,\,\,\,6}{\os{\backslash\,/}{4}}\,\os{8}{\os{|}{7}})_3\,
(7\!\!\!\!\!{\os{10\,\,11\,\,12}{\os{\backslash\,|\,/}{9}}}\!\!\!\!\!1)~~~~~~~~~~~~~~~~~~(11,9,7,5)
\ee
\be
(\os{3}{\os{|}{2}}\,\os{5\,\,\,\,6}{\os{\backslash\,/}{4}}\!7\,8)_3\,
(8\!\!\!\!\!{\os{10\,\,11\,\,12}{\os{\backslash\,|\,/}{9}}}\!\!\!\!\!1)~~~~~~~~~~~~~~~~~~(11,9,6,4)
\ee
\be
(2345)_2\,(\!\!\os{6\,\,\,\,7}{\os{\backslash\,/}{8}}9\!\os{11\,\,\,12}{\os{\backslash\,/}{10}}\!\!)
~~~~~~~~~~~~~~~~~~~(11,8,6,4)
\ee
\be
(\os{3}{\os{|}{2}}\,{\os{5\,\,6\,\,7}{\os{\backslash\,|\,/}{4}}}\!8)\,(8\,9\!\os{11\,\,\,12}{\os{\backslash\,/}{10}}\!1)_3
~~~~~~~~~~~~~~~~(11,8,6,4)
\ee
\be
(6\,7\!\!\os{\,9\,\,\,10}{\os{\backslash\,/}{8}}\,\os{12}{\os{|}{11}})_3\,
(11\,1\!\!{\os{3\,\,4\,\,5}{\os{\backslash\,|\,/}{2}}}\!\!)~~~~~~~~~~~~~~~(11,9,6,4)
\ee
\be
(\os{7}{\os{|}{6}}\os{\,9\,\,\,10}{\os{\backslash\,/}{8}}\!\!11\,\,12)_3\,
(12\,\,1\!\!{\os{3\,\,4\,\,5}{\os{\backslash\,|\,/}{2}}}\!\!)~~~~~~~~~~~~~(10,8,6,4)
\ee
\be
(\!\!\os{2\,\,\,\,3}{\os{\backslash\,/}{4}}\!5\,6)\,
(6\!\!\!{\os{\,8\,\,9\,\,10}{\os{\backslash\,|\,/}{7}}}\,\os{12}{\os{|}{11}})\,(11\,1\,4)~~~~~~~~~\,(11,9,7,4)
\ee
\be
(2\,3\!\os{5\,\,\,\,6}{\os{\backslash\,/}{4}}\!\!)\,(789\,10)_2(11\,\,12\,\,1)_2~~~~~~~~~\,(11,9,7,5)
\ee
\be
(23)\,(\!\!\os{4\,\,\,\,5}{\os{\backslash\,/}{6}}7\!\os{\,9\,\,\,10}{\os{\backslash\,/}{8}}\!\!\!)\,(11\,\,12\,\,1)_2
~~~~~~~~~~~~~~(11,9,6,4)
\ee
\be
(4\,5\!\os{7\,\,\,\,8}{\os{\backslash\,/}{6}}\,\os{10}{\os{|}{9}})_3\,
(9\os{12\,\,\,1\,}{\os{\backslash\,/}{11}}\,\os{3}{\os{|}{2}})~~~~~~~~~~~~~~~~~~\,(11,9,7,4)
\ee
\be
(\os{5}{\os{|}{4}}\,\os{7\,\,\,\,8}{\os{\backslash\,/}{6}}\!9\,10)_3\,
(10\os{12\,\,\,1\,}{\os{\backslash\,/}{11}}\,\os{3}{\os{|}{2}})~~~~~~~~~~~~~~~~(10,8,6,4)
\ee
\be
(234)\,(4\!\!{\os{6\,\,7\,\,8}{\os{\backslash\,|\,/}{5}}}\,\os{10}{\os{|}{9}})\,
(9\os{12\,\,\,1\,}{\os{\backslash\,/}{11}}2)~~~~~~~~~~~~~~(11,9,7,5)
\ee
\be
(23)(4567)_2\,(\!\os{8\,\,\,\,9}{\os{\backslash\,/}{10}}\,11\,\,\os{1}{\os{|}{12}})~~~~~~~~~~~~~~(10,8,6,4)
\ee
\be
(45)(6789)_2\,(10\,\,11\,\,\os{1}{\os{|}{12}}\,\,\os{3}{\os{|}{2}})_3~~~~~~~~~~(10,8,6,4)
\ee

\subsection{$n\!=\!13$}

\be
(2345)_2(6789)_2(10\,\,11\,\,12\,\,13)_2~~~~~~~~~~~~~~(12,10,8,6,4)
\ee
\be
(\os{3}{\os{|}{2}}\,{\os{5\,\,6\,\,7}{\os{\backslash\,|\,/}{4}}}\,\os{9}{\os{|}{8}})\,
(8\!\!\!\!{\os{11\,\,12\,\,13}{\os{\backslash\,|\,/}{10}}}\!\!\!\!1)~~~~~~~~~~~~~~~~~~~~~~~~~~~~~~(12,10,8,6,4)
\ee
\be
(\os{7}{\os{|}{6}}{\os{\,\,9\,\,10\,\,11}{\os{\backslash\,|\,/}{8}}}\,\os{13}{\os{|}{12}})\,
(12\,\,1\!\!{\os{3\,\,4\,\,5}{\os{\backslash\,|\,/}{2}}}\!\!)~~~~~~~~~~~~~~~~~~~~~~~~~(12,10,8,6,4)
\ee
\be
(23)(4567)_2(89\,10\,\,11)_2(12\,\,13\,\,1)_2~~~~~~~~~\,(12,10,8,6,4)
\ee
\be
(\os{5}{\os{|}{4}}\,{\os{7\,\,8\,\,9}{\os{\backslash\,|\,/}{6}}}\,\os{11}{\os{|}{10}})\,
(10\os{13\,\,\,1\,}{\os{\backslash\,/}{12}}\,\os{3}{\os{|}{2}})~~~~~~~~~~~~~~~~~~~~~~~~~~(12,10,8,6,4)
\ee

\newpage

\end{document}